\documentclass[11pt]{article}  
 \pdfoutput=1 
\usepackage{jheppub}\usepackage{titlesec}
\usepackage{tikz, amsmath}
\usetikzlibrary{calc}
\usepackage{bbm} 
\usepackage{xcolor}
\usepackage{dsfont}
\usepackage{subfigure}
\usepackage{hyperref}  
\usepackage{microtype}   
\usepackage{caption}
\usepackage{amsmath}
\usepackage{float}
\usepackage{graphics}
\usepackage{psfrag}
\usepackage{color}
\usepackage{slashed}
\usepackage{subfigure}
\usepackage{graphicx}
\usepackage{array,multirow}
\usepackage{amsfonts} 
\usepackage{amssymb}
\usepackage{mathtools}
\usepackage{amsmath}
\usepackage{amsfonts}
\usepackage{mathrsfs}
\usepackage{multirow}
\usepackage{feynmp}
\usepackage{array}
\usepackage{textcomp} 
\usepackage{phaistos} 
\usepackage[utf8]{inputenc}
\usepackage{pifont} 
\usepackage{pgf} 

\usepackage{subfigure}
\usepackage{multirow}
   \usepackage{tabu}

\newcommand{\beq}{\begin{eqnarray}}
\newcommand{\eeq}{\end{eqnarray}}

\newcommand{\bmp}{\noindent\begin{minipage}{16cm}}
\newcommand{\emp}{\end{minipage}\vskip 7mm} 

\usepackage{dcolumn}
\usepackage{bm}
\usepackage{slashed}

\usepackage{epsfig}

\usepackage[ margin=5pt, font=normalsize,labelfont=bf,justification=raggedright]{caption}

\usepackage{hyperref}

\newcommand{\bea}{\begin{eqnarray}}
\newcommand{\eea}{\end{eqnarray}}

\newcommand{\ba}{\begin{eqnarray}}
\newcommand{\ea}{\end{eqnarray}}

\newcommand{\normord}[1]{:\mathrel{\mspace{1mu}#1\mspace{1mu}}:}
       
\title{\Large Coherence in scattering of massive \\ weakly interacting  neutral particles off nuclei}
\author[1]{V.A.~Bednyakov}
\emailAdd{bedny@jinr.ru}
\affiliation[1]{Dzhelepov Laboratory of Nuclear Problems, JINR, 141980, Dubna, Russia}

\abstract{ The paper presents a novel approach to the description of the nonrelativistic weak interaction of a massive neutral particle (lepton) and a nucleus, in which the latter retains its integrity. The cross section of such a process is a sum of the elastic (or coherent) contribution, when the nucleus remains in its original state, and the inelastic (incoherent) contribution, when the nucleus is in an excited state. Smooth transition from elastic scattering to inelastic scattering is governed by the dependence of the nuclear form factors on the momentum transferred to the nucleus. The intensity of the weak interaction is set by the parameters that determine the contributions to the probability amplitude from the scalar products of the leptonic and nucleon currents. The resulting expressions are of interest, at least in the problem of direct detection of neutral massive weakly interacting particles of dark matter, since in this case, in contrast to the generally accepted approach, both elastic and inelastic processes are simultaneously considered. It is shown that the presence of the inelastic contribution accompanied by emission of characteristic radiation (photons) from the deexcitation of the nucleus turns out to be decisive when the coherent cross section is strongly suppressed or cannot be detected. The former takes place if the corresponding interaction constant is close to zero or if the momentum transferred to the nucleus is too large and the coherence condition is not met. When the measurable recoil energy of the nucleus is below the detection threshold, the coherent cross section cannot be seen at all. In this situation, “inelastic”\/ photons are the only detectable signal of the interaction between dark matter particles and matter. Therefore in order to extract maximum information about dark matter particles, one should plan experiments aimed at the direct detection of dark matter particles in a setting that allows one to detect both the recoil energy of the nucleus and the gamma quanta from the deexcitation of the nucleus.}

\begin{document}
\maketitle

\section{\large  Introduction: from neutrino to massive neutral lepton}
\label{1chiA-IntroDuction} 
In \cite{Bednyakov:2018mjd,Bednyakov:2019dbl,Bednyakov:2021ppn}, an approach to description of the 
neutrino–nucleus $\nu A\to \nu A^{(*)}$ and antineutrino–nucleus $\bar\nu A\to \bar\nu A^{(*)}$
interactions was formulated and detailed for the case where the target nucleus $A^{}$ 
may remain in its initial state or undergo a transition to the excited state $A^{*}$ (with its integrity retained). 
The approach relies on the description of the nucleus as a bound state of its constituent nucleons based on the 
multiparticle wave function of the nucleus in the general form.
\par 
It was shown that an elastic interaction which preserves the initial nuclear quantum state leads to quadratic enhancement 
of the corresponding observed cross section in terms of the number of nucleons. 
At the same time, the total cross section of all other (possible) inelastic processes accompanied by a change in 
the nuclear quantum state generally demonstrates only linear dependence on the number of nucleons. 
In addition, it was demonstrated that the behavior of elastic and inelastic cross sections is determined by the
 factors   $|F_{p/n}(\bm{q})|^2$ and  $(1-|F_{p/n}(\bm{q})|^2)$, respectively, where  $F_{p/n}(\bm{q})$  is the nuclear form factor of the proton/neutron normalized to unity. 
 These form factors govern a smooth transition from the regime of elastic (coherent) (anti)neutrino–nucleus scattering to 
 the inelastic (incoherent) regime. 
 In the general case, where the three-momentum $\bm{q}$   transferred to the nucleus is neither small nor large, the coherent and incoherent contributions to the total cross section should be considered simultaneously.

It was also noted that owing to their common nature resulting from weak neutral currents, 
the elastic and inelastic neutrino processes $\nu(\bar\nu) A$ are experimentally indistinguishable 
when the only observable is the target nucleus recoil energy. 
Therefore, in experiments aimed (at rather high energies) to study coherent (anti)neutrino scattering, there can be an incoherent background indistinguishable from the signal when the $\gamma$ deexciting the nucleus are undetectable. 
For example, if in the COHERENT experiment 
\cite{Akimov:2017ade} with the ${}^{133}\text{Cs}$  nucleus at the (anti)neutrino energy of
30–50 MeV nuclear deexcitation $\gamma$ quanta fail to be detected, the data from this 
experiment contain the desired CE$\nu$NS events with an unavoidable inelastic (incoherent) admixture at a level of 15–20\%.
\par
On the other hand, the incoherent contribution can be directly measured (estimated) by purposeful detection of photons emitted by target nuclei excited in inelastic processes \cite{Bednyakov:2021bty}
 with the number of these photons being proportional to the inelastic-to-elastic $\nu A$ scattering ratio. 
These photons should be time-correlated with the incident neutrino beam and have a range of energies typical of the nucleus $A$, which are usually much higher than the nuclear recoil energy, thus noticeably simplifying their detection. 
Simultaneous detection of the recoil energy and the nuclear deexcitation 
$\gamma$ energy will make it possible to separate the “pure” CE$\nu$NS contribution, 
and, consequently, to study the nuclear structure and carry out the precision search for new physics in neutrino processes. 
\par
Obviously, the approach to the scattering of neutrinos and antineutrinos off nuclei used in 
\cite{Bednyakov:2018mjd,Bednyakov:2019dbl,Bednyakov:2021ppn,Bednyakov:2021bty}
can be generalized to the case where a massive neutral particle is considered instead of a massless (anti)neutrino 
and where the rather weak interaction of this particle with nucleons of the nucleus is described by the 
phenomenological Lagrangian taking into account possible Lorentz structures (scalar, pseudoscalar, vector and axial vector). 
The relevant importance of considering this case follows, for example, from the necessity of correctly understanding the balance of elastic and inelastic interaction in the problem of direct detection of massive weakly interacting neutral particles of dark matter via their (nonrelativistic) scattering from target nuclei of the corresponding detector.
\par
Note that CE$\nu$NS-events considered in 
\cite{Bednyakov:2018mjd,Bednyakov:2019dbl,Bednyakov:2021ppn,Bednyakov:2021bty}
are an irremovable background for experiments aimed at direct detection of dark matter particles under terrestrial conditions (see, for example, \cite{Bednyakov:2015uoa,Papoulias:2018uzy,Boehm:2018sux,Cooley:2021rws,Cebrian:2022brv}, 
since the main and so far, the only signature of both processes is the target nucleus recoil energy. 
For this reason, it seems expedient to consider these two processes, which proved to be interrelated, within a single approach.
\par
In addition, the above-mentioned analysis of coherence–incoherence balance in these processes at neutrino energies of 30 to 50 MeV suggests that the role of inelastic (incoherent) channels in the interaction of massive neutral particles with nuclei could well be more significant, especially in view of new physics effects.
\par
The long absence of positive results\footnote{
Except the DAMA/LIBRA results\cite{Bernabei:2022loo}, which are almost ignored without adequate refutation.}
from a wide range of dark matter search experiments gave rise to a lot of new, 
sometimes exotic suggestions about both the possible composition of dark matter itself 
\cite{Hurtado:2020vlj,Du:2020ldo,Baryakhtar:2020rwy,Majumdar:2021vdw,Afek:2021vjy}, its unusual properties \cite{Giudice:2017zke,Zurowski:2020dxe,Wang:2021jic,Feng:2021hyz,Emken:2021vmf,Granelli:2022ysi,Filimonova:2022pkj,Bell:2022yxn},  and about more sophisticated methods of its detection
\cite{Tsuchida:2019hhc,Coskuner:2021qxo,Boos:2022gtt,Flambaum:2022zuq,Fan:2022uwu,Blanco:2022cel,Billard:2022cqd,Araujo:2022wjh}.  
 Nevertheless, due to its exceptional significance
 \cite{Bednyakov:2015uoa,Bednyakov:2020njj,Slatyer:2021qgc,Cebrian:2022brv,Aboubrahim:2022lwb}, 
 it seems somewhat untimely to bury in oblivion the traditional method of directly detecting massive weakly 
 interacting dark-matter particles without critical analysis of fundamental assumptions and details accepted in this method.
 \par
Thus, the first objective of this work is to generalize the above {\em massless neutrino}\/ approach 
\cite{Bednyakov:2018mjd,Bednyakov:2019dbl,Bednyakov:2021ppn,Bednyakov:2021bty}
to the case of interaction between {\em massive}\/ neutral weakly
interacting $\chi$-particles and nuclei $\chi A\to \chi A^{(*)}$ in {\em the nonrelativistic} approximation. 
The second objective is to investigate those regions of parameters (including kinematic ones) 
where the inelastic (incoherent) process $\chi A\to \chi A^{*}$ can noticeably compete with the 
elastic (coherent) process $\chi A\to \chi A^{}$, which is traditionally believed to be generally dominant. 
The presence of regions where the inelastic channel plays a leading part, especially in view of the fact that the character of 
the $\chi A$ interaction is usually beyond the Standard Model, may explain “blindness”\/ of dark matter detectors 
tuned to search $\chi A$ scattering events only in the elastic channel.

\section{\large Kinematics and cross section of elastic and inelastic ${\chi A}$ scattering}
\label{2chiA-Kinematics}
\begin{figure}[h]
\includegraphics[scale=1.3]{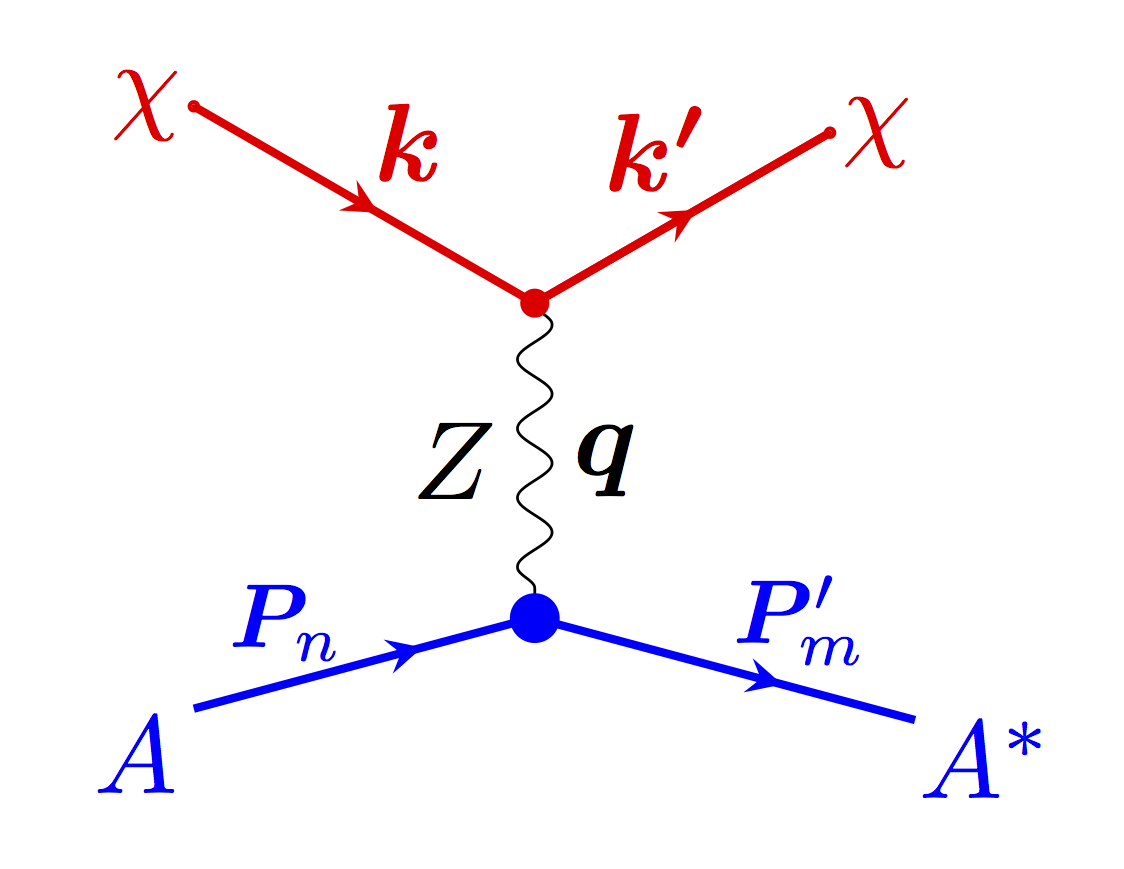} 
\includegraphics[scale=0.3]{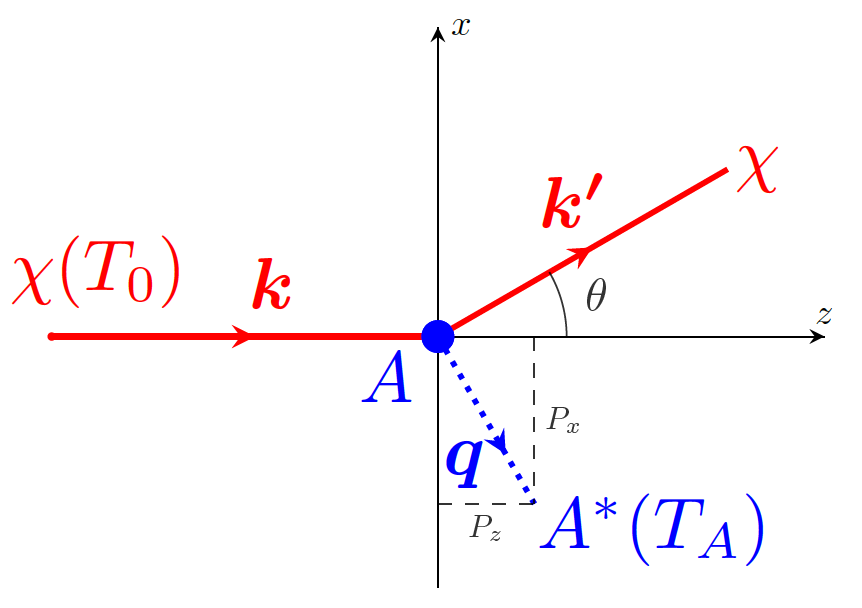}
\caption{\small “Diagram"\/ of the $\chi A$ interaction due to, for example, the neutral $Z$ boson exchange (left).
Kinematics of this process in the lab frame where the nucleus $A$ is at rest (right).} 
\label{fig:DiagramCENNS}
\end{figure}

In the interaction of two particles with the formation of two particles ($2\to 2$ process): 
\begin{equation} \label{eq:2DM-Kinematics-2-to-2-Definition}
\chi(k)+A(P_n)\rightarrow \chi(k' )+A^{(*)}(P'_m),
\end{equation}
the four-momenta of the incident and outgoing neutral massive leptons ($\chi$ particles) are denoted by 
$k=(k_0=E_\chi,\bm{k})$ and $k'=(k'_0=E_\chi',\bm{k}')$,
and the four-momenta of the initial and final nuclear states are
denoted by $P_n=(P^0_n,\bm{P}_n)$ and $P'_m=(P^0_m, \bm{P}_m)$  (see Fig.~\ref{fig:DiagramCENNS}, left). 
The total energy of the nuclear state $|P_n\rangle$  is $P_n^0 = E_{\bm{P}}+\varepsilon_n$,
where $\varepsilon_n$  is the internal energy of the $n$-th nuclear quantum state. 
The expression for the cross section of process (\ref{eq:2DM-Kinematics-2-to-2-Definition}) has the form (see, for example, 
\cite{Tanabashi:2018oca,Peskin:1995ev,Bilenky:1995zq,Bednyakov:2021ppn,Bednyakov:2018mjd})
\footnote{The squared matrix element $|\mathcal{M}_{mn}|^2$ usually does not depend on the $\phi$-angle. 
Integration is performed over it.}:
 \begin{equation}\label{eq:2DM-Kinematics-CrossSection-2-to-2-Definintion}
d\sigma =-\frac{\delta(k_0+P^0_n-k'_0-P^0_m)|{\cal M}|^2|\bm{k}'|^2 d|\bm{k}'| d\cos\theta}{32\pi k'_0 P^0_m \sqrt{(k P_n)^2-m_\chi^2 m_A^2}}.
\end{equation}
If the $\chi$ particle with mass $m_\chi$  and momentum $\bm{k}$ impinges on the nucleus $A$ at rest along the $z$ axis 
and moves away with the momentum $\bm{k}'$ at an angle $\theta$ with the $x$ axis (the $y$ axis can always be chosen to be 
perpendicular to the scattering plane), the four-momenta can be written as
\begin{eqnarray*}
k&=&\Big(k_0=\sqrt{m^2_\chi+|\bm{k}|^2}, 0, 0, k_z=|\bm{k}|\Big),   \qquad 
P_n =  \Big(P_n^0=m_A +\varepsilon_n, 0, 0, 0\Big),   \\
k'&=&\Big(k'_0=\sqrt{m^2_\chi +  |\bm{k}'|^2}, k'_x= |\bm{k}'|\sin\theta, 0, k'_z =  |\bm{k}'|\cos\theta\Big),  \\
P'_m&=&\Big(P_m^0=\varepsilon_m + \sqrt{m_A^2 + (P_m^x)^2 + (P_m^z)^2}, P_m^x, 0, P_m^z\Big)= \\ 
&=& \Big(P_m^0=\varepsilon_m + \sqrt{m_A^2 + \bm{q}^2}, - |\bm{k}'|\sin\theta, 0,  |\bm{k}|-  |\bm{k}'|\cos\theta\Big),
\end{eqnarray*} 
where $m_A$  is the mass of the nucleus $A$, and $\varepsilon_m$ is the 
excitation energy of the $m$-th level (state) of this nucleus. 
Here allowances are made for conservation of the momentum components along the $z$ axis: 
$ k'_z + P_m^z = k_z$ (or $P_m^z = |\bm{k}|-  |\bm{k}'|\cos\theta$) and the $x$ axis:
$k'_x + P_m^x = 0$ (or $P_m^x= -  |\bm{k}'|\sin\theta$),  
and for the fact that the four-momentum transferred to the nucleus $q =(q_0, \bm{q})$ is related to these quantities as follows:
\begin{eqnarray} \nonumber \label{eq:2DM-Kinematics-momentum-transfer}
q^2&\equiv &(k-k')^2 = 2\big(m^2_\chi- (kk')\big) = 
2 \big(m^2_\chi -\sqrt{(m^2_\chi +  |\bm{k}'|^2)(m^2_\chi +  |\bm{k}|^2)}  + |\bm{k}| |\bm{k}'|\cos\theta\big), 
\\ q_0 &=& k_0 -k'_0 = P^0_m - P^0_n = \Delta\varepsilon_{mn}+ T_A, 
\\ \bm{q}^2 &=& (\bm{k}-\bm{k}')^2 = (-|\bm{k}'|\sin\theta)^2+(|\bm{k}|-  |\bm{k}'|\cos\theta)^2=
 |\bm{k}|^2 + |\bm{k}'|^2 - 2 |\bm{k}| |\bm{k}'|\cos\theta. 
\nonumber\end{eqnarray} 
The kinetic energy of the recoil nucleus motion $T_A$ is defined as
\begin{equation} \label{eq:2DM-Kinematics-Recoil-kinetic-energy}
T_A=\sqrt{m^2_A+\bm{q}^2}-m_A
.\end{equation}
The energy conservation law from (\ref{eq:2DM-Kinematics-CrossSection-2-to-2-Definintion}), i.e., the equality $k_0 + P_n^0=k'_0 + P_m^0$, can be recast in the form
\begin{equation}\label{eq:2DM-Kinematics-Kinematics-EnergyConservation}
k_0+m_A-\Delta\varepsilon_{mn}=\sqrt{m^2_\chi +  |\bm{k}'|^2}+ \sqrt{m^2_A+|\bm{k}'|^2+|\bm{k}|^2  - 2 |\bm{k}| |\bm{k}'|\cos\theta} 
, \end{equation} 
where the difference of the energies of the nuclear states  $|m\rangle$ and $|n\rangle$  is denoted as follows:
\begin{equation}\label{eq:2DM-Kinematics-delta_Epsilon}
\Delta\varepsilon_{mn} \equiv \varepsilon_m-\varepsilon_n.
\end{equation}
Equation (\ref{eq:2DM-Kinematics-Kinematics-EnergyConservation}) in the laboratory frame gives dependence of the momentum $|\bm{k}'|$ on the angle  $\theta$  between the vectors $\bm{k}$ and $\bm{k}'$. 
\par
Since in the laboratory frame the nucleus before the interaction is supposed to be at rest and in a certain quantum state $|n\rangle$ and considering the above kinematic relations, cross section (\ref{eq:2DM-Kinematics-CrossSection-2-to-2-Definintion}) can be written as
 \begin{equation} \label{eq:2DM-Kinematics-CrossSection-DM-n-2-DM-m}
\frac{d^2\sigma_{mn}}{d|\bm{k}'| d\cos\theta}
= \frac{-|{i\cal M}_{mn}|^2 |\bm{k}'|^2 }{2^5\pi\sqrt{m^2_\chi +  |\bm{k}'|^2}}
\frac{\delta\Big(k_0-\sqrt{m^2_\chi +  |\bm{k}'|^2}-\Delta\varepsilon_{mn}-T_A(|\bm{k}'|,\cos\theta)\Big)}
{(m_A+\varepsilon_m+T_A(|\bm{k}'|)) \sqrt{k^2_0(m_A+\varepsilon_n)^2 -m_\chi^2 m_A^2}}.
\end{equation}
Here the following expressions for the square of the scalar product are used:
$$(k P_n)^2=(k_0\, P^0_n)^2=k^2_0(m_A+\varepsilon_n)^2, \quad k_0=(m^2_\chi +  |\bm{k}|)^{1/2}, 
\ \text{~and~} \ q_0= \Delta\varepsilon_{mn}+ T_A(|\bm{k}'|, \cos\theta).$$ 
From the energy conservation delta function of (\ref{eq:2DM-Kinematics-CrossSection-DM-n-2-DM-m}), 
it also follows that
 \begin{equation}\label{eq:2chiA-Kinematics-T_A-vs-T_0-etc}
T_A(|\bm{k}'|,\cos\theta)=k_0-k'_0-\Delta\varepsilon_{mn}
=\sqrt{m^2_\chi +  |\bm{k}|}-\sqrt{m^2_\chi +  |\bm{k}'|^2}-\Delta\varepsilon_{mn}.
\end{equation}
As is known \cite{Bednyakov:2015uoa,Bertone:2004pz,Drukier:1986tm,Freese:1987wu,Lewin:1996rx}, 
the velocity of massive dark-matter particles near the Earth is estimated at
about $10^{-3}$  of the speed of light, i.e.,
 $ |\bm{v}|={|\bm{k}|}/{m_\chi} \simeq 10^{-3}c\simeq 300\div 400~$km/s. 
 Therefore, the kinetic energy of these particles when incident on the target,
\begin{equation}\label{eq:2chiA-Kinematics-T_0}
T_0\equiv \dfrac{|\bm{k}|^2}{2 m_\chi} =\dfrac{|\bm{k}|^2}{m^2_\chi}\dfrac{m_\chi}{2 }
\simeq 10^{-6}  \dfrac{m_\chi c^2}{2 },   
\end{equation}
ranges from 1 keV to 10 MeV, because it is now believed that the mass of a nonrelativistic particle of
(cold) dark matter is between few GeV/$c^2$ and a few tens of TeV/$c^2$.
In addition, for the square of the momentum transferred to the nucleus $q^2$ there is an approximation
\begin{eqnarray*}
q^2&=& - 2m^2_\chi \Bigg(\sqrt{\Big(1 +  \frac{|\bm{k}'|^2}{m^2_\chi}\Big)\Big(1 + \frac{|\bm{k}|^2}{m^2_\chi}}\Big)   -1 - \frac{ |\bm{k}| |\bm{k}'|}{m^2_\chi}\cos\theta\Bigg)  \simeq - \bm{q}^2  ,\\ \bm{q}^2 &=& (\bm{k}-\bm{k}')^2 = |\bm{k}|^2 + |\bm{k}'|^2 - 2 |\bm{k}| |\bm{k}'|\cos\theta \simeq  2 m_A T_A. 
\end{eqnarray*}  
The latter approximation follows from (\ref{eq:2DM-Kinematics-Recoil-kinetic-energy}) if $\bm{q}^2\ll m^2_A$, 
which holds (very well) in problems of direct detection of dark matter, where it is usually assumed that 
$m_A\ge 10$ GeV$/c^2$, and nuclear recoil energies are
no higher than 100–150 keV, i.e., $T_A \le 1.5\times 10^{-4}$ GeV. 
A typical excitation energy difference is in the region of a few hundreds of keV, i.e., being undoubtedly
$\Delta\varepsilon_{mn}\le 1\,$MeV.
\par
Thus, the problem of searching for and detecting dark-matter particles using 
$\chi A$ scattering involves the following typical scales of quantities:
\begin{eqnarray}\nonumber \label{eq:2DM-Kinematics-DDMD-scales}
\frac{|\bm{k}|}{m_\chi} \simeq 10^{-3}, &\quad& \Delta\varepsilon_{mn}\le 1 \text{~MeV}, \quad 
10 \le m_A \le 100 \text{~GeV}/c^2, \quad 1<m_\chi< 10^4 \text{~GeV}/c^2
,\\ && T_A \le 150 \text{~keV}, \quad  |\bm{q}| \le 0.2   \text{~GeV}/c,  \ \quad  1 \text{~keV} \le T_0 \le 10 \text{~MeV}
.\end{eqnarray}
Therefore, it is sufficient to use the nonrelativistic approximation here. 
It also seems applicable to problems of detecting candidates for dark-matter particles produced at accelerators
\cite{Krnjaic:2022ozp,Boos:2022gtt,Kim:2017qdi} or accelerated in space 
\cite{Bardhan:2022ywd,CDEX:2022fig,Xia:2022tid,Granelli:2022ysi,Wang:2021jic}, 
when at their rather high energies $T_0$ the momentum $|\bm{q}|$ transferred to the nucleus 
is still insufficient for complete breakup of the nucleus.
\par
The $\chi A\to \chi A^{(*)}$ scattering cross section (\ref{eq:2DM-Kinematics-CrossSection-DM-n-2-DM-m}) in the nonrelativistic approximation, i.e., at
$$k_0=(m^2_\chi +  |\bm{k}|^2)^{1/2}\simeq m_\chi + \dfrac{|\bm{k}|^2}{2 m_\chi} = m_\chi + T_0, 
\quad  \sqrt{m^2_\chi +  |\bm{k}'|^2}\simeq m_\chi + \dfrac{|\bm{k}'|^2}{2 m_\chi},\quad \text{and}$$ 
\begin{equation}\label{eq:2DM-Kinematics-Kinetic-energy-of-nucleus-via-k-k'-nonrel}
T_A(|\bm{k}'|, \cos\theta) \simeq  \dfrac{\bm{q}^2}{2 m_A} = \dfrac{|\bm{k}'|^2+|\bm{k}|^2  - 2 |\bm{k}| |\bm{k}'|\cos\theta}{2m_A} 
, \end{equation} 
takes the form
\begin{equation}\label{eq:2DM-Kinematics-CrossSection-DM-n-2-DM-m-nonrel} 
\frac{d^2\sigma_{mn}}{d|\bm{k}'| d\cos\theta}\!
= \frac{-|{i\cal M}_{mn}|^2 |\bm{k}'|^2 }{2^5\pi m_A |\bm{k}|\sqrt{1 + \dfrac{m_\chi}{m_A }\dfrac{\varepsilon_n}{T_0} }} \frac{\delta\Big(T_0\!-\!\dfrac{|\bm{k}'|^2}{2 m_\chi}\!-\!\Delta\varepsilon_{mn}\!-\! \dfrac{|\bm{k}'|^2\!+\!|\bm{k}|^2\! -\!2 |\bm{k}| |\bm{k}'|\cos\theta}{2m_A}
\Big)}{\big(m_A+\varepsilon_m+T_A(|\bm{k}'|, \cos\theta)\big)
\Big(m_\chi + \dfrac{|\bm{k}'|^2}{2 m_\chi} \Big)}. \qquad
\end{equation}
Here a transition is made from the relativistic expression for the initial particle flux
\begin{eqnarray}\label{eq:2DM-Kinematics-Flux-rel}
\sqrt{w}\equiv \sqrt{k^2_0(m_A+\varepsilon_n)^2 -m_\chi^2 m_A^2}
= m_A |\bm{k}|  \Big[1+\dfrac{\varepsilon_n }{m_A}\big(2+\dfrac{\varepsilon_n}{ m_A}\big)
+\dfrac{m^2_\chi \varepsilon_n }{|\bm{k}|^2 m_A} \big(2 +\dfrac{\varepsilon_n}{m_A}\big) \Big]^{1/2}
\end{eqnarray}  
to its nonrelativistic version in the form
\begin{equation}\label{eq:2DM-Kinematics-Flux-unrel}
\sqrt{w} =  \sqrt{(m_\chi +T_0)^2(m_A+\varepsilon_n)^2-m_\chi^2 m_A^2}\simeq 
m_A |\bm{k}|\sqrt{1 + \frac{m_\chi}{m_A }\frac{\varepsilon_n}{T_0}} = m_A |\bm{k}|.
\end{equation}
The last equality in (\ref{eq:2DM-Kinematics-Flux-unrel})  is valid if the nuclear state $|n \rangle$ 
is the ground state with the minimal energy, i.e.,  $\varepsilon_n=0$.
On the one hand, the nuclear recoil kinetic energy $T_A$ from (\ref{eq:2chiA-Kinematics-T_A-vs-T_0-etc})
 enters into the relation
\begin{equation}\label{eq:2chiA-Kinematics-T_A-vs-T_0-etc-nonrel}
T_A(|\bm{k}'|,\cos\theta)=T_0 - \dfrac{|\bm{k}'|^2}{2 m_\chi} - \Delta\varepsilon_{mn}
.\end{equation}
On the other hand, according to (\ref{eq:2DM-Kinematics-Kinetic-energy-of-nucleus-via-k-k'-nonrel}), it is a function of 
two independent variables $|\bm{k}'|$ and $\cos\theta$, and, therefore, the energy-conserving delta function in 
 (\ref{eq:2DM-Kinematics-CrossSection-DM-n-2-DM-m-nonrel}) 
also simultaneously depends on these two variables. 
The expression for the nonrelativistic cross section (\ref{eq:2DM-Kinematics-CrossSection-DM-n-2-DM-m-nonrel}),
 like the initial formula (\ref{eq:2DM-Kinematics-CrossSection-DM-n-2-DM-m}), can be integrated either over 
 $|\bm{k}'|$ or over $\cos\theta$ using this delta function 
 (connecting the independent $|\bm{k}'|$ and $\cos\theta$). 
 As a result, the delta function disappears, and there remains only one independent variable that determines the 
 differential cross section for the process of our interest. 
 It becomes more convenient to use the energy conservation delta function for integrating the differential cross section 
 (\ref{eq:2DM-Kinematics-CrossSection-DM-n-2-DM-m-nonrel}) over $\cos\theta$. 
 When $|\bm{k}'|$ is not yet connected to $\cos\theta$, the delta function from 
 \ref{eq:2DM-Kinematics-CrossSection-DM-n-2-DM-m-nonrel}) has the form
\begin{equation}\label{eq:2DM-Kinematics-Delta-function-4-cosT}
\delta\Big(T_0\!-\dfrac{|\bm{k}'|^2}{2 m_\chi}\!-\Delta\varepsilon_{mn}\!-\dfrac{|\bm{k}'|^2\!+|\bm{k}|^2\!  - 2 |\bm{k}| |\bm{k}'|\cos\theta}{2m_A} \Big)\equiv \delta(f(\cos\theta))=\frac{m_A\delta(\cos\theta-\cos\theta_i)}{|\bm{k}'| |\bm{k}|}
,\end{equation}
\begin{equation}\label{eq:2DM-Kinematics-CosT-vs-k-prime-from-Delta-function}
\text{where}\qquad
\cos\theta_i =-\dfrac{2m_A\Big(T_0- \dfrac{|\bm{k}'|^2}{2 m_\chi}-\Delta\varepsilon_{mn}\Big)
-|\bm{k}'|^2-|\bm{k}|^2}{2 |\bm{k}| |\bm{k}'|}\qquad 
\end{equation} 
is the solution of the nonrelativistic equation for energy conservation $ f(\cos\theta_i)=0$. 
“Inverse”\/ of formula (\ref{eq:2DM-Kinematics-CosT-vs-k-prime-from-Delta-function}) has the form
\begin{eqnarray*} 
|\bm{k}'| =|\bm{k}| \frac{r \cos\theta_i + \sqrt{1-r^2 \sin^2\theta_i-\alpha (1+r) }}{1+r}
, \quad \text{where}\quad  
r=\dfrac{m_\chi}{m_A}, \quad \alpha=\dfrac{\Delta\varepsilon_{mn}}{T_0} 
.\end{eqnarray*}
Integration of the cross section (\ref{eq:2DM-Kinematics-CrossSection-DM-n-2-DM-m-nonrel})  
over $\cos\theta$  using (\ref{eq:2DM-Kinematics-Delta-function-4-cosT}) leads to the expression
\begin{eqnarray} \label{eq:2DM-Kinematics-dSigma-po-dk-prime-nonrel} 
\frac{d\sigma_{mn}}{d|\bm{k}'|}  &=& 
 \frac{1}{2^5\pi |\bm{k}|^2\sqrt{1 + \dfrac{m_\chi}{m_A }\dfrac{\varepsilon_n}{T_0} }}
\frac{-|{i\cal M}_{mn}|^2 |\bm{k}'| }{(m_A+\varepsilon_m+T_A(|\bm{k}'|, \cos\theta_i))}
\frac{1}{\Big(m_\chi + \dfrac{|\bm{k}'|^2}{2 m_\chi} \Big) }
, \end{eqnarray}
where dependence of $\cos\theta_i$ on  $|\bm{k}'|$ is given by formula 
(\ref{eq:2DM-Kinematics-CosT-vs-k-prime-from-Delta-function}).
\par
The next step in the transformation of the formula for the cross section is a transition to differentiation
with respect to the observable $\dfrac{d \sigma_{mn}}{d T_A} = \dfrac{d \sigma_{mn}}{d|\bm{k}'|} \dfrac{d |\bm{k}'|}{d T_A} $; i.e.,
the transition Jacobian $\dfrac{d |\bm{k}'|}{d T_A} $ should be found. 
This can be done in two ways. 
One is to explicitly (and tediously) calculate the kinetic energy derivative from formula (\ref{eq:2DM-Kinematics-Kinetic-energy-of-nucleus-via-k-k'-nonrel})
\begin{equation}\label{eq:2DM-Kinematics-Diff-Kinetic-energy-of-nucleus-via-k-k'-nonrel}
\dfrac{d T_A(|\bm{k}'|, \cos\theta(|\bm{k}'|))}{d |\bm{k}'|} 
= \dfrac{1}{m_A} \Big\{|\bm{k}'| - |\bm{k}|\cos\theta - |\bm{k}| |\bm{k}'| \dfrac{d \cos\theta}{d |\bm{k}'| } \Big\}
\end{equation}
by substituting into (\ref{eq:2DM-Kinematics-Diff-Kinetic-energy-of-nucleus-via-k-k'-nonrel})
expression (\ref{eq:2DM-Kinematics-CosT-vs-k-prime-from-Delta-function})  for $\cos\theta_i$ as a function of $|\bm{k}'|$. 
The other is to simply differentiate the energy conservation law (\ref{eq:2chiA-Kinematics-T_A-vs-T_0-etc-nonrel})
written as
\begin{equation}\label{eq:2DM-Kinematics-Chi-prime-energy-nonrel}
\dfrac{|\bm{k}'|^2}{2 m_\chi}=T_0-\Delta\varepsilon_{mn}-T_A.
\end{equation}
In both cases, there arises a simple formula
\begin{equation} \label{eq:2DM-Kinematics-Diff-Kinetic-energy-of-nucleus-via-k-k'-nonrel-final}
\dfrac{d T_A(|\bm{k}'|)}{d |\bm{k}'|} = -\dfrac{|\bm{k}'|}{m_\chi}.
\end{equation}
As a result, in view of 
(\ref{eq:2DM-Kinematics-dSigma-po-dk-prime-nonrel}), (\ref{eq:2DM-Kinematics-Chi-prime-energy-nonrel}) and
(\ref{eq:2DM-Kinematics-Diff-Kinetic-energy-of-nucleus-via-k-k'-nonrel-final}) 
the $\chi A_n\to \chi A_m $ cross section (\ref{eq:2DM-Kinematics-CrossSection-DM-n-2-DM-m-nonrel}) 
(in the nonrelativistic approximation) takes the form
\begin{eqnarray}                                 \label{eq:2DM-Kinematics-dSigma-po-d-T-A-nonrel}
\dfrac{d \sigma_{mn}}{d T_A} \big(\chi A_n\to \chi A_m\big) 
&=&\dfrac{|{i\cal M}_{mn}|^2  }{2^6\pi T_0 m_\chi m_A } C^{\rm nonrel}_{mn}(T_A) , 
\end{eqnarray}
which involves the introduced (kinematic) coefficient of about a unity
\begin{eqnarray}\label{eq:2DM-Kinematics-dSigma-po-d-T-A-nonrel-Coeff}
C^{\rm nonrel}_{mn}(T_A)&=& \dfrac{1}{\sqrt{1 + \dfrac{m_\chi}{T_0}\dfrac{\varepsilon_n}{m_A } }}
\frac{1}{1+\dfrac{\varepsilon_m+T_A}{m_A}}
\frac{1}{1  + \dfrac{T_0-\Delta\varepsilon_{mn}-T_A}{m_\chi} } 
\equiv C^{}_{2,nm}
\simeq O(1). \qquad
\end{eqnarray}
Considering relations  (\ref{eq:2DM-Kinematics-DDMD-scales}), 
it can be shown that this coefficient is really little different from unity and almost independent of either 
$T_A$, or subscripts $n$  $m$.
\par
In the experimental situation (or in calculations of the expected cross section) the initial external quantity
is recoil energy $T_A$. 
Therefore, specifying a certain value of $T_A$, initial $\chi$-particle energy $T_0$, and nuclear
characteristic  $\Delta\varepsilon_{mn}$, one finds from the energy 
conservation law (\ref{eq:2chiA-Kinematics-T_A-vs-T_0-etc-nonrel})
 the kinetic energy of the outgoing  $\chi$ particle  (\ref{eq:2DM-Kinematics-Chi-prime-energy-nonrel}) or 
  $|\bm{k}'(T_A)|^2 = 2 m_\chi (T_0-\Delta\varepsilon_{mn}-T_A)$. 
With {\em this}\/ known value of $|\bm{k}'|^2$ and “its generating”\/ value of $T_A$, 
one derives from the definition of the nonrelativistic nuclear recoil energy, 
formula (\ref{eq:2DM-Kinematics-Kinetic-energy-of-nucleus-via-k-k'-nonrel}), 
an expression for the outgoing angle of the $\chi$-particle in the lab frame
as a function of $T_A$, $\Delta\varepsilon_{mn}$ , and $T_0$ as follows:
\begin{eqnarray} \label{eq:2chiA-Kinematics-CosT-from-k2prime-and-T_A}
 \cos\theta(T_A) &=&  \dfrac{|\bm{k}'|^2+|\bm{k}|^2-2 m_A T_A}{2 |\bm{k}|  |\bm{k}'|}  = \dfrac{m_\chi( 2 T_0- \Delta\varepsilon_{mn})  - T_A (m_\chi +m_A)}{2 m_\chi\sqrt{ T_0 (T_0-\Delta\varepsilon_{mn}-T_A)}} 
.\end{eqnarray}

\section{\large The amplitude of ${\chi}$ particle-nucleus scattering}
\label{3chiA-ScatteringAmplitude}
The formalism of obtaining the amplitude of the scattering of a massive weakly interacting neutral ${\chi}$ particle from a nucleus 
(as a compound complex system) is a generalization of the approach 
\cite{Bednyakov:2018mjd,Bednyakov:2019dbl,Bednyakov:2021ppn}
first proposed in \cite{Bednyakov:2018mjd} for the description of the neutrino-nucleus scattering.
\par
To construct the scattering amplitude (Fig~\ref{fig:DiagramCENNS}, left), 
we will describe the initial and the final nucleus (as a bound compound system) by the wave function $|P_{l}\rangle$. 
It corresponds to a nucleus with four-momentum $P_{l}$, in a certain $l$-th internal quantum state  ($l=n,m$) 
and is a superposition of free nucleons $|\{p\}\rangle$, “weighted” with
the general wave function of the bound state $\tilde{\psi}'_n(\{p\})$.
The latter is a product of the wave function $\widetilde{\psi}_n(\{p{^\star}\})$, 
describing the internal structure of the nucleus in its rest frame (the corresponding momenta are marked by
the superscript $\star$)  and the wave function $\Phi_n(p)$ 
responsible for the motion of a nucleus as a whole with
$\bm{p}=\sum^A_{i=1}\bm{p}_i$ and the nuclear spin projection  $s$
\begin{equation}		\label{eq:3chiA-ScatteringAmplitude-1}
\widetilde{\psi}'_n(\{p\}) = \widetilde{\psi}_n(\{p{^\star}\})\Phi_n(p), 
\quad \text{where} \quad p=(\bm{p},s).
\end{equation}
The “internal”\/ wave function depends on $A-1$ three-momenta since the sum of all three-momenta 
is equal to the total momentum of a nucleus as an entire object. 
Thus, for the state $|P_n\rangle$  we will use the (antinsymmetrized) expression
\begin{equation}\label{eq:3chiA-ScatteringAmplitude-nucleus_state}
|P_n\rangle=\int\left(\prod^{A}_{i}d\widetilde{\bm{p}}^\star_i\right)\frac{\widetilde{\psi}_n(\{p^\star\})}{\sqrt{{A!}}} \Phi_n(p)|\{p^\star\}\rangle, \quad\text{where}\quad  d\bm{\widetilde{p}}^\star_i \equiv \frac{d\bm{p}^\star_i}{(2\pi)^3 \sqrt{2E_{\bm{p}^\star_i}}}
, \end{equation}
and the function
\begin{equation}\label{eq:3chiA-ScatteringAmplitude-Outer-Nuclear-state}
\Phi_n(p)=(2\pi)^3\sqrt{2P_n^0}\delta^3(\bm{p}-\bm{P})
\end{equation}
corresponds to a nucleus with a certain three-momentum $\bm{P}$  and energy 
$P^0_n=E_{\bm{p}}+\varepsilon_n$, involving the nuclear
excitation energy $\varepsilon_n$.
The symbol $\{p^\star\}$ denotes the notation  $\{p^\star\}\equiv (p^\star_1\dots p^\star_n)$, 
where $p^\star_i$  is the four-momentum of the $i$-th nucleon in the center-of-mass
system of the nucleus (at rest).
\par 
For nuclear states $|n\rangle$ describing a nucleus at rest in the $n$-th internal quantum state (nth level), 
the conventional normalization condition is
\begin{equation} \label{eq:3chiA-ScatteringAmplitude-nucleus_state_norm}
\langle m|n\rangle \equiv\int\left(\prod^{A}_{i}\frac{d\bm{p}^\star_i}{(2\pi)^3}\right)\widetilde{\psi}_n(\{p^\star\})\widetilde{\psi}^*_m(\{p^\star\})(2\pi)^3 \delta^3(\sum^A_{i=1} \bm{p}^\star_i)
=\delta_{mn}. 
\end{equation}
For the nuclear wave functions  $|P_n\rangle$ from (\ref{eq:3chiA-ScatteringAmplitude-nucleus_state}), it
gives a simple normalization condition
\begin{equation}\label{eq:3chiA-ScatteringAmplitude-nucleus_state_norm_1}
\langle P'_m|P_n\rangle = (2\pi)^3 2P^0_n \delta^3(\bm{P}-\bm{P}')\delta_{nm}.
\end{equation}
The state $|n\rangle$ satisfying (\ref{eq:3chiA-ScatteringAmplitude-nucleus_state_norm})
can be formally defined as
\begin{equation}\label{eq:3chiA-ScatteringAmplitude-n_state}
|n\rangle = \int \left(\prod_{i=1}^{A} d\widetilde{\bm{p}}_i^\star\right) \frac{\widetilde\psi_n(\{p{^\star}\}) }{\sqrt{A!}} \Bigg[(2\pi)^3\delta^3\left(\sum_{i=1}^A\bm{p}_i^\star\right)\Bigg]^{1/2}|\{p^\star\}\rangle.
\end{equation}
Based on the wave function (\ref{eq:3chiA-ScatteringAmplitude-nucleus_state}), one can calculate the amplitude (probability) 
of the scattering of a massive neutral $\chi$ particle from a nucleus as a system of mutually interacting nucleons, 
assuming that the external interaction is between the (pointlike) $\chi$ particle and the structureless nucleons. 
It is an acceptable hypothesis, since the $\chi$ particle with an energy below 50–100 MeV is usually incapable of 
\/“penetrating into a nucleon and seeing quarks.” 
Therefore, the effective four-fermion interaction Lagrangian can be used with a sufficient accuracy 
\cite{Bednyakov:2018mjd,Bednyakov:2019dbl,Bednyakov:2021ppn}.
\par
In the problem of direct detection of dark-matter particles, 
the effective Lagrangian involving only the axial-vector (or spin-dependent) and scalar (spin-independent) interactions 
between the $\chi$ particle and nucleons is normally used (see, for example, 
\cite{Freese:1987wu,Jungman:1996df,Bertone:2004pz,Bednyakov:2015uoa,Vergados:1996hs}). 
It can be written as a product of two currents
\begin{equation}		\label{eq:3chiA-ScatteringAmplitude-CohDM-Effective-Axial-and-Scalar-chi-N-Lagrangian}
\mathcal{L}(x)=\dfrac{G_F}{\sqrt 2} L_\mu(x)H^\mu(x).
\end{equation}
Here the current operators of the $\chi$ particle and the nucleons of the nucleus are given 
in terms of normal products of quantum-field operators $\psi_\chi(x)$ and $\psi_{n,p}(x)$ 
\begin{eqnarray}			\label{eq:3chiA-ScatteringAmplitude-Lepton-S+A-Operator}
L_\mu(x)&=&\chi_A \normord{ \overline{\psi}_\chi(x)\gamma_\mu\gamma_5\psi_\chi(x)}
+\chi_S \normord{ \overline{\psi}_\chi(x)\psi_\chi(x)},  \\ 						
\label{eq:3chiA-ScatteringAmplitude-Nucleon-S+A-Operator}
H^\mu(x)&=&\sum_{f=n,p}\big[ h_A^f \normord{ \overline{\psi}_f(x)\gamma^\mu\gamma_5\psi_f(x)}
+ h^f_S \normord{ \overline{\psi}_f(x) \psi_f(x)} \big], 
\end{eqnarray}
The effective coupling constants $$c_{A,S}^f \equiv \chi^{}_{A,S} h^f_{A,S}$$  
specify the intensity of the interaction between $\chi$ particles and nucleons (relative to the Fermi constant $G_F$).
In Section \ref{43chiA-CrossSection-via-ScalarProducts}, 
the effective Lagrangian (\ref{eq:3chiA-ScatteringAmplitude-CohDM-Effective-Axial-and-Scalar-chi-N-Lagrangian}) 
 is generalized to the case of other admissible Lorentz-invariant structures.
\par 
With the Lagrangian of the form (\ref{eq:3chiA-ScatteringAmplitude-CohDM-Effective-Axial-and-Scalar-chi-N-Lagrangian}), the $\mathbb{S}$-matrix $\langle P'_m,k'|\mathbb{S}|P_n,k\rangle$
describing the probability for the transition of the nucleus and the $\chi$ particle
from the initial state $|P_n,k\rangle$  to the final state $\langle P'_m,k'|$
due to their interaction is written in the first order with respect to the Fermi constant $G_F$ in a standard form
\begin{equation}		\label{eq:3chiA-ScatteringAmplitude-CohDM-A+S-MatrixElevent} 
\langle P'_m,k'|\mathbb{S}|P_n,k\rangle= (2\pi)^4\delta^4(q+P_n-P'_m)i\mathcal{M}_{mn}
=\dfrac{i G_{\rm F}}{\sqrt{2}} \int d^4 x \, H^\mu_{nm}(x) \,  L_\mu^\chi(x)
, \end{equation} 
where $H^\mu_{mn}(x)\equiv \langle P'_m| H^\mu(x)|P_n\rangle$  is the amplitude of the
probability (matrix element) of the transition of the nucleus from the state $|P_n\rangle$ to the state 
$\langle P'_m|$ due to the hadronic current of the form $H^\mu(x)$.
Substituting operator expressions for the currents
$H^\mu_{nm}(x)$ and $L_\mu^\chi(x)$ into the right-hand side of 
(\ref{eq:3chiA-ScatteringAmplitude-CohDM-A+S-MatrixElevent}),
considering the explicit form of the nuclear functions  (\ref{eq:3chiA-ScatteringAmplitude-nucleus_state}), performing the appropriate calculations, taking the integral over $d x^4$ responsible for energy
conservation at the $\chi$-particle-nucleon level, and
introducing the designation for the leptonic current (or $\chi$ particle current)
\begin{equation}\label{eq:3chiA-ScatteringAmplitude-CohDM-LeptonicCurrent}
 l_\mu(k',k,s',s) \equiv \overline{u}_\chi(\bm{k}',s') O_{\mu} u_\chi (\bm{k},s)
,\end{equation}
we arrive at the following expression for the matrix element from 
(\ref{eq:3chiA-ScatteringAmplitude-CohDM-A+S-MatrixElevent}):
 \begin{eqnarray}  \label{eq:3chiA-ScatteringAmplitude-M-Element-via-H-and-L}
i \mathcal{M}_{mn}=\dfrac{i G_{\rm F}}{\sqrt{2}}  \sqrt{4P^{0'}_m P^0_n}\, \,  l_\mu(k',k,s',s)\, h^\mu_{mn} (\bm{q}) 
.\end{eqnarray}
The hadronic current $h^\mu_{mn}(\bm{q})=\langle m|H^\mu(0)|n\rangle$
in (\ref{eq:3chiA-ScatteringAmplitude-M-Element-via-H-and-L}) 
defined in terms of nuclear state functions of the nucleus at rest has the form
\footnote{Derivation of formula (\ref{eq:3chiA-ScatteringAmplitude-h-mu-mn-with-delta-p_k})  
does not depend on the explicit form of combinations of $\gamma$-matrices $O^\mu_k$ 
(see \cite{Bednyakov:2018mjd,Bednyakov:2019dbl,Bednyakov:2021ppn}). 
}:  
\begin{eqnarray}\nonumber \label{eq:3chiA-ScatteringAmplitude-h-mu-mn-with-delta-p_k}
h^\mu_{mn}(\bm{q})&=& \sum^A_{k} 
\frac{ \overline{u}(\bm{\bar{p}}^\star_k+\bm{q},r'_{k})\, O^\mu_k\,  u(\bm{\bar{p}^{\star}}_k,r_{k})}
{\sqrt{4E_{\bm{\bar{p}}^\star_k}E_{\bm{\bar{p}}^\star_k+\bm{q}}}} 
\times\\&&\times 
\int \prod^{A}_{i=1}\frac{d\bm{p}^\star_i   \delta\big(f(\bm{p^{\star}_k})\big)}{(2\pi)^3}
\widetilde{\psi}^{*}_m(\{p^{(k)}_\star\},\bm{p^{\star}}_k\!+\!\bm{q} )\widetilde{\psi}_n(\{p^\star\}) 
(2\pi)^{3} \delta^3(\sum^A_{i=1}\bm{p}^\star_i)
, \end{eqnarray}
where  $\bm{\bar{p}^{\star}_k}(\bm{q})$ is the solution of the equation $f(\bm{\bar{p}^{\star}_k}) = 0$,
which depends on $\bm{q}$ and is determined by the condition
\begin{equation}\label{eq:3chiA-ScatteringAmplitude-Energy-and-Identity-Conservation}
 \delta\big(f(\bm{p^{\star}_k})\big) \equiv \delta( - T_A - \Delta\varepsilon_{mn}+\sqrt{m^2+ {\bm{p^{\star}_k}}^2}-\sqrt{m^2+ (\bm{p^{\star}}_k+\bm{q})^2})\ne 0.
\end{equation} 
It arises from the necessity that the energy conservation
law be simultaneously obeyed at the levels of the
nucleus and a single nucleon. Technically, it is because
the $q_0$-dependent delta function responsible for energy
conservation at the level of the entire nucleus 
$$\delta(q_0+P_{0,n}-P'_{0,m}) = \delta(q_0 - T_A - \Delta\varepsilon_{mn})
=\delta\big(q_0 - \frac{\bm{q}^2}{2m_A} - \Delta\varepsilon_{mn}\big), 
$$  
which is on the left in (\ref{eq:3chiA-ScatteringAmplitude-CohDM-A+S-MatrixElevent}), should be “balanced” by
the delta function of energy conservation at the level of a single (active) nucleon
\footnote{Hereinafter, unless stated otherwise, the proton mass is taken to
be equal to the neutron mass $m=m_p=m_n$.} 
$$\delta(q_0+p^{\star}_{0,k}-p^{'\star}_{0,k})= \delta(q_0+ \sqrt{m^2+  {\bm{p^{\star}_k}}^2} -  \sqrt{m^2+ (\bm{p^{\star}}_k+\bm{q})^2}), 
$$
which naturally arises on the right in (\ref{eq:3chiA-ScatteringAmplitude-CohDM-A+S-MatrixElevent}) 
after integration over $d x^4$ and depends on the same variable $q_0$.  
Therefore, relation (\ref{eq:3chiA-ScatteringAmplitude-Energy-and-Identity-Conservation}) 
 is the condition for energy conservation and preservation of {\em  integrity of the
nucleus}\/  met simultaneously
\begin{equation}\label{eq:3chiA-ScatteringAmplitude-Energy-and-Identity-Conservation-1}
\displaystyle\sqrt{m^2+(\bar{\bm{p}}+\bm{q})^2}-\sqrt{m^2+\bar{\bm{p}}^2}
=\frac{\bm{q}^2}{2m_A}+\Delta\varepsilon_{mn} = T_A+\Delta\varepsilon_{mn}, 
\end{equation}
that “chooses” the active nucleon momentum $\bar{\bm{p}}= ( p_L,  p_T)$
dependent on $\bm{q}$ in the form \cite{Bednyakov:2018mjd,Bednyakov:2019dbl,Bednyakov:2021ppn}:
\begin{equation}\label{eq:3chiA-ScatteringAmplitude-Energy-and-Identity-Conservation-Solution}
 p_L = -\frac{|\bm{q}|}{2}\Bigg[1-\sqrt{\beta}\sqrt{1+\frac{4m_{T}^2}{\bm{q}^2(1-\beta)}}\Bigg], \quad \text{where} \quad \beta=\frac{(T_A+\Delta\varepsilon_{mn})^2}{\bm{q}^2}, \quad  m_{T}^2 = m^2+p_T^2
.  \end{equation} 
According to the condition for preservation of integrity of the nucleus 
(\ref{eq:3chiA-ScatteringAmplitude-Energy-and-Identity-Conservation-1}), 
in the rest frame of the
nucleus the active nucleon “meets” the $\chi$ particle with the nonzero momentum $p_l = (m,0,0,p_L)$  from 
(\ref{eq:3chiA-ScatteringAmplitude-Energy-and-Identity-Conservation-Solution}), 
incident on the nucleus (along the $z$ axis), and therefore
the $s$ invariant in the rest frame of the nucleus (lab frame) should be recalculated
\begin{eqnarray}\label{eq:42chiA-CrossSection-ScalarProducts-ChiEta-All-s-with-T_A-recoil} 
s&=& (k_l+p_l)^2  
m^2_\chi +m^2 + 2 m m_\chi\Bigg\{ \sqrt{1 + \dfrac{2 T_0}{m_\chi }} 
\sqrt{1+ \dfrac{p_L^2}{m^2} } - \dfrac{|\bm{k}^l_\chi |}{m_\chi} \dfrac{p_L}{m}\Bigg\}.
\end{eqnarray}
In other words, the invariant energy $s$, in terms of which scalar products are calculated in the lepton–
nucleon center-of-mass system, depend not only on the kinetic energy of the incident $\chi$ particle 
$T_0$ but also on the kinetic energy of the nuclear recoil $T_A$, 
since $\bm{q}^2\simeq 2 m_A T_A$. 
Note that this correction is relatively small, because $\dfrac{p_L}{m} \le  0.1$
\cite{Bednyakov:2018mjd,Bednyakov:2019dbl,Bednyakov:2021ppn} and is ignored in what follows.
\par
Arguments of the wave functions of the bound nuclear states $\widetilde{\psi}_m^*(\{p^{(k)}_\star\} )$
and $\widetilde{\psi}_n(\{p_\star\})$ in (\ref{eq:3chiA-ScatteringAmplitude-h-mu-mn-with-delta-p_k})  are
expressions defined as $\{p_\star\}=(p_1^\star\dots p_A^\star)$, where the $i$-th element is a pair 
$p_i^\star \equiv (\bm{p}_i^\star,r_i)$ of the three-momentum (in the rest frame of the nucleus) and the spin of
the $i$-th nucleon. 
The expression $\{p^{(k)}_\star\}$  coincides with $\{p_\star\}$ except for the $k$-th element which is 
$(\bm{p}_{k}^\star+\bm{q},s_k)$,
where $\bm{q}$  is the $\chi$ particle momentum transferred (to the nucleus).
\par
The hadronic current $h^\mu_{mn}(\bm{q})$ defined by formula
(\ref{eq:3chiA-ScatteringAmplitude-h-mu-mn-with-delta-p_k}) is a sum of terms like 
$\bar{u}(\bm{p}_k+\bm{q},s_k) O^\mu_k u(\bm{p}_k,r_k)$ corresponding
to scattering of the $\chi$ particle by the single $k$-th nucleon, which had the three-momentum $\bm{p}_k$ 
and the spin projection on a certain preferred direction $r_k$ before the interaction. 
The amplitude of the probability to find this nucleon in the initial nuclear state $|P_n\rangle$
with exactly these quantum numbers is governed by
the wave function $\tilde{\psi}_n(\{p_\star\})$ given in the momentum
representation and depending on the nucleon momenta in the rest frame of the nucleus. 
A scattered
(“interaction-stricken,” active) nucleon has a three-momentum $ \bm{p}_k+\bm{q}$
and, generally, an arbitrary spin projection on the preferred direction $s_k$. 
The corresponding amplitude of the probability to find the nucleon in the final nucleus $|P_m\rangle$ with exactly these quantum numbers is given by the expression $\tilde{\psi}^*_m(\{p^{(k)}_\star\})$. 
All three-momenta of the other spectator nucleons (not participating in the interaction with the
$\chi$ particle) remain unchanged and drop from consideration due to the normalization rules of nuclear wave functions.
\par
To transform the hadronic current  (\ref{eq:3chiA-ScatteringAmplitude-h-mu-mn-with-delta-p_k}), 
we assume, as in \cite{Bednyakov:2018mjd,Bednyakov:2019dbl,Bednyakov:2021ppn}, 
that the wave function $\widetilde{\psi}_n$  can be written as a product of the momentum 
$\widetilde{\psi}_n$  and spin $\chi_n$ components
\begin{equation}\label{eq:3chiA-ScatteringAmplitude-CohDM-factorize_spin}
\widetilde{\psi}_n(\{p^\star\}) =\widetilde{\psi}_n(\{\bm{p}_\star\})\chi_n(\{r\}),
\end{equation}
the former being dependent on the momentum variables  $\{\bm{p}_\star\}=(\bm{p}_1^\star\dots{} \bm{p}_A^\star)$, and the latter on the spin variables $\{r\}=(r_1\dots r_A)$. 
It is evident from (\ref{eq:3chiA-ScatteringAmplitude-h-mu-mn-with-delta-p_k}) that it involves products of spin functions
\begin{equation}				\label{eq:3chiA-ScatteringAmplitude-spin-products}
\lambda^{mn}(r',r) \equiv \lambda^{mn}_{r'r} \equiv \chi^*_m(\{r^{(k)}\})\chi_n(\{r\})
, \end{equation}
where $\{r^{(k)}\}$ coincides with $\{r\}$, except for the $k$-th element, which is equal to $r'_k$.
\par 
Obviously, after the interaction of the active ($k$-th) nucleon with the $\chi$ particle its spin state either
does not change (index $r'_k=r_k$) or changes ($r'_k \ne r_k$).
In the former case, according to the normalization condition 
(\ref{eq:3chiA-ScatteringAmplitude-nucleus_state_norm}), the product of spin functions 
(\ref{eq:3chiA-ScatteringAmplitude-spin-products}) takes the form  $\chi_m^{*}(\{r\})\chi_n(\{r\}) = \delta_{nm}$, 
which means that the nucleus did not change, i.e., after the interaction it remained in its initial state  ($m=n$). 
In the latter case, the product of spin functions (\ref{eq:3chiA-ScatteringAmplitude-spin-products}) 
will be different from zero only if the index $r'_k$, not equal to exactly $r_k$ corresponds to the definition 
of the spin state of $|m\rangle $ the nucleus  $A$
\footnote{To be in correspondence to \cite{Bednyakov:2018mjd,Bednyakov:2019dbl,Bednyakov:2021ppn}, 
the product $\chi^*_m(\{r^{(k)}\})\chi_n(\{r\})$, like $\chi_n^{*}(\{r\})\chi_n(\{r\})$, will be considered equal to
unity. Validity of this assumption is a subject for a separate study.}. 
In other words, in the cases of our interest the product of the spin wave functions
from (\ref{eq:3chiA-ScatteringAmplitude-spin-products})   
may be thought of as being almost independent of the indices $m,n$: 
\begin{equation} \label{eq:3chiA-ScatteringAmplitude-CohDM-spin_functions_norm2}
\lambda^{mn}(r',r)= \delta_{mn}\delta_{r'r}+(1-\delta_{mn})(1-\delta_{r'r}) 
. \end{equation} 
Considering these assumptions and $\bar{\bm{p}}= f(\bm{q})$, from
(\ref{eq:3chiA-ScatteringAmplitude-Energy-and-Identity-Conservation}), 
expression  (\ref{eq:3chiA-ScatteringAmplitude-h-mu-mn-with-delta-p_k}) is recast as
\begin{equation}	\label{eq:3chiA-ScatteringAmplitude-h-mu-mn-with-delta-p_k-spin-factorization}
\begin{aligned}
h^\mu_{mn}(\bm{q})& =  \sum_{k=1}^{A}  \frac{\bar{u}(\bar{\bm{p}}+\bm{q},r'_k) O^\mu_k u(\bar{\bm{p}},r_k)}{\sqrt{4E_{\bar{\bm{p}}}E_{\bar{\bm{p}}+\bm{q}}}} 
\lambda^{mn}(r',r)  \times 
\\ & ~~ \times\int \Bigg[\prod^{A}_{i=1}\frac{d\bm{p}^\star_i  }{(2\pi)^3}\Bigg] \delta\big(f(\bm{p^{\star}_k})\big)
\widetilde{\psi}_m^*(\{\bm{p}^{(k)}_\star\},\bm{p^{\star}}_k\!+\!\bm{q} ) \widetilde{\psi}_n(\{\bm{p}_\star\})
(2\pi)^3 \delta^3\big(\sum_{l=1}^A \bm{p}^\star_l\big) 
. \end{aligned}
\end{equation}
The multidimensional integral in this formula can be written as a matrix element of the operator 
$\hat{\bm{X}}_k$ that implements the three-dimensional shift of the
$k$-th nucleon
\begin{equation}\label{eq:3chiA-ScatteringAmplitude-f^k_mn-definition} 
f^k_{mn}(\bm{q})\equiv\langle m|e^{i\bm{q}\hat{\bm{X}}_{k}}|n\rangle  =
\int \Bigg[\prod^{A}_{i=1}\frac{d\bm{p}^\star_i  }{(2\pi)^3}\Bigg] \delta\big(f(\bm{p^{\star}_k})\big) 
\widetilde{\psi}_m^*(\{\bm{p}^{(k)}_\star\}) \widetilde{\psi}_n(\{\bm{p}_\star\})
(2\pi)^3 \delta^3\big(\sum_{l=1}^A \bm{p}^\star_l\big), \qquad
\end{equation}
where the delta function $\delta\big(f(\bm{p^{\star}_k})\big)$ ensures preservation of the integrity of the nucleus after the shift of the active $k$-th-nucleon momentum by the operator  $\hat{\bm{X}}_k$
due to an external action.
\par
As a result, in view of (\ref{eq:3chiA-ScatteringAmplitude-h-mu-mn-with-delta-p_k-spin-factorization}) and 
(\ref {eq:3chiA-ScatteringAmplitude-f^k_mn-definition}), 
one obtains the following expression for the matrix element 
(\ref{eq:3chiA-ScatteringAmplitude-M-Element-via-H-and-L})
defining the probability of the $\chi_s A_n \to \chi_{s'} A_m$ process:
\begin{equation}			\label{eq:3chiA-ScatteringAmplitude-CohDM-matrix_element}
i\mathcal{M}^{s' s,r'r }_{mn}(\bm{q}) = i\frac{G_F}{\sqrt{2}} \frac{m_A}{m}C_{1,mn}^{1/2} \sum_{k=1}^{A} 
 f^k_{mn}(\bm{q})  \lambda^{mn}(r',r) (l_{s's},h^k_{r'r}),  \quad\text{where}
\end{equation}
\begin{equation}\label{eq:3chiA-ScatteringAmplitude-CohDM-ScalarProduct-lh}
(l_{s's}, h^k_{r'r})\equiv l_\mu(k',k,s',s)\  \bar{u}(\bar{\bm{p}}+\bm{q},r'_k) O^\mu_k u(\bar{\bm{p}},r_k)
\end{equation}
is the scalar product of the lepton and $k$-th-nucleon currents, which involves all the specificity of the interaction
between them. 
Formula (\ref{eq:3chiA-ScatteringAmplitude-CohDM-matrix_element}) takes into account
that the general kinematic factor, which arises as we
progress, can be rewritten by “separating from it” the
leading factor  $m_A/m$, where $m$ and $m_A$  are the nucleon
and nuclear masses, respectively, and introducing a correction  $C_{1,mn}$
that is close to unity with a good accuracy,
\begin{equation}	\label{eq:3chiA-ScatteringAmplitude-CohDM-C1_def}
C_{1,mn}^{1/2} =  \sqrt{\frac{P^0_n}{m_A}\frac{P^{'0}_m}{m_A}\frac{m}{E_{\bar{\bm{p}}}}\frac{m}{E_{\bar{\bm{p}}+\bm{q}}}}   \sim O(1).
\end{equation}
Thus, the matrix element, or the probability amplitude,
for the process  $\chi A_n\to \chi A^{(*)}_m$ (\ref{eq:3chiA-ScatteringAmplitude-CohDM-matrix_element})
is a sum of individual lepton-nucleon probability amplitudes
proportional to the corresponding scalar products of
currents $(l_{s's}, h^k_{r'r})$ weighted with two factors, each of which is little different from unity. 
$\lambda^ {mn}(r',r)$  (see. (\ref{eq:3chiA-ScatteringAmplitude-CohDM-spin_functions_norm2}))
is almost independent of the nuclear states $|n\rangle$ and $|m\rangle$, 
but it “controls” (as will be shown below) the initial $r$ and final $r'$ values of the (doubled) projection
of the spin of the active nucleon of a nucleus on a preselected axis. 
The other factor $f^k_{mn}(\bm{q})$ defines the nucleon form factor of the nuclear transition from the
$|n\rangle$-state to the $|m\rangle$-state.
\par
The expression $\langle m|e^{i\bm{q}\hat{\bm{X}}_{k}}|n\rangle$ from 
(\ref{eq:3chiA-ScatteringAmplitude-f^k_mn-definition}) is a generalization of the factor  $e^{i\bm{q} \bm{x}_k}$
used by Freedman \cite{Freedman:1973yd}. 
The fundamental difference between $\langle m|e^{i \bm{q}\hat{\bm{X}}_k} |n\rangle$ and $e^{i\bm{q}\bm{x}_k}$
is that when $e^{i\bm{q}\bm{x}_k}$ “works,” the $k$-th nucleon is explicitly considered to have a quite definite position in the nucleus specified by the coordinate $\bm{x}_k$ (“nailed” at the point $\bm{x}_k$). 
However, scattering by one free nucleon cannot lead to the nucleus moving as a single entity,
since nucleons should be bound to one another, otherwise the nucleus will break up. 
The wave functions of the bound nuclear state used here do not rely on the assumption that spatial positions of nucleons in a nucleus are fixed, and therefore the matrix element  
$\langle m|e^{i \bm{q}\hat{\bm{X}}_k} |n\rangle$ 
does not depend on the number $k$
\cite{Bednyakov:2018mjd,Bednyakov:2019dbl,Bednyakov:2021ppn}.
Moreover, $f^k_{mn}(\bm{q})=\langle m|e^{i\bm{q}\hat{\bm{X}}_{k}}|n\rangle$ is the key quantity for
establishing interrelation of the coherent and incoherent
regimes in the $\chi$-nucleus scattering (Section \ref{41chiA-CrossSection-Coh-vs-InCoh}).

\section{\large Cross sections of ${\chi}$ particle-nucleus scattering}\label{4chiA-CrossSection}  
\subsection{\normalsize\em Coherent and Incoherent contributions
to the  ${\chi A\to \chi A^{(*)}}$ scattering cross section} \label{41chiA-CrossSection-Coh-vs-InCoh} 
 The  observed differential cross section for the process $\chi A\to \chi A^{(*)}$ can be obtained by averaging the
differential cross section that determines the nuclear transition from the state $|n\rangle$ 
 to the state $|m\rangle$ (formulas (\ref{eq:2DM-Kinematics-dSigma-po-d-T-A-nonrel}) and 
(\ref{eq:2DM-Kinematics-dSigma-po-d-T-A-nonrel-Coeff})) over all potentially possible initial (internal)
states $|n\rangle$ of the target nucleus and summing over all
allowed final states $|m\rangle$ of the recoil nucleus
\begin{equation}\label{eq:41chiA-CrossSection-Coh-vs-InCoh-CrossSection-definition}
\frac{d\sigma}{dT_A}(\chi A \to \chi A^{(*)}) = \sum_{n,m}\omega_n \frac{d\sigma_{mn}}{dT_A}(\chi A \to \chi A^{(*)}), \quad \text{where}\quad  \sum_n \omega_n=1 
\end{equation}
is the total sum of probabilities of all possible initial states of the nucleus $A$. 
The matrix element of the $\chi A \to \chi A^{(*)}$ process in (\ref{eq:2DM-Kinematics-dSigma-po-d-T-A-nonrel}) 
was given by formula (\ref{eq:3chiA-ScatteringAmplitude-CohDM-matrix_element}). 
With summation taken in it over “internal” spin indices of the (active) nucleon and with “external” spin
indices of the $\chi$ particle retained, formula (\ref{eq:3chiA-ScatteringAmplitude-CohDM-matrix_element})  takes the form
\begin{eqnarray}\label{eq:41chiA-CrossSection-MatrixElement-with-s-sprime}
i\mathcal{M}_{mn}^{s's}(\bm{q})  = i\frac{G_F}{\sqrt{2}} \frac{m_A}{m}C_{1,mn}^{1/2} \sum_{k=1}^{A} \sum_{r',r} f^k_{mn}(\bm{q}) (l_{s's},h^k_{r'r}) \lambda^{mn}(r',r) 
. \end{eqnarray}
Recall that a similar matrix element for the (anti)neutrino–nucleus scattering was considered
\cite{Bednyakov:2018mjd,Bednyakov:2019dbl,Bednyakov:2021ppn} 
at fixed values of lepton helicity (spin projection on momentum direction). 
Since the neutrino helicity is always negative, $s'=s=-1$, and the antineutrino helicity 
is always positive, $s'=s=+1$, the subscripts $s', s$ have always been specified for the (anti)neutrino
matrix element like  (\ref{eq:41chiA-CrossSection-MatrixElement-with-s-sprime}) “from the outside.” 
By following this logic, the form   (\ref{eq:41chiA-CrossSection-MatrixElement-with-s-sprime})
was determined for the matrix element. 
It corresponds to the $\chi A$ interaction when the $\chi$ particle has fixed values of its initial $s$
 and final $s'$ spin projection on a certain preferred direction.
Summation over the active nucleon spin projections $\sum_{r'r}$  is made “inside” the matrix element (at the
level of nucleon amplitudes) rather than at the level of the square of the matrix element, as is usually done for
real (or potentially detectable) final-state particles.
 This summation is justified by impossibility of determining the active nucleon spin before and after interaction.
For example, the matrix element $i\mathcal{M}_{mn}^{-+}$ describes a process in which the 
 $\chi$ particle has the spin projection $s=+1$ upon incidence and $s'=-1$ for the outgoing one. 
 Since we cannot know (even in general) the spin projection of the active nucleon when it
“meets” and “sends away” the particle, we have to take into consideration all possible spin projections of
this active nucleon, i.e., to sum over $r=\pm$ and $r'=\pm$.
\par
Substituting (\ref{eq:41chiA-CrossSection-MatrixElement-with-s-sprime}) 
into cross-section expression  (\ref{eq:41chiA-CrossSection-Coh-vs-InCoh-CrossSection-definition}),
we obtain
\begin{eqnarray}\label{eq:41chiA-CrossSection-Coh-vs-InCoh-CrossSection-with-s-sprime-and-Tmns}
\frac{d\sigma_{s's}}{d T_A}(\chi A \to \chi A^{(*)})
 &=& \dfrac{G^2_F  m_A}{2^7\pi m^2 T_0 m_\chi} \Big[T^{s's}_{m=n} + T^{s's}_{m\ne n}\Big],  \quad\text{where}
\end{eqnarray} 
\begin{eqnarray} \label{eq:41chiA-CrossSection-Coh-vs-InCoh-CrossSection-Term-nn-definition-with-s-sprime}
T^{s's}_{m=n}\!\!&=&\!\!\sum^A_{k,j}\sum_{n}\omega_n\Big[ C^{}_{1,nn}C^{\rm nonrel}_{nn}\, f^k_{nn}f^{j*}_{nn}
\sum_r  (l_{s's},h^k_{rr})\sum_{x} (l_{s's},h^j_{xx})^{*}\Big], 
\\ 
\label{eq:41chiA-CrossSection-Coh-vs-InCoh-CrossSection-Term-mn-definition-with-s-sprime}
T^{s's}_{m\ne n}\!\!&=&\!\!\sum^A_{k,j}\sum_{n}\omega_n  \Big[ \sum_{m\ne n} C^{}_{1, mn}C^{\rm nonrel}_{mn}\, f^k_{mn} f^{j*}_{mn}  \sum_{r'r}
\lambda^{mn}_{r'r}
(l_{s's}, h^k_{r'r})
\Big(\sum_{x'x}\lambda^{mn}_{x'x}(l_{s's}, h^j_{x'x}) \Big)^{\dag} \Big].\qquad  
\end{eqnarray}
Here $T_{m=n}$  is the contribution to the cross section that corresponds to preservation of the initial nuclear
state while the spin projection of the interacting (active) nucleon does not change. 
According to (\ref{eq:3chiA-ScatteringAmplitude-CohDM-spin_functions_norm2}), $\lambda^{nn}({r',r})=\delta_{r'r}$, and spin amplitudes drop from further consideration. 
The crucial part in (\ref{eq:41chiA-CrossSection-Coh-vs-InCoh-CrossSection-with-s-sprime-and-Tmns})  
is played by the scalar products of four-vectors of the leptonic and
nucleon currents $(l_{s's},h^{p/n}_{r'r})$ defined by formua (\ref{eq:3chiA-ScatteringAmplitude-CohDM-ScalarProduct-lh}).
Introducing the notations
\begin{eqnarray}\label{eq:41chiA-CrossSection-Coh-vs-InCoh-g-factors}
g^{p/n}_\text{coh}= C^{p/n}_{1,nn}  C^{\rm nonrel}_{nn} \qquad\text{and}\qquad 
g^{p/n}_\text{inc}= C^{p/n}_{1,mn} C^{\rm nonrel}_{mn} 
,\end{eqnarray}
and considering the form of formulas (\ref{eq:2DM-Kinematics-dSigma-po-d-T-A-nonrel-Coeff}) and 
(\ref{eq:3chiA-ScatteringAmplitude-CohDM-C1_def}),  we can assume with a good degree of confidence that
these quantities have constant values of about unity and are almost independent of the nuclear recoil 
energy $T_A$  and nuclear indices $n, m$. 
In view of this fact, summation over the index  $n$ in 
(\ref{eq:41chiA-CrossSection-Coh-vs-InCoh-CrossSection-Term-nn-definition-with-s-sprime}) 
gives rise to form factors averaged over all initial nuclear states, i.e.,
\begin{equation}\label{eq:41chiA-CrossSection-Coh-vs-InCoh-form-factors-with-s-sprime}
\sum_n \omega_n f_{nn}^k f_{nn}^{j*}= \left\{\begin{matrix}
|F_{p/n}(\bm{q})|^2,                   &\text{ when }(k,j)=(p,p) \text{~~or when~~} (k,j)=(n,n);\\
F_{p}(\bm{q})F_{n}^*(\bm{q}),   &\text{ when }(k,j) = (p,n);\\
F_{n}(\bm{q})F_{p}^*(\bm{q}),   &\text{ when }(k,j)= (n,p).\\
\end{matrix}\right.
\end{equation}
In other words, the left-hand side of this relation is equal to the square of the modulus of the nuclear proton
or neutron form factor  $|F_{p/n}(\bm{q})|^2$ if the summation indices $k$ and $j$ correspond to either a pair of protons or a pair of neutrons. 
If these indices do not simultaneously correspond to a pair of identical nucleons, we
have a product of the nuclear proton and neutron form
factors in the form $F_{p}(\bm{q})F_{n}^*(\bm{q})$ or $F_{n}(\bm{q})F_{p}^*(\bm{q})$. 
Therefore, expression (\ref{eq:41chiA-CrossSection-Coh-vs-InCoh-CrossSection-Term-nn-definition-with-s-sprime}) 
can be written as a square of the modulus of the sum of the proton and neutron contributions
\begin{eqnarray}\nonumber
\label{eq:41chiA-CrossSection-Coh-vs-InCoh-cross_section_coherent_term-1-with-s-sprime}
T^{s's}_{m=n}(\bm{q}) &=&\Big|\sum^Z_{k}\sqrt{g^p_\text{coh}}\sum_{r}  (l_{s's},h^p_{rr}(\bm{q}))F_p(\bm{q})
	 +\sum^N_{j}\sqrt{g^n_\text{coh}} \sum_{r}  (l_{s's},h^n_{rr}(\bm{q}))F_n(\bm{q})\Big|^2
= \\&&= \Big|\sum_{f=p,n}\sqrt{g^f_\text{coh}}\sum^{A_f}_{k=1}\sum_{r} (l_{s's},h^f_{rr}(\bm{q}))F_f(\bm{q})\Big|^2.
\end{eqnarray}
Here  $A_f$ denotes the total number of nucleons of the type $f=p, n$  in the nucleus.
\par 
The second term in the cross section formula (\ref{eq:41chiA-CrossSection-Coh-vs-InCoh-CrossSection-with-s-sprime-and-Tmns}), $T_{m\ne n}$, specified by 
(\ref{eq:41chiA-CrossSection-Coh-vs-InCoh-CrossSection-Term-mn-definition-with-s-sprime}), involves summation over the nuclear indices $m,n$. 
To make this summation, we make use of the condition 
(\ref{eq:3chiA-ScatteringAmplitude-CohDM-spin_functions_norm2})  of $\lambda^{mn}(r',r)$ 
being independent of the indices $m,n$ when  ${m\ne n}$. 
We will (temporarily) consider \cite{Bednyakov:2018mjd,Bednyakov:2019dbl,Bednyakov:2021ppn}, 
 that they are equal to their averages for protons and neutrons and that the following
“normalization condition” applies to them:
\begin{equation}\label{eq:41chiA-CrossSection-Coh-vs-InCoh-NuclearSpinAmplitudeProduct}
\lambda^{mn}_{r'r} \simeq \lambda^{p/n}_{r'r} \quad \text{and}\quad \lambda^{f}_{r'r} [\lambda^{f}_{x'x}]^* \equiv \delta_{r'x'} \delta_{rx}|\lambda^{f}_{r'r}|^2. 
\end{equation} 
Then summation over $m,n$ can be made in 
(\ref{eq:41chiA-CrossSection-Coh-vs-InCoh-CrossSection-Term-mn-definition-with-s-sprime}),
since the products of the spin wave functions $\lambda^{f}_{r'r} [\lambda^{f}_{x'x}]^*$ 
can be taken outside the sum sign.
\par
If the indices $k$ and $j$ in 
(\ref{eq:41chiA-CrossSection-Coh-vs-InCoh-CrossSection-Term-mn-definition-with-s-sprime})
 “indicate” the same nucleon, e.g., the proton, summation yields
\begin{eqnarray}\label{eq:41chiA-CrossSection-Coh-vs-InCoh-cross_section_incoherent_term1}
\sum_n \omega_n\sum_{m\ne n} f_{mn}^k f_{mn}^{k*} &=&\sum_n \omega_n \Big[\sum_{m} f_{mn}^k f_{mn}^{k*} -f_{nn}^k f_{nn}^{k*}\Big] = \\ &=&\sum_n \omega_n \Big[ \langle n|e^{i\bm{q}\bm{X}_k}\sum_{m}|m\rangle\langle m|e^{-i\bm{q}\bm{X}_k}|n\rangle\Big] -|F_{p}(\bm{q})|^2 =1-|F_{p}(\bm{q})|^2
\nonumber .\end{eqnarray}
If $k\ne j$, but they still indicate protons (index $p$), it can be written that
 \begin{equation}		 \label{eq:41chiA-CrossSection-Coh-vs-InCoh-cross_section_incoherent_term2}
 \sum_n \omega_n\sum_{m\ne n} f_{mn}^k f_{mn}^{j*}  = \langle\text{cov}(e^{i\bm{q}\hat{\bm{X}}_k}, e^{-i\bm{q}\hat{\bm{X}}_j})\rangle_p, 
 \end{equation}
 where the covariation operator of the shift operators  $e^{-i\bm{q}\hat{\bm{X}}_j}$ 
 and $e^{i\bm{q}\hat{\bm{X}}_k}$  with respect to the state $|n\rangle$  is introduced in the form
 \begin{equation} \label{eq:41chiA-CrossSection-Coh-vs-InCoh-nn-Covariance}
\text{cov}_{nn}(e^{i\bm{q}\hat{\bm{X}}_k},e^{-i\bm{q}\hat{\bm{X}}_j}) \equiv \langle n|e^{i\bm{q}\hat{\bm{X}}_k}\, e^{-i\bm{q}\hat{\bm{X}_j}} |n\rangle - \langle n|e^{i\bm{q}\hat{\bm{X}}_k}|n\rangle \langle n|e^{-i\bm{q}\hat{\bm{X}}_j}|n\rangle.
\end{equation}
The averaging symbol $\langle\dots\rangle$ in 
(\ref{eq:41chiA-CrossSection-Coh-vs-InCoh-cross_section_incoherent_term2})
is defined as
\begin{equation}\label{eq:41chiA-CrossSection-Coh-vs-InCoh-Averaged-Covariance}
\langle\text{cov}(e^{i\bm{q}\hat{\bm{X}}_k},e^{-i\bm{q}\hat{\bm{X}}_j})\rangle_p 
\equiv \sum_n \omega_n \text{cov}_{nn}(e^{i\bm{q}\hat{\bm{X}}_k},e^{-i\bm{q}\hat{\bm{X}}_j}).
\end{equation}
Expression (\ref{eq:41chiA-CrossSection-Coh-vs-InCoh-Averaged-Covariance})
 becomes zero at both small and
large transferred momenta
 \begin{equation} \label{eq:041chiA-CrossSection-Coh-vs-InCoh-cross_section_incoherent_term3}
\lim_{\bm{q}\to 0}\langle\text{cov}(e^{-i\bm{q}\hat{\bm{X}}_j},e^{i\bm{q}\hat{\bm{X}}_k})\rangle_p = 0,\quad 
 \lim_{\bm{q}\to \infty}\langle\text{cov}(e^{-i\bm{q}\hat{\bm{X}}_j},e^{i\bm{q}\hat{\bm{X}}_k})\rangle_p=0.
\end{equation}
A similar consideration applies to neutrons and is also generalized to the common case of protons and
neutrons. If, for example, $k$ corresponds to a proton
and $j$ to a neutron, ($k\ne j$), their covariation is
 \begin{equation} \label{eq:41chiA-CrossSection-Coh-vs-InCoh-cross_section_incoherent_term4}
 \sum_n \omega_n\sum_{m\ne n} f_{mn}^k f_{mn}^{j*} =
 \langle\text{cov}(e^{i\bm{q}\hat{\bm{X}}_k},e^{-i\bm{q}\hat{\bm{X}}_j})\rangle_{pn}.
 \end{equation}
It is believed that when correlations between nucleons in a nucleus are rather weak, covariation functions
 like (\ref{eq:41chiA-CrossSection-Coh-vs-InCoh-cross_section_incoherent_term2})
 can be ignored. 
 For example, in nuclear shell models where multiparticle wave functions of nuclei
are constructed in the form of a product of single-particle
wave functions \cite{Blokhintsev:1963,Bohr:1974}, covariation 
 (\ref{eq:41chiA-CrossSection-Coh-vs-InCoh-cross_section_incoherent_term2}) 
  identically vanishes. 
 Counting in favor of smallness of covariation  
 (\ref{eq:41chiA-CrossSection-Coh-vs-InCoh-cross_section_incoherent_term2}) 
 is the fact that the inelastic scattering cross section can be approximated with a good accuracy
by the linear dependence on the number of scatterers. 
 \par
 Thus, covariation contributions to the observed cross section $\displaystyle\frac{d\sigma}{d T_A}$ like 
 (\ref{eq:41chiA-CrossSection-Coh-vs-InCoh-cross_section_incoherent_term2}) and 
(\ref{eq:41chiA-CrossSection-Coh-vs-InCoh-cross_section_incoherent_term4}) 
 are ignored below, since all covariation contributions are taken to be zero.
This simplifies further calculations.
 \par 
Considering the above assumptions, relations 
 (\ref{eq:41chiA-CrossSection-Coh-vs-InCoh-cross_section_incoherent_term1}),
and the assumption that at $k\ne j$ all correlators of the type of 
 (\ref{eq:41chiA-CrossSection-Coh-vs-InCoh-cross_section_incoherent_term2}) and 
(\ref{eq:41chiA-CrossSection-Coh-vs-InCoh-cross_section_incoherent_term4}) vanish
 $$\langle\text{cov}(e^{i\bm{q}\hat{\bm{X}}_k},e^{-i\bm{q}\hat{\bm{X}}_j})\rangle_{p}
 =\langle\text{cov}(e^{i\bm{q}\hat{\bm{X}}_k},e^{-i\bm{q}\hat{\bm{X}}_j})\rangle_{n}
 =\langle\text{cov}(e^{i\bm{q}\hat{\bm{X}}_k},e^{-i\bm{q}\hat{\bm{X}}_j})\rangle_{pn} =0, $$
the term (\ref{eq:41chiA-CrossSection-Coh-vs-InCoh-CrossSection-Term-mn-definition-with-s-sprime})
can be finally written as a sum over protons and neutrons
\begin{eqnarray}\label{eq:41chiA-CrossSection-Coh-vs-InCoh-CrossSection-Term-mn-with-s-sprime-final}
T^{s's}_{m\ne n} &=& \sum_{f=p,n}g^f_\text{inc}\big[1-|F_f(\bm{q})|^2\big]
\sum^{A_f}_{k=1} \sum_{r'r}|\lambda^{f}_{r'r}|^2\big| (l_{s's}, h^f_{r'r}(\bm{q})) \big|^2
.\end{eqnarray}
Thus, the measured differential cross section 
(\ref{eq:41chiA-CrossSection-Coh-vs-InCoh-CrossSection-with-s-sprime-and-Tmns})
for the process $\chi A\to \chi A^{(*)}$ can be written in the form
of two fundamentally different terms
\begin{eqnarray}\label{eq:41chiA-CrossSection-Coh-vs-InCoh-CrossSection-TwoTerms-with-s-sprime}
\frac{d\sigma_{s's}}{d T_A}(\chi A\to \chi A^{(*)})&=&
\Big[\dfrac{G^2_F  m_A}{2^7\pi m^2 T_0 m_\chi}\Big]\Big\{T^{s's}_{m=n} + T^{s's}_{m\ne n}\Big\}
\equiv  \frac{d\sigma^{s's}_\text{inc}}{d T_A} + \frac{d\sigma^{s's}_\text{coh}}{d T_A},
\quad \text{where}\\ \nonumber    
\frac{d\sigma^{s's}_\text{inc}}{d T_A} &=& c_A(T_0, m_A, m_\chi) g_\text{i}\!\sum_{f=n,p}\sum_{k=1}^{A_f} \sum_{r'r}  |\lambda^f_{r'r}|^2 |(l_{s's},h^f_{r'r}(\bm{q}))|^2[1-|F_f(\bm{q})|^2], 
\\ \nonumber \frac{d\sigma^{s's}_\text{coh}}{d T_A} &=& c_A(T_0, m_A, m_\chi)
g_\text{c}\Big|\sum_{f=n,p}\sum_{k=1}^{A_f} \sum_{r} (l_{s's}, h^f_{rr}(\bm{q}))F_f(\bm{q})\Big|^2
.\end{eqnarray}
Kinematic correction coefficients $g_\text{c/i}$  in expressions
 (\ref{eq:41chiA-CrossSection-Coh-vs-InCoh-CrossSection-TwoTerms-with-s-sprime}) are determined by averaging parameters from (\ref{eq:41chiA-CrossSection-Coh-vs-InCoh-g-factors}) 
 over nucleons and factoring them out from the double summation  $\sum_{nm}$. 
 Squares of nuclear form factors $|F_{p/n}|^2$  for protons and neutrons are defined by
 (\ref{eq:41chiA-CrossSection-Coh-vs-InCoh-form-factors-with-s-sprime}).
In formulas (\ref{eq:41chiA-CrossSection-Coh-vs-InCoh-CrossSection-TwoTerms-with-s-sprime})  
the following universal common factor is introduced:
\begin{eqnarray}  \label{eq:43chiA-CrossSection-via-ScalarProducts-c_A-definition}
c_A \equiv c_A(T_0, m_A, m_\chi) &\equiv&  \dfrac{G^2_F  m_A}{2^6\pi m^2 (2T_0 m_\chi)}
=\dfrac{G^2_F  m_A}{4\pi }\dfrac{1}{2^4 m^2 |\bm{k}^l_\chi|^2 },\quad 
\end{eqnarray}
where the dependence on the initial energy $T_0$  of the $\chi$ particle incident on the nucleus 
at rest is explicitly separated out, and the intensity of the hypothetical interaction
of the $\chi$ particle with nucleons (proportional to $G^2_F$) is “hidden deep” in the corresponding scalar products of currents.
\par
When obtaining the main formulas of this section
(\ref{eq:41chiA-CrossSection-Coh-vs-InCoh-CrossSection-TwoTerms-with-s-sprime}), 
a simplifying assumption was made that all correlation contributions like 
(\ref{eq:41chiA-CrossSection-Coh-vs-InCoh-cross_section_incoherent_term2})  and  
(\ref{eq:41chiA-CrossSection-Coh-vs-InCoh-cross_section_incoherent_term4}) can be neglected. 
Another assumption was the use of the normalization
condition (\ref{eq:3chiA-ScatteringAmplitude-CohDM-spin_functions_norm2}) for spin amplitudes.
Note that the first and second terms in 
(\ref{eq:41chiA-CrossSection-Coh-vs-InCoh-CrossSection-with-s-sprime-and-Tmns}) and 
(\ref{eq:41chiA-CrossSection-Coh-vs-InCoh-CrossSection-TwoTerms-with-s-sprime})
correspond to the elastic and inelastic $\chi$-nucleus interactions, respectively. 
The first term (when $\sum_{n=m}$) contains both indices $k$ and $j$, 
which leads to quadratic dependence of the cross section on the number of nucleons in the nucleus. 
The second term (when $\sum_{n\ne m}$) is proportional to  $\delta_{kj}$ in good approximation, which automatically leads to linear dependence of the cross section on the number of nucleons in the nucleus. 
Summation at the level of amplitudes of scattering by individual nucleons (which is traditionally
associated with the notion of coherence) is explicitly
seen in the first term of (\ref{eq:41chiA-CrossSection-Coh-vs-InCoh-CrossSection-TwoTerms-with-s-sprime}). Here each nucleon (proton and/or neutron) enters the sum being weighted
with its corresponding form factor $F_{p/n}(\bm{q})$. 
This term corresponds to the case where the nucleus does not
change its spin state (spin projection on a given direction),
which is evident from the character of summation
over spin projections of nucleons participating in
the interaction, i.e., $\sum_r (l, h^{p/n}_{rr})$. 
All active nucleons retain the spin orientation after the interaction ($r'\equiv r$).
\par 
The second term in (\ref{eq:41chiA-CrossSection-Coh-vs-InCoh-CrossSection-TwoTerms-with-s-sprime}) 
is traditionally associated with the incoherent contribution to the cross section.
It involves the square of the product of the nuclear
spin amplitudes  $|\lambda^{p/n}_{r'r}|^2$, which, according to the
adopted condition  (\ref{eq:3chiA-ScatteringAmplitude-CohDM-spin_functions_norm2}), 
should be equal to unity with a good accuracy
\cite{Bednyakov:2018mjd,Bednyakov:2019dbl,Bednyakov:2021ppn}
\begin{equation}\label{eq:41chiA-CrossSection-Coh-vs-InCoh-AllNuclearSpinsEquialTo-1}|\lambda^{p/n}_{r'r}|^2\to 1.\end{equation}
Thus, differential cross sections 
(\ref{eq:41chiA-CrossSection-Coh-vs-InCoh-CrossSection-with-s-sprime-and-Tmns}) 
for the scattering of a massive $\chi$ particle from a nucleus $\chi_s A\to \chi_{s'} A^{(*)}$
(with its spin index changed from $s$ to $s'$) in terms of scalar products of leptonic and nucleon
currents  $(l_{s's},h^f_{r'r})$ take the form
\begin{eqnarray}  \nonumber
\label{eq:41chiA-CrossSection-Coh-vs-InCoh-CrossSections-via-ScalarProducts-with-s-sprime}
\frac{d\sigma^{s's}_\text{inc}}{d T_A} &=& 
c_A(T_0, m_A, m_\chi) g_\text{i} \sum_{f=p,n}[1-|F_f(\bm{q})|^2]\sum^{A_f}_{k=1}
\Big[ \sum^{}_{r'=\pm}|(l_{s's}, h^{\eta,f}_{r'+})|^2+\sum^{}_{r'=\pm}|(l_{s's},h^{\eta,f}_{r'-})|^2 \Big], \qquad 
\\ 
\frac{d\sigma^{s's}_\text{coh}}{d T_A} &=&
c_A(T_0, m_A, m_\chi) g_\text{c}
\Big| \sum_{f=p,n}F_f(\bm{q}) \sum^{A_f}_{k=1}[(l_{s's},h^{\eta,f}_{++})+(l_{s's},h^{\eta,f}_{--})] \Big|^2
,\end{eqnarray}
where sums over initial nucleon spin projections on the $\chi$ particle motion direction 
are explicitly separated out, which is marked by the index $\eta$ in hadronic currents in scalar products. 
Since scalar products $(l_{s's},h^f_{r'r})$ depend only on the type of the active nucleon (index $f$
indicates a proton or a neutron) and do not depend on the active nucleon number in the nucleus (summation
index $k$), simple summation over this index can be taken in formulas 
(\ref{eq:41chiA-CrossSection-Coh-vs-InCoh-CrossSections-via-ScalarProducts-with-s-sprime}), 
i.e., separate summation over all protons and all neutrons in the nucleus (considering their spin projections). 
Then formulas (\ref{eq:41chiA-CrossSection-Coh-vs-InCoh-CrossSections-via-ScalarProducts-with-s-sprime})
take the form
\begin{eqnarray}  \nonumber
\label{eq:41chiA-CrossSection-Coh-vs-InCoh-CS-with-s-sprime-for-proton-neutron}
\frac{d\sigma^{s's}_\text{inc}}{d T_A} &=& c_A(T_0, m_A, m_\chi) g_\text{i} \sum_{f=p,n}[1-|F_f(\bm{q})|^2]
\Big[A^f_+  \sum^{}_{r'=\pm}|(l_{s's}, h^{\eta,f}_{r'+})|^2+A^f_- \sum^{}_{r'=\pm}|(l_{s's},h^{\eta,f}_{r'-})|^2 \Big], \qquad 
\\ 
\frac{d\sigma^{s's}_\text{coh}}{d T_A} &=&c_A(T_0, m_A, m_\chi) g_\text{c}
\Big| \sum_{f=p,n}F_f(\bm{q})  [A^f_+(l_{s's},h^{\eta,f}_{++})+ A^f_-(l_{s's},h^{\eta,f}_{--})] \Big|^2
, \end{eqnarray}
where  $A^f_\pm$ is the number of $f$-type nucleons ($f=p,n$) with the spin projection $\pm 1$ on the preferred direction 
(e.g., of the arrival of $\chi$ particle).
\par
To finish derivation of the formula for the differential cross section of the process $\chi A\to \chi A^{(*)}$, 
one should have, according to
(\ref{eq:41chiA-CrossSection-Coh-vs-InCoh-CS-with-s-sprime-for-proton-neutron}), 
explicit expressions for scalar products $(l_{s's},h^{p/n}_{r'r})$ of interacting leptonic and nucleon currents. 
These quantities were obtained in \cite{Bednyakov:2021pgs} and are given in Section
\ref{42chiA-CrossSection-ScalarProducts-ChiEta-All}.

\subsection{\normalsize\em Set of scalar products for ${\chi A\to \chi A^{(*)}}$ scattering} 
\label{42chiA-CrossSection-ScalarProducts-ChiEta-All} 
In this section, we give scalar products for all combinations of $\chi$-lepton and nucleon currents corresponding
to all possible lepton helicities and all possible nucleon spin projections on the specified direction.
All scalar products are calculated in the lepton and (active) nucleon center-of-mass system (c.m.s.) in
the so-called mixed $\chi\eta$-basis, where $\chi$ particle currents are taken in the helicity basis 
(by analogy with neutrino), and nucleon currents are taken in the $\eta$-basis 
(or $\sigma_3$-basis, i.e., quantized with respect to the direction of the momentum of the incident particle).
The procedure for obtaining these scalar products is detailed in \cite{Bednyakov:2021pgs}.
They are needed for self-consistent calculations
of coherent and incoherent $\chi A\to \chi A^{(*)}$ scattering
cross sections and are defined by the expressions
\begin{equation*}
(l^{i}_{s's} \, h^{k}_{r'r})  \equiv \sum^{4}_{\mu,\nu} J^{i,\mu}_{s's}(\bm{k}')\, g_{\mu\nu}\, J^{k,\nu}_{r'r}(\bm{p}'),
\end{equation*} 
where subscripts $s$ and $r$  denote fixed values of the initial lepton and nucleon spin states, 
and subscripts  $s'$ and $r'$ denote the respective final states. 
Indices $i$ and $k$ designate vector ($J^{v,\mu}\equiv V^\mu$), 
axial vector ($J^{a,\mu}\equiv A^\mu$), scalar ($J^{s}\equiv S$), 
and pseudoscalar ($J^{p}\equiv P $) 
leptonic (argument $\bm{k}'$) and nucleon (argument $\bm{p}'$) currents. 
For example, the expression for the
scalar product of the axial-axial currents is
\begin{equation}\label{eq:42chiA-CrossSection-ScalarProducts-ChiEta-All-A-A} 
(l^{a}_{s's} \, h^{a}_{r'r})  = \sum^{}_{\mu,\nu} A^\mu_{s's}(\bm{k}')\, g_{\mu\nu} \, A^\nu_{r'r}(\bm{p}'). 
\end{equation}
Scalar products are expressed in terms of nucleon and $\chi$ particle masses
$$m, \quad m_\chi, \quad \text{and \ parameters} \quad 
\lambda_\pm=\sqrt{E_p\pm m}, \qquad \xi_\pm=\sqrt{E_\chi\pm m_\chi} ,$$
and the c.m.s. angle of elastic scattering of a $\chi$ lepton from a nucleon, 
i.e., the angle $\theta$  between the direction of the initial lepton momentum  $\bm{k}$ 
and the direction of the final lepton momentum $\bm{k}'$, 
where $\bm{k}+\bm{p}=\bm{k}'+\bm{p}'=0$; 
in addition, the following are valid:
\begin{eqnarray}\nonumber \label{eq:42chiA-CrossSection-ScalarProducts-ChiEta-All-Energies-etc} 
E_\chi &\equiv& \sqrt{m^2_\chi+|\bm{k}|^2} = \sqrt{m^2_\chi+|\bm{k}'|^2} = \frac{s+m^2_\chi-m^2}{2\sqrt{s}}, \\
E_p &\equiv& \sqrt{m^2+|\bm{p}|^2} =\sqrt{m^2+|\bm{p}'|^2}  = \frac{s+m^2-m^2_\chi}{2\sqrt{s}}
, \\ |\bm{p}| &=& \sqrt{E^2_p-m^2} = \lambda_+ \lambda_-  
=|\bm{k}| = \sqrt{E^2_\chi- m^2_\chi} = \xi_+ \xi_- 
= \dfrac{\lambda(s, m^2, m^2_\chi)}{2\sqrt{s}} 
\nonumber
.\end{eqnarray}
Here  $p$ and $k$  are the four-momenta of the nucleon and the $\chi$ particle
\footnote{The kinematic $\lambda$-function is defined by the expression
$\lambda^2(s, m^2, m^2_\chi)\equiv 
\big(s-(m+m_\chi)^2\big)\big(s-(m-m_\chi)^2\big)$.}.
The invariant square of the c.m.s. total energy has the form
\begin{eqnarray}\label{eq:42chiA-CrossSection-ScalarProducts-ChiEta-s-in-CMS-lepton-nucleon}
s&=&(p+k)^2 = (p_0+k_0)^2 - (\bm{p}+\bm{k})^2 \big|_{\text{c.m.s.}}= (k_0+p_0)^2 = (E_\chi+ E_p)^2
 \end{eqnarray}
In the lab frame, where the nucleon is at rest, $p_l =(m, \bm{p}_l=\bm{0})$, it is
\begin{eqnarray}\label{eq:42chiA-CrossSection-ScalarProducts-ChiEta-s-in-CLab-lepton-nucleon}
s&=& (k_l+p_l)^2\big|_{\text{Lab}} = m^2_\chi +m^2 + 2k_lp_l 
=m^2_\chi +m^2 + 2 m\sqrt{m^2_\chi + [\bm{k}^l_\chi]^2} 
.\end{eqnarray}
From (\ref{eq:42chiA-CrossSection-ScalarProducts-ChiEta-s-in-CMS-lepton-nucleon})
and (\ref{eq:42chiA-CrossSection-ScalarProducts-ChiEta-s-in-CLab-lepton-nucleon})
there follows dependence of the momentum $\bm{k}$ and kinetic energy $T_* = \dfrac{|\bm{k}^2|}{2 m_\chi}$
of the incident $\chi$ particle in the c.m.s. on its momentum
$\bm{k}^l_\chi \equiv \bm{k}^{}_l$  in the lab frame
\begin{equation}\label{A1DM-CrossSectionsEvaluations-Helper-k-from-Lab-to-CMS}
|\bm{k}|=   |\bm{k}^l_\chi| \dfrac{m}{\sqrt{s} } \quad \text{and}\quad  T_* = T_0 \dfrac{m^2}{s}, 
\quad \text{where}\quad T_0 = \dfrac{|\bm{k}^l_\chi|^2}{2 m_\chi} 
.\end{equation}
The scattering angle appears in the three-momentum transferred to the nucleon
\begin{equation}\label{eq:42chiA-CrossSection-ScalarProducts-ChiEta-q2}
\bm{q}^2=(\bm{k}-\bm{k}')^2 = |\bm{k}|^2 + |\bm{k}'|^2 - 2 |\bm{k}| |\bm{k}'|\cos\theta
= 2 |\bm{k}|^2(1-\cos\theta) \equiv \bm{q}^2_{\max} \sin^2\dfrac{\theta}{2}. 
\end{equation}
Hence, in view of the nonrelativistic approximation $\bm{q}^2 \simeq  2 m_A T_A$,  there follows
\begin{equation}\label{eq:42chiA-CrossSection-ScalarProducts-ChiEta-sin2-via-TTmax}
\sin^2\dfrac{\theta}{2} = \dfrac{\bm{q}^2}{\bm{q}^2_{\max}}
\simeq \dfrac{T_A^{}}{T_A^{\max}},
\quad \text{where}\quad T_A^{\max}  \simeq \dfrac{4 m_\chi m^2 T_0}{s m_A}
.\end{equation}
Then, knowing $m_\chi$, nucleon mass $m$, and $T_0$ in the lab frame, we can find $s$
\begin{equation} \label{A1DM-CrossSectionsEvaluations-Helper-s-invariant-in-CLab}
s= (k_l+p_l)^2\big|_{\text{Lab}} =m^2_\chi +m^2 + 2 m m_\chi \sqrt{1 + \dfrac{2 T_0}{m_\chi }}, 
\end{equation}
and calculate (in the lepton and nucleon c.m.s.) all necessary energies and momenta 
(\ref{eq:42chiA-CrossSection-ScalarProducts-ChiEta-All-Energies-etc}). 
After that, considering the determined angle $\theta$ 
(\ref{eq:42chiA-CrossSection-ScalarProducts-ChiEta-sin2-via-TTmax}) 
and the parameters $T_0$  (initial kinetic energy of the $\chi$ particle) 
and $T_A$  (kinetic energy of the nuclear recoil) specified by the lepton-nucleus scattering conditions, 
the kinematics of the (internal) elastic 2-in-2 scattering of a lepton by a nucleon in their c.m.s. 
and then all scalar products are completely determined.
\par
Below, expressions for all scalar products from \cite{Bednyakov:2021pgs} are given. 
According to (\ref{eq:42chiA-CrossSection-ScalarProducts-ChiEta-All-A-A}), a set of scalar products of
axial vector currents $(l^a,h^a)$ has the form
\begin{eqnarray*}
(l^a_{\pm\pm}\, h^a_{\pm\pm}) &=& -  4\cos\frac{\theta}{2} \Big[\xi_+ \xi_-  \lambda_+ \lambda_- \cos^2\frac{\theta}{2} + (m_\chi + \xi^2_-)(m + \lambda^2_- \cos^2\frac{\theta}{2}) \Big]
, \\ (l^a_{\pm\pm}\, h^a_{\mp\mp})&=& + 4 \cos\frac{\theta}{2} \Big[\xi_+ \xi_-\lambda_+ \lambda_-  \cos^2\frac{\theta}{2}   +(m_\chi + \xi^2_-) \big(m + \lambda^2_- (1 + \sin^2\frac{\theta}{2}) \big) \Big]
,\\  (l^a_{\pm\pm}\, h^a_{\pm\mp})&=& \mp 4 \sin\frac{\theta}{2} \Big[ \xi_+ \xi_-  \lambda_+ \lambda_- \cos^2\frac{\theta}{2}    +(m_\chi + \xi^2_-)\big(2m + \lambda^2_- (1+\sin^2\frac{\theta}{2}\big) \Big]e^{\mp i\phi}
, \\ (l^a_{\pm\pm}\, h^a_{\mp\pm}) &=&  \mp 4 \sin\frac{\theta}{2}\Big[\xi_+ \xi_-  \lambda_+ \lambda_-  \cos^2\frac{\theta}{2} + (m_\chi + \xi^2_-) \lambda^2_- \cos^2\frac{\theta}{2} \Big]  e^{\pm i\phi}
,\\  (l^a_{\pm\mp}\, h^a_{\mp\mp})&=& \mp 4m_\chi\big(m + \lambda^2_- \cos^2\frac{\theta}{2}\big) \sin\frac{\theta}{2}e^{\mp i\phi} 
,\\ (l^a_{\pm\mp}\, h^a_{\pm\pm})&=& \mp 4m_\chi \sin\frac{\theta}{2} \big(\lambda^2_-  \cos^2\frac{\theta}{2} - m\big) e^{\mp i\phi}
, \\ (l^a_{\pm\mp}\, h^a_{\mp\pm}) &=& - 4m_\chi \cos\frac{\theta}{2} \big(2m + \lambda^2_- \sin^2\frac{\theta}{2}\big)
, \quad  (l^a_{\pm\mp}\, h^a_{\pm\mp})=  + 4m_\chi \lambda^2_-  \cos\frac{\theta}{2}\sin^2\frac{\theta}{2}  e^{\mp 2 i \phi}
.\end{eqnarray*}
In the nonrelativistic approximation
\footnote{Formally due to the condition $\xi^2_- =E_\chi -m_\chi \to 0\quad \text{and} \quad \lambda^2_- =E-m\to 0$.}
these scalar products are greatly simplified
\begin{eqnarray}\nonumber\label{eq:42chiA-CrossSection-ScalarProducts-ChiEta-Axial-Nonrel}
  (l^a_{\pm\pm}\, h^a_{\pm\pm})&=&  - m^2_c ,\qquad (l^a_{\pm\pm}\, h^a_{\mp\mp})=  + m^2_c
, \qquad (l^a_{\pm\mp}\, h^a_{\mp\pm}) =  - 2 m^2_c
, \\(l^a_{+-}\, h^a_{\mp\mp})&=&  \mp m^2_s e^{-i \phi} ,\quad (l^a_{-+}\, h^a_{\mp\mp})=  \mp  m^2_s e^{+i \phi} ,\quad (l^a_{\pm\pm}\, h^a_{\pm\mp})=  \mp 2 m^2_s e^{\mp i \phi}  
\nonumber ,\\ (l^a_{\pm\pm}\, h^a_{\mp\pm})&=&(l^a_{\pm\mp}\, h^a_{\pm\mp})= 0
.\end{eqnarray}
Here the following convenient designations for fur- ther use are introduced:
\begin{equation}\label{eq:42chiA-CrossSection-ScalarProducts-ChiEta-All-nonrel-parameters} 
\displaystyle  m^2_c \equiv  4 m_\chi m  \cos\frac{\theta}{2},  \quad m^2_s \equiv  4 m_\chi m  \sin\frac{\theta}{2}.
\end{equation}
A set of scalar products of vector currents 
$ (l^{v}_{s's} \, h^{v}_{r'r}) = V^{\mu} _{s's}(\bm{k}') g_{\mu\nu} V^{\nu }_{r'r}(\bm{p}')$ 
in the nonrelativistic approximation is
\begin{eqnarray*}  
(l^v_{\pm\pm} h^v_{\pm\pm})&=& (l^v_{\pm\pm} h^v_{\mp\mp})= m^2_c 
, \quad  (l^v_{+-} h^v_{\mp\mp})= + m^2_s e^{-i\phi} ,\quad  (l^v_{-+} h^v_{\mp\mp})= - m^2_s e^{+i\phi}
,\\ (l^v_{\pm\pm} h^v_{+-})&=&  (l^v_{\mp\mp} h^v_{-+})=  
 (l^v_{+-} h^v_{\mp\pm})= (l^v_{-+} h^v_{\pm\mp}) = 0
.\end{eqnarray*} 
Scalar products of vector-axial and axial vector currents in the nonrelativistic approximation com- pletely vanish
$$(l^v_{s's}\, h^a_{r'r})_{\text{nonrel}}  = V^{\mu}_{s's}(\bm{k}') \,g_{\mu \nu}\, A^{\nu}_{r'r}(\bm{p}') \simeq 0.$$
$$ (l^a_{s's}\, h^v_{r'r})_{\text{nonrel}}  = A^{\mu}_{s's}(\bm{k}') \,g_{\mu \nu}\, V^{\nu}_{r'r}(\bm{p}') \simeq 0.$$
In the nonrelativistic approximation, only scalar products of scalar currents
$(l^s_{s's}\, h^s_{r'r})^{} = S_{s's}(\bm{k}') S_{r'r}(\bm{p}')$ corresponding to preservation of the nucleon 
spin projection on the chosen direction “survive,” while all the other products of scalars vanish
\begin{eqnarray}\nonumber \label{eq:42chiA-CrossSection-ScalarProducts-ChiEta-Scalar-Nonrel}
(l^s_{\pm\pm}\, h^s_{\pm\pm})^{}&=& (l^s_{\pm\pm}\, h^s_{\mp\mp})^{}= m^2_c
, \quad (l^s_{+-}\, h^s_{\mp\mp})^{}=   +m^2_s e^{-i\phi},  \quad (l^s_{-+}\, h^s_{\mp\mp})^{}=  -m^2_se^{+i\phi}
, \\ (l^s_{-+}\, h^s_{\pm\mp})^{}&=&(l^s_{+-}\, h^s_{\mp\pm})^{}   =
(l^s_{\mp\mp}\, h^s_{+-})^{}=(l^s_{\mp\mp}\, h^s_{-+})^{}= 0 
.\end{eqnarray}
These expressions entirely coincide with the similar ones for vector currents. All scalar products of scalar- pseudoscalar currents, pseudoscalar-scalar currents, and pseudoscalar currents vanish in the nonrelativistic approximation
\begin{eqnarray*}
(l^s_{s's}\, h^p_{r'r})^{} &=& S_{s's}(\bm{k}') P_{r'r}(\bm{p}')=(l^p_{s's}\, h^s_{r'r})_{\text{nonrel}} =0
, \\ (l^p_{s's}\, h^s_{r'r})^{} &=& P_{s's}(\bm{k}') S_{r'r}(\bm{p}')=(l^s_{s's}\, h^p_{r'r})_{\text{nonrel}} = 0
,\\ (l^p_{s's}\, h^p_{r'r})^{} &=& P_{s's}(\bm{k}') P_{r'r}(\bm{p}')=(l^p_{s's}\, h^p_{r'r})_{\text{nonrel}} = 0
.\end{eqnarray*}
Scalar products of the leptonic ($\chi$ particle) and nucleon currents for any spin projections of the nucleon 
 ($r',r=\pm1$)  and the $\chi$ particle ($s',s=\pm1$, 
 which correspond to the neutral weak current interaction of a massive neutral $\chi$ particle 
 with a nucleon, are defined as \cite{Bednyakov:2021pgs}
\begin{eqnarray}\label{eq:42chiA-CrossSection-ScalarProducts-ChiEta-All-WeakGeneralCurrent-definition} 
(l^w_{s's}, h^{w,f}_{r'r})   &=& \alpha_f (l^v_{s's}\, h^v_{r'r}) + \beta_f (l^v_{s's}\, h^a_{r'r}) 
+ \gamma_f (l^a_{s's}\, h^v_{r'r})  +  \delta_f  (l^a_{s's}\, h^a_{r'r}) ,
\end{eqnarray} 
\begin{eqnarray*}
\text{where} \quad  \alpha_f = \chi_V h^f_V  = + g_V^f ,\quad  \beta_f = \chi_V h^f_A = - g_A^f
, \quad \gamma_f = \chi_A  h^f_V = - g_V^f ,\quad \delta_f = \chi_A h^f_A = + g_A^f
. \end{eqnarray*}
The index $f$ denotes a neutron or a proton. 
On the right are the values of these constants for the neutrino Standard Model (since in the neutrino
$\chi_V  = - \chi_A = 1$)). 
In the nonrelativistic approximation, scalar products for the weak neutral current of a massive 
particle (\ref{eq:42chiA-CrossSection-ScalarProducts-ChiEta-All-WeakGeneralCurrent-definition})  have the form
\begin{eqnarray} \nonumber \label{eq:42chiA-CrossSection-ScalarProducts-ChiEta-All-Weak-Nonrel}
(l^w_{\pm\pm}, h^{w,f}_{\pm\pm}) &=&   m^2_c (\alpha_f - \delta_f)
, \quad (l^w_{\pm\pm}, h^{w,f}_{\mp\mp}) =  m^2_c (\alpha_f + \delta_f)
,\quad  (l^w_{\pm\mp}, h^{w,f}_{\mp\pm}) = - 2 m^2_c \delta_f
,\\ (l^w_{\pm\mp}, h^{w,f}_{\mp\mp}) &= &  \pm m^2_s e^{\mp i\phi} (\alpha_f-\delta_f)  
, \quad  (l^w_{\pm\mp}, h^{w,f}_{\pm\pm}) = \pm m^2_s e^{\mp i\phi} (\alpha_f+\delta_f)
, \\  (l^w_{\pm\pm}, h^{w,f}_{\pm\mp})  &= &  \mp 2m^2_s e^{\mp i \phi} \delta_f 
,\quad (l^w_{\pm\pm}, h^{w,f}_{\mp\pm}) =  (l^w_{\pm\mp}, h^{w,f}_{\pm\mp})^{}\simeq  0
.  \nonumber  \end{eqnarray}
The set of scalar products corresponding to the $\chi$ particle-nucleon interaction, which is given in this section, is a basis for calculating coherent (elastic) and incoherent (inelastic) cross sections of the $\chi$ particle–nucleus interaction in the nonrelativistic approximation.

\subsection{\normalsize\em Expansion of scalar products in cross sections of ${\chi A\to \chi A^{(*)}}$ scattering} 
\label{43chiA-CrossSection-via-ScalarProducts} 
According to the results of Section \ref{41chiA-CrossSection-Coh-vs-InCoh}, 
the expression for the observed differential cross section
$\displaystyle \frac{d\sigma}{dT_A}(\chi A\to \chi A^{(*)})$  involves the incoherent (inelastic)
and coherent (elastic) terms. 
They are given by formulas (\ref{eq:41chiA-CrossSection-Coh-vs-InCoh-CS-with-s-sprime-for-proton-neutron})
written below in the following form:
\begin{eqnarray} \nonumber\label{eq:43chiA-CrossSection-via-ScalarProducts-in-Apm}
\frac{d\sigma^{s's}_\text{coh}(\bm{q})}{g_\text{c}c_A d T_A} &=& 
\Big| \sum_{f=p,n}F_f(\bm{q})  [A^f_+(l_{s's},h^{\eta,f}_{++})+ A^f_-(l_{s's},h^{\eta,f}_{--})] \Big|^2
,\\
\frac{d\sigma^{s's}_\text{inc}(\bm{q})}{g_\text{i} c_A d T_A} &=& 
\sum_{f=p,n}[1-|F_f(\bm{q})|^2] \Big[A^f_+  \sum^{}_{r'=\pm}|(l_{s's}, h^{\eta,f}_{r'+})|^2+A^f_- \sum^{}_{r'=\pm}|(l_{s's},h^{\eta,f}_{r'-})|^2 \Big]   
.\end{eqnarray}
Here $A^f_\pm$  is the number of $f$-type nucleons with the  spin projection  $\pm 1$ on the preferred direction 
(e.g., that of the $\chi$ particle arrival), and the universal common factor 
$c_A \equiv c_A(T_0, m_A, m_\chi) $ from (\ref{eq:43chiA-CrossSection-via-ScalarProducts-c_A-definition})
is introduced for convenience. 
Helicities of the $\chi$ particle (in a more general form, spin projections on a certain direction) in the initial $s$ and final $s'$ 
states are considered to be fixed in (\ref{eq:43chiA-CrossSection-via-ScalarProducts-in-Apm}). 
Later, averaging (summation) can be performed over them.
\par
Formulas  (\ref{eq:43chiA-CrossSection-via-ScalarProducts-in-Apm})  can be conveniently written in terms
of the total number of  $f$-type nucleons, $A_f$, and the difference in the number of nucleons,  $\Delta A_f $, 
having a positive and a negative spin projection on the preferred direction. With a simple transformation
\begin{equation}\label{eq:43chiA-CrossSection-via-ScalarProducts-to-A_f-and-DeltaA_f}
 A^f_\pm =\dfrac12 (A_f\pm \Delta A_f),
\quad\text{where}\quad  A_f\equiv A^f_+ + A^f_- \quad\text{and}\quad   \Delta A_f \equiv A^f_+ - A^f_-
,\end{equation}  
cross sections  (\ref{eq:43chiA-CrossSection-via-ScalarProducts-in-Apm})
take a key form for further consideration
\begin{eqnarray} \nonumber  \label{eq:43chiA-CrossSection-via-ScalarProducts-both-CS-main} 
\frac{d\sigma^{s's}_\text{coh}(\bm{q})}{g_\text{c} c_A  d T_A}&=& 
 \Big| \sum^{}_{f=p,n} F_{f}(\bm{q}) \dfrac{A_f}{2}  Q^f_{s's}  \Big|^2
,\\  \frac{d\sigma^{s's}_\text{inc}(\bm{q})}{g_\text{i} c_A  d T_A} &=& 
 \sum_{f=p,n}  \big[1-F^2_f(\bm{q})\big] \dfrac{ A_f  }{2}   \Big[ \, Q^{s's}_+  + \dfrac{\Delta A_f }{A_f } \, Q^{s's}_-  \Big]
.\end{eqnarray}
Thus, the coherent (elastic) and incoherent (inelastic) $\chi A$ cross sections are defined by the following 
respective combinations of scalar products:
\begin{equation}\label{eq:43chiA-CrossSection-via-ScalarProducts-CohCS-factors}
Q_f^{s's}\equiv Q^f_{s's}\equiv \hat Q^{s's}_{+} + \dfrac{\Delta A_f}{A_f} \hat Q^{s's}_{-} , 
\quad \text{where} \quad  \hat Q^{s's}_{\pm}\equiv (l_{s's},h^{f}_{++}) \pm (l_{s's},h^{f}_{--}) 
,\end{equation}
\begin{equation}\label{eq:43chiA-CrossSection-via-ScalarProducts-InCohCS-factors}
\text{and}\quad  Q^{s's}_\pm\equiv 
\sum^{}_{r'=\pm}|(l_{s's}, h^{f}_{r'+})|^2 \pm  \sum^{}_{r'=\pm}|(l_{s's},h^{f}_{r'-})|^2
.\end{equation}
In  (\ref{eq:43chiA-CrossSection-via-ScalarProducts-InCohCS-factors}), 
the superscripts and the subscript respectively correspond to the $\chi$-particle spin projections 
and the nucleon spin projection on the preferred direction.
It is seen that $Q^{s's}_+$  is the total sum of squares of scalar products over all nucleon spin projections 
for a fixed pair of lepton helicities $s'$ and $s$. 
As a rule, if the nucleus has the total zero spin, then $ \Delta A_f= 0$.
Note also nearly always  ${\Delta A_f }\ll {A_f }$ holds (except probably
for the lightest nuclei).
\par
Since the emphasis in this work is on the nonrelativistic approximation, 
recall that the nonrelativistic approximation corresponds to the case when the condition 
of the smallness of the kinetic energy of the $\chi$ particle compared to its rest energy  (at $c=1$) is satisfied, i.e.,
\begin{equation}\label{eq:43chiA-CrossSection-via-ScalarProducts-NonREL-definition}
T_0\equiv \dfrac{|\bm{k}^l_\chi |^2}{2 m_\chi} \simeq 10^{-6} c^2  m_\chi  \ll m_\chi \quad \text{or}\quad 
\dfrac{T_0}{ m_\chi}  \simeq \dfrac{ |\bm{k}^l_\chi|^2}{ m^2_\chi} \simeq 10^{-6} \ll 1. 
\end{equation}
In this approximation, the square of the total energy of the lepton and the (active) nucleon in the lab frame 
(where the nucleus/nucleon is at rest) has the form
$$ s =  m^2_\chi +m^2 + 2 m m_\chi\sqrt{1 + [\bm{k}^l_\chi]^2/m^2_\chi} \equiv (m_\chi +m) ^2 + 2 m T_0. $$ 
Here the second term proportional to $T_0$, is much smaller than the first one 
(and is only retained as a leading-order contribution with respect to the smallness parameter $T_0/m$).

\subsubsection{\normalsize\em Scalar ${\chi A}$  interaction}
By scalar interaction is meant interaction of a $\chi$ particle with nucleons that leads to scalar products of the scalar 
leptonic and scalar nucleon currents in the form
\begin{equation}\label{eq:43chiA-CrossSection-via-ScalarProducts-Scalar-ScalarProduct-deff}
(l^s_{s's}, h^{s,f}_{r'r})=c_{S}^f\, \overline{u}_\chi (s',k') u_\chi (s,k) \times \overline{u}_f (r',p') u_f (r,p) = c_{S}^f\, S_{s's}(\bm{k}') S_{r'r}(\bm{p}'), 
\end{equation}
where effective coupling constants $\ c_{S}^f \equiv \chi^{}_{S} h^f_{S}$ from 
(\ref{eq:3chiA-ScatteringAmplitude-Lepton-S+A-Operator}) and
(\ref{eq:3chiA-ScatteringAmplitude-Nucleon-S+A-Operator})
specify intensity of this interaction (in terms of $G_{\rm F}$). 
In the nonrelativistic approximation, these scalar products are given by formulas 
(\ref{eq:42chiA-CrossSection-ScalarProducts-ChiEta-Scalar-Nonrel}).
Calculation of coherent cross sections requires the following combinations of scalar products:
\begin{eqnarray*}
 \hat Q^{\mp\mp}_{+}&=& (l_{\mp\mp},h^{n}_{++}) + (l_{\mp\mp},h^{n}_{--})= m^2_c + m^2_c = 2 m^2_c
, \\  \hat Q^{\mp\mp}_{-}&=& (l_{\mp\mp},h^{n}_{++})- (l_{\mp\mp},h^{n}_{--})= m^2_c - m^2_c =0 
, \\   \hat Q^{\mp\pm}_{+}&=& (l_{\mp\pm},h^{n}_{++}) + (l_{\mp\pm},h^{n}_{--}) = \mp m^2_se^{\pm i\phi} + (\mp m^2_se^{\pm i\phi}) = \mp 2m^2_se^{\pm i\phi}
, \\  \hat Q^{\mp\pm}_{-}&=& (l_{\mp\pm},h^{n}_{++})- (l_{\mp\pm},h^{n}_{--}) = \mp m^2_se^{\pm i\phi} 
 - (\mp m^2_se^{\pm i\phi}) =0
.\end{eqnarray*}
With these combinations, coherent scattering cross sections are calculated from 
(\ref{eq:43chiA-CrossSection-via-ScalarProducts-both-CS-main}) and
(\ref{eq:43chiA-CrossSection-via-ScalarProducts-CohCS-factors}) in terms of the quantities
\begin{eqnarray*}
Q_f^{\mp\mp} = \hat Q^{\mp\mp}_{+} + \dfrac{\Delta A_f}{A_f}  \hat Q^{\mp\mp}_{-}  = 2 m^2_c 
\quad\text{and}\quad Q_f^{\mp\pm} = \hat Q^{\mp\pm}_{+} + \dfrac{\Delta A_f}{A_f}  \hat Q^{\mp\pm}_{-}
= \mp 2m^2_se^{\pm i\phi}
.\end{eqnarray*}
A set of nonrelativistic cross sections for the coherent $\chi A$ scattering in the scalar-current interaction 
channel with the $\chi$-particle helicity unchanged (first formula)  and helicity flip (second formula) can be written as
\begin{eqnarray}\nonumber\label{eq:42chiA-CrossSectionsEvaluations-Scalar-ChiEta-nonrel-CohCS}
\frac{d\sigma^{\mp\mp}_\text{coh}(\bm{q})}{d T_A} &=&c_A    \big| \sum^{}_{f=p,n} F_{f}(\bm{q}) {A_f} c_S^{f} m^2_c   \big|^2 = \dfrac{G^2_F  m_A }{4\pi}\cos^2\frac{\theta}{2}  \dfrac{m^2_\chi  }{|\bm{k}^l_\chi|^2}
 \big[ \sum^{}_{f=p,n} F_{f}(\bm{q}) {A_f} c_S^{f} \big]^2
,\\ \frac{d\sigma^{\mp\pm}_\text{coh}(\bm{q})}{ d T_A} &=& c_A \big| \sum^{}_{f=p,n} F_{f}(\bm{q}) {A_f} c_S^{f} m^2_s \big|^2 =\dfrac{G^2_F  m_A }{4\pi}\sin^2\frac{\theta}{2}  \dfrac{m^2_\chi  }{|\bm{k}^l_\chi|^2 }
 \big[ \sum^{}_{f=p,n} F_{f}(\bm{q}) {A_f} c_S^{f} \big]^2,\qquad 
 \\ \dfrac{d\sigma^{\text{total}}_\text{coh}(\bm{q})}{d T_A}&=&
\dfrac12 \sum^{}_{s's}\dfrac{d\sigma^{s's}_\text{coh}}{d T_A}
 =\dfrac{G^2_F  m_A }{4\pi} \dfrac{m^2_\chi  }{|\bm{k}^l_\chi|^2 }
 \big[ \sum^{}_{f=p,n} F_{f}(\bm{q}) {A_f} c_S^{f} \big]^2
\nonumber.\end{eqnarray}
The last formula gives the total (averaged over initial lepton helicities and summed over final lepton helicities) 
coherent cross section for the $\chi A$ scattering due to interaction of scalar currents in the nonrelativistic limit. 
Here we use
\begin{equation}\label{43chiA-CrossSection-via-ScalarProducts-c_A+}
c_A (4m_\chi m)^2 = (4m_\chi m)^2 \dfrac{ G^2_F  m_A}{2^6\pi m^2 |\bm{k}^l_\chi|^2 }
 = \dfrac{ G^2_F  m_A}{4\pi}\dfrac{m^2_\chi }{|\bm{k}^l_\chi|^2 },
\end{equation}  and it is considered that  $g_\text{c}\simeq 1$.
\par 
Incoherent $\chi A$-cross sections are defined by general formulas 
 (\ref{eq:43chiA-CrossSection-via-ScalarProducts-both-CS-main}) and 
 (\ref{eq:43chiA-CrossSection-via-ScalarProducts-InCohCS-factors}). 
They depend only on squares of scalar products of scalar currents through the parameters
\begin{eqnarray*}
Q^{s's}_\pm &\equiv&  S_+^{s's} \pm S_-^{s's},  \quad \text{where} \quad  
S_\pm^{s's} \equiv \sum^{}_{r'=\pm}|(l_{s's}, h^{f}_{r'\pm})|^2 = |(l_{s's}, h^{f}_{+\pm})|^2+|(l_{s's}, h^{f}_{-\pm})|^2
.\end{eqnarray*}
In the nonrelativistic limit, for the squares of these scalar products we have
 \begin{eqnarray}\nonumber\label{eq:42chiA-CrossSectionsEvaluations-Scalar-ChiEta-nonrel-ScalarProducts^2}
|(l^s_{--}\, h^s_{--})|^2&=& |(l^s_{++}\, h^s_{--})|^2=|(l^s_{--}\, h^s_{++})|^2=|(l^s_{++}\, h^s_{++})|^2 =  m^4_c
, \\ |(l^s_{+-}\, h^s_{--})|^2&=&  |(l^s_{+-}\, h^s_{++})|^2 =|(l^s_{-+}\, h^s_{--})|^2=|(l^s_{-+}\, h^s_{++})|^2 =  m^4_s
, \\ \nonumber  |(l^s_{-+}\, h^s_{+-})|^2&=&|(l^s_{+-}\, h^s_{-+})|^2  =
|(l^s_{-+}\, h^s_{-+})|^2 =|(l^s_{+-}\, h^s_{+-})|^2 =  0 , \\\nonumber
|(l^s_{--}\, h^s_{+-})|^2&=&|(l^s_{++}\, h^s_{+-})|^2=|(l^s_{--}\, h^s_{-+})|^2=|(l^s_{++}\, h^s_{-+})|^2 = 0 \nonumber
.\end{eqnarray}
Then one obtains that $\ Q^{\mp\mp}_+ = 2 m^4_c,\ Q^{\mp\pm}_+ = 2 m^4_s, \ Q^{\mp\mp}_- = Q^{\mp\pm}_- = 0.$
As a result, the set of cross sections for incoherent $\chi A$ scattering due to interaction of scalar currents 
with $\chi$-particle helicity unchanged (first formula) and helicity flip (second formula) 
according to the formula in (\ref{eq:43chiA-CrossSection-via-ScalarProducts-both-CS-main}), 
and the total $\chi A$ cross section (last line) are as follows:
\begin{eqnarray}\nonumber \label{eq:42chiA-CrossSectionsEvaluations-Scalar-ChiEta-nonrel-InCohCS}
\frac{d\sigma^{\pm\pm}_\text{inc}(\bm{q})}{d T_A}  &=& c_A   \sum_{f=p,n} A^f  \big[1-F^2_f(\bm{q})\big] [c_S^f]^2 m^4_c   = \dfrac{G^2_F  m_A }{4\pi}\cos^2\frac{\theta}{2}  \dfrac{m^2_\chi  }{|\bm{k}^l_\chi|^2}
 \sum_{f=p,n} A^f  [c_S^f]^2\big[1-F^2_f(\bm{q})\big]
  , \\ \nonumber
\frac{d\sigma^{\pm\mp}_\text{inc}(\bm{q})}{d T_A} &=& c_A   \sum_{f=p,n}A^f \big[1-F^2_f(\bm{q})\big]  [c_S^f]^2m^4_s = \dfrac{G^2_F  m_A }{4\pi}\sin^2\frac{\theta}{2}  \dfrac{m^2_\chi  }{|\bm{k}^l_\chi|^2}
 \sum_{f=p,n} A^f [c_S^f]^2 \big[1-F^2_f(\bm{q})\big]
,\\
 \dfrac{d\sigma^{\text{total}}_\text{inc}}{d T_A}&=&\dfrac12 \sum_{s',s} \dfrac{d\sigma^{s's}_\text{inc}}{d T_A}= \dfrac{G^2_F  m_A }{4\pi}  \dfrac{m^2_\chi  }{|\bm{k}^l_\chi|^2} \sum_{f=p,n} A^f [c_S^f]^2 \big[1-F^2_f(\bm{q})\big]  
.\end{eqnarray}
Here relation (\ref{43chiA-CrossSection-via-ScalarProducts-c_A+}) is considered, and we take $g_\text{i} \simeq 1$.
\par
Formulas (\ref{eq:42chiA-CrossSectionsEvaluations-Scalar-ChiEta-nonrel-CohCS}) and 
(\ref{eq:42chiA-CrossSectionsEvaluations-Scalar-ChiEta-nonrel-InCohCS}) 
give a general form of the total set of expressions for cross sections of the $\chi A$ scattering due to the interaction 
of the scalar leptonic and scalar nucleon currents in the nonrelativistic approximation. 
The total (experimentally measured) cross section of this interaction is expressed in terms 
of the sum of the coherent and incoherent cross sections
\begin{eqnarray}\label{eq:42chiA-CrossSectionsEvaluations-Scalar-ChiEta-nonrel-Full}
\dfrac{d\sigma^{\text{scalar}}_{\text{nonrel}}}{d T_A}(\chi A \to \chi A^*)&=&\dfrac12\sum^{}_{s's}\dfrac{d\sigma^{s's}_\text{inc}}{d T_A} + \dfrac12\sum^{}_{s's}\dfrac{d\sigma^{s's}_\text{coh}}{d T_A}
=\\&=& \dfrac{G^2_F  m_A }{4\pi}  \dfrac{m^2_\chi  }{|\bm{k}^l_\chi|^2}\Big\{\sum_{f=p,n} A^f [c_S^f]^2 \big[1-F^2_f(\bm{q})\big]+ \big[ \sum^{}_{f=p,n} F_{f}(\bm{q}) {A_f} c_S^{f} \big]^2 \Big\} 
\nonumber. \end{eqnarray}
On the assumption of identity of nuclear proton and neutron form factors, $F_{p}(\bm{q})=F_{n}(\bm{q})\equiv F(\bm{q})$, 
and isoscalar character of the scalar interaction, $c_S^n=c_S^p\equiv c_S^{}$, 
simple formulas for measured cross sections are obtained, which involve terms proportional
to $A^2=[A_p+A_n]^2$ and $A=A_p+A_n$, respectively,
\begin{eqnarray*}\nonumber
\dfrac{d\sigma^{\text{total}}}{d T_A}(\chi A \to \chi A^*)&=&
\dfrac{G^2_F  m_A }{4\pi}c_S^2  \dfrac{m^2_\chi  }{|\bm{k}^l_\chi|^2}
\Big\{ F^2(\bm{q}) A^2 + [1-F^2(\bm{q})] A  \Big\}
,\\ \dfrac{d\sigma^{\mp\mp}}{d T_A}(\chi A \to \chi A^*)&=&\cos^2\frac{\theta}{2}
\dfrac{d\sigma^{\text{total}}}{d T_A}(\chi A \to \chi A^*) 
, \\ \dfrac{d\sigma^{\mp\pm}}{d T_A}(\chi A \to \chi A^*)&=&\sin^2\frac{\theta}{2}
\dfrac{d\sigma^{\text{total}}}{d T_A}(\chi A \to \chi A^*) 
 .\end{eqnarray*}
The ratio between the total (differential) cross sections for the incoherent 
 (\ref{eq:42chiA-CrossSectionsEvaluations-Scalar-ChiEta-nonrel-InCohCS}) and coherent 
 (\ref{eq:42chiA-CrossSectionsEvaluations-Scalar-ChiEta-nonrel-CohCS}) 
$\chi A$ scattering due to scalar interaction in the nonrelativistic approximation has a simple form
\begin{eqnarray}\label{eq:43chiA-CrossSectionsEvaluations-Scalar-ChiEta-nonrel-InCoh-2-Coh-total-ratio}
R_{\rm scalar}(m_A, T_A)\equiv  \dfrac{\dfrac{d\sigma^{\text{total}}_\text{inc}}{d T_A}(\chi A\to \chi A^{*})}{\dfrac{d\sigma^{\text{total}}_\text{coh}}{d T_A}(\chi A\to \chi A^{})} &=&\dfrac{1-F^2(\bm{q})}{F^2(\bm{q})A}
.\end{eqnarray}
This ratio is obviously equal to unity, i.e., the incoherent and coherent contributions to the cross section
are equal when $F^2(\bm{q})=(A+1)^{-1}.$ 
For example, in the case of a hydrogen target ($A=1$),  a “drop” to $1/2$ for
the square of the nuclear form factor is sufficient for the scalar-isoscalar 
coherent contribution to the cross section to become equal to the incoherent contribution.

\subsubsection{\normalsize\em  Axial vector  ${\chi A}$ interaction}
By axial vector (or axial) interaction is meant interaction of a $\chi$ particle with nucleons that leads to the 
following form of scalar products (\ref{eq:42chiA-CrossSection-ScalarProducts-ChiEta-All-A-A}):
\begin{eqnarray}\label{eq:43chiA-CrossSection-via-ScalarProducts-Axial-ScalarProduct-deff}
(l^a_{s's}, h^{a,f}_{r'r})\!=\!\chi^{}_{A} \overline{u}_\chi (s',k')\gamma_\mu\gamma_5 u_\chi (s,k)
h^f_{A} \overline{u}_f (r',p') \gamma^\mu\gamma_5 u _f (r,p)
\!=\!c_{A}^f \sum^{}_{\mu,\nu}\!A^\mu_{s's}(\bm{k}')g_{\mu\nu}A^\nu_{r'r}(\bm{p}') ,\quad 
\end{eqnarray}
where the effective coupling constants $\ c_{A}^f \equiv \chi^{}_{A} h^f_{A}$ specify intensity of this interaction. 
All necessary scalar products are given in Section \ref{42chiA-CrossSection-ScalarProducts-ChiEta-All}.
\par 
In the nonrelativistic approximation, combinations 
(\ref{eq:43chiA-CrossSection-via-ScalarProducts-CohCS-factors})  of scalar products of the axial 
$\chi A$ interaction from (\ref{eq:42chiA-CrossSection-ScalarProducts-ChiEta-Axial-Nonrel}) 
involved in the coherent cross sections are
\begin{eqnarray*}
\hat Q^{\mp\mp}_{+}&=&  (l^a_{\mp\mp},h^{a}_{++}) + (l^a_{\mp\mp},h^{a}_{--}) = \pm m^2_c +(\mp m^2_c) = 0
,\\ \hat Q^{\mp\mp}_{-}&=& (l^a_{\mp\mp},h^{a}_{++})- (l^a_{\mp\mp},h^{a}_{--}) = \pm m^2_c - (\mp m^2_c) = \pm 2 m^2_c 
,\\  \hat Q^{\mp\pm}_{+}&=& (l^a_{\mp\pm},h^{a}_{++}) + (l^a_{\mp\pm},h^{a}_{--})=  m^2_s e^{\pm i \phi}
+ (-  m^2_s e^{\pm i \phi}) = 0 
,\\  \hat Q^{\mp\pm}_{-}&=& (l^a_{\mp\pm},h^{a}_{++})- (l^a_{\mp\pm},h^{a}_{--})=  m^2_s e^{\pm i \phi} - (-  m^2_s e^{\pm i \phi}) = 2 m^2_s e^{\pm i \phi}
.\end{eqnarray*}
Hence, for all possible lepton helicity combinations we have four quantities from 
 (\ref{eq:43chiA-CrossSection-via-ScalarProducts-CohCS-factors}) 
\begin{eqnarray*}
Q_f^{\mp\mp} &=& \pm  2 m^2_c\frac{\Delta A_f}{A_f}  \quad 
\text{and}\quad Q_f^{\mp\pm}=  2 m^2_s e^{\pm i \phi} \frac{\Delta A_f}{A_f} 
. \end{eqnarray*}
As a result, coherent $\chi A$ cross sections (\ref{eq:43chiA-CrossSection-via-ScalarProducts-both-CS-main})
due to interaction of axial currents with the 
$\chi$-particle helicity unchanged (upper formula) and helicity flip (middle formula), 
and the total coherent cross section (lower formula) in the nonrelativistic approximation take the form
\begin{eqnarray}\nonumber\label{eq:43chiA-CrossSectionsEvaluations-Axial-ChiEta-nonrel-CohCS}
\frac{d\sigma^{\mp\mp}_\text{coh}(\bm{q})}{g_\text{c} d T_A}&=& 
 c_A m^4_c \Big[\sum^{}_{f=p,n} c_{A}^f F_{f} \Delta A_f  \Big]^2 = \cos^2\frac{\theta}{2}  
\dfrac{ G^2_F  m_A}{4\pi}\dfrac{m^2_\chi }{|\bm{k}^l_\chi|^2 } 
\big[c_{A}^p F_p(\bm{q}) \Delta A_p +c_{A}^n F_n(\bm{q}) \Delta A_n \big]^2 
, \\  \frac{d\sigma^{\mp\pm}_\text{coh}(\bm{q})}{g_\text{c} d T_A}&=& \sin^2\frac{\theta}{2}
\dfrac{ G^2_F  m_A}{4\pi}\dfrac{m^2_\chi }{|\bm{k}^l_\chi|^2 } 
\big[ c_{A}^p F_p(\bm{q}) \Delta A_p +c_{A}^n F_n(\bm{q}) \Delta A_n \big]^2 
,\\. \dfrac{d\sigma^{\text{axial}}_\text{coh}(\bm{q})}{g_\text{c} d T_A}&=&\dfrac12 \sum\dfrac{d\sigma^{s's}_\text{coh}}{g_\text{c} d T_A}
= \dfrac{ G^2_F  m_A}{4\pi}\dfrac{m^2_\chi }{|\bm{k}^l_\chi|^2 }  \big[\sum^{}_{f=p,n} c_{A}^f \Delta A_f  F_{f}(\bm{q}) \big]^2
\nonumber .\end{eqnarray}
In the formula for incoherent cross sections 
(\ref{eq:43chiA-CrossSection-via-ScalarProducts-both-CS-main}) and 
(\ref{eq:43chiA-CrossSection-via-ScalarProducts-InCohCS-factors})
 in the nonrelativistic approximation there only remains dependence on the following squares of scalar products:
\begin{eqnarray*} 
|(l^a_{++}\, h^a_{--})|^2&=&|(l^a_{--}\, h^a_{++})|^2 =
|(l^a_{++}\, h^a_{++})|^2 = |(l^a_{--} h^a_{--}) |^2 = m^4_c
,\\|(l^a_{++}\, h^a_{-+})|^2 &=&|(l^a_{--}\, h^a_{+-})|^2 =
|(l^a_{+-}\, h^a_{+-})|^2 = |(l^a_{-+}\, h^a_{-+})|^2 = 0 
, \\|(l^a_{+-}\, h^a_{--})|^2 &=&|(l^a_{-+}\, h^a_{++})|^2= 
|(l^a_{+-}\, h^a_{++})|^2 =|(l^a_{-+}\, h^a_{--})|^2  = m^4_s
, \\ |(l^a_{++}\, h^a_{+-})|^2&=&|(l^a_{--}\, h^a_{-+})|^2  = 4 m^4_s 
,\qquad  |(l^a_{+-}\, h^a_{-+})|^2 = |(l^a_{-+}\, h^a_{+-})|^2 = 4 m^4_c
, \end{eqnarray*} 
based on which expressions (\ref{eq:43chiA-CrossSection-via-ScalarProducts-InCohCS-factors})
take the form
\begin{eqnarray*}
Q^{\mp\mp}_+ &=& 2 (m^4_c + 2 m^4_s), \quad Q^{\mp\pm}_+ = 2(2 m^4_c+  m^4_s),
\quad Q^{\mp\mp}_- = \pm 4 m^4_s,  \quad Q^{\mp\pm}_-  = \mp 4 m^4_c
.\end{eqnarray*}
As a result, cross sections of the incoherent $\chi A$ interaction 
(\ref{eq:43chiA-CrossSection-via-ScalarProducts-both-CS-main})
due to axial currents with no change in the $\chi$-particle helicity (first formula) 
and helicity flip (second formula) and the total $\chi A$ cross section (last formula) 
in the nonrelativistic approximation are expressed as follows:
\begin{eqnarray} \nonumber\label{eq:43chiA-CrossSectionsEvaluations-Axial-ChiEta-nonrel-InCohCS}
\frac{d\sigma^{\mp\mp}_\text{inc}(\bm{q})}{d T_A}&=& 
\dfrac{ G^2_F  m_A}{4\pi}\dfrac{m^2_\chi }{|\bm{k}^l_\chi|^2 } g_\text{i}  
\sum_{f=p,n} [c^f_A]^2 A_f \big[1-F^2_f(\bm{q})\big]
\Big[1+\sin^2\frac{\theta}{2} \pm \sin^2\frac{\theta}{2} \dfrac{2\Delta A_f }{A_f } \Big]
, \\ \frac{d\sigma^{\mp\pm}_\text{inc}(\bm{q})}{d T_A}&=&
\dfrac{ G^2_F  m_A}{4\pi}\dfrac{m^2_\chi }{|\bm{k}^l_\chi|^2 } g_\text{i}
\sum_{f=p,n} [c^f_A]^2 A_f\big[1-F^2_f(\bm{q})\big]
\Big[1+\cos^2\frac{\theta}{2}\mp \cos^2\frac{\theta}{2} \dfrac{2\Delta A_f }{A_f } \Big]
,\qquad \\ \dfrac{d\sigma^{\text{axial}}_\text{inc}}{d T_A} &=& 
 \dfrac12 \sum_{s',s} \dfrac{d\sigma^{s's}_\text{inc}}{d T_A}= 
\dfrac{ G^2_F  m_A}{4\pi}\dfrac{m^2_\chi }{|\bm{k}^l_\chi|^2 } 
g_\text{i} \sum_{f=p,n} [c^f_A]^2 A_f\big[1-F^2_f(\bm{q})\big] \, 3
 \nonumber .\end{eqnarray}
 The experimentally measured total cross section of the $\chi A$ scattering due to interaction of pure 
 axial leptonic and nucleon currents has the form of a sum of two terms
 \begin{eqnarray}\label{eq:43chiA-CrossSectionsEvaluations-Axial-ChiEta-nonrel-Full}
\dfrac{d\sigma^{\text{axial}}_{\text{nonrel}}}{d T_A}(\chi A \to \chi A^*)&=&\dfrac12\sum^{}_{s's}\dfrac{d\sigma^{s's}_\text{inc}}{d T_A} + \dfrac12 \sum^{}_{s's}\dfrac{d\sigma^{s's}_\text{coh}}{d T_A} 
=\\&=& \dfrac{G^2_F  m_A }{4\pi}  \dfrac{m^2_\chi  }{|\bm{k}^l_\chi|^2}
\Big\{3 g_\text{i}  \sum_{f=p,n} [c^f_A]^2  A_f\big[1-F^2_f(\bm{q})\big]
+  g_\text{c} \big[\sum^{}_{f=p,n} c_{A}^f \Delta A_f  F_{f}(\bm{q}) \big]^2 \Big\} 
\nonumber. \end{eqnarray}
The ratio of the axial total cross sections for the incoherent and coherent $\chi A$ scattering 
in the nonrelativistic approximation is
 \begin{eqnarray}\label{eq:43chiA-CrossSectionsEvaluations-Axial-ChiEta-nonrel-InCoh-2-Coh-total-ratio}
\dfrac{\dfrac{d\sigma^{\text{total}}_\text{inc}}{d T_A}(\chi A\to \chi A^{*})}{\dfrac{d\sigma^{\text{total}}_\text{coh}}{d T_A}(\chi A\to \chi A)}
&=&3 \dfrac{g_\text{i}}{g_\text{c}} 
\dfrac{\sum_{f=p,n} [c^f_A]^2  A_f\big[1-F^2_f(\bm{q})\big]}{\big[\sum^{}_{f=p,n} c_{A}^f \Delta A_f  F_{f}(\bm{q}) \big]^2 }
\simeq \dfrac{ 3A [1-F^2(\bm{q})]}{F^2(\bm{q})  \big[ \Delta A_p +\Delta A_n \big]^2}
.\qquad 
\end{eqnarray}
The last approximation in this formula corresponds to the assumption that nuclear proton and neutron form factors are identical, 
i.e., $F_{p}(\bm{q})=F_{n}(\bm{q})\equiv F(\bm{q})$, 
and the axial vector interaction under consideration is isoscalar, i.e., $c_A^n=c_A^p$ 
(and also $g_\text{i}\simeq g_\text{c}$). 
\par 
It is evident from relation (\ref{eq:43chiA-CrossSectionsEvaluations-Axial-ChiEta-nonrel-InCoh-2-Coh-total-ratio})
that in the case of pure axial interaction of leptonic and nucleon currents in the nonrelativistic approximation, 
the inelastic (incoherent) cross section can be consider\-ably larger than the elastic (coherent) one, 
except probably for very light nuclei with a large spin, when $\Delta A_f\simeq A_f$. 
The coherent cross section completely vanishes for spin-zero nuclei. 
The quantity $\Delta \equiv \Delta A_p +\Delta A_n$ 
actually plays the role of the total nuclear spin, thus governing the effect of coherence. 
Therefore, it is only in the region of (quite) small transfers where the form factor is little different from 
unity that the inelastic cross section can be considered suppressed. 
Even when $|F_{f}(\bm{q})|^2 \simeq 1/2 $, the 
inelastic cross section dominates over the elastic one by almost two orders of magnitude.
\par
Concluding this section, we note that unlike the case in the nonrelativistic approximation
(\ref{eq:43chiA-CrossSectionsEvaluations-Axial-ChiEta-nonrel-CohCS}), 
in the relativistic case, coherent $\chi A$ cross sections for pure axial interaction do not disappear for spin-zero nuclei
(more specifically, when $\Delta A_f=0$).

\subsubsection{\em Weak ${\chi A}$ interaction} 
Scalar products $(l^w_{s's}, h^{w}_{r'r})$ corresponding to the interaction of a massive
neutral $\chi$ particle with a nucleon in the neutral weak current channel were given by expression 
(\ref{eq:42chiA-CrossSection-ScalarProducts-ChiEta-All-WeakGeneralCurrent-definition}). 
In the nonrelativistic approximation, they are expressed by formulas 
(\ref{eq:42chiA-CrossSection-ScalarProducts-ChiEta-All-Weak-Nonrel}).
\par
According to the definition of the coherent $\chi A$ scattering cross sections 
(\ref{eq:43chiA-CrossSection-via-ScalarProducts-both-CS-main}), 
it is necessary to calculate auxiliary quantities (\ref{eq:43chiA-CrossSection-via-ScalarProducts-CohCS-factors}),  
which, in this approximation, are as follows:
\begin{eqnarray*}
\hat Q^{\mp\mp}_{+}&=& (l^w_{\mp\mp},h^{w}_{++}) + (l^w_{\mp\mp},h^{w}_{--})
=  m^2_c (\alpha \pm \delta) + m^2_c (\alpha \mp \delta) = + 2 m^2_c \alpha
,\\
\hat Q^{\mp\mp}_{-}&=& (l^w_{\mp\mp},h^{w}_{++}) - (l^w_{\mp\mp},h^{w}_{--})
= m^2_c (\alpha \pm \delta) - m^2_c (\alpha \mp \delta) = \pm  2 m^2_c \delta
,\\
\hat Q^{\mp\pm}_{+}&=& (l^w_{\mp\pm},h^{w}_{++})+ (l^w_{\mp\pm},h^{w}_{--})
= \mp m^2_s e^{\pm i\phi} (\alpha\mp \delta) + (\mp m^2_s  e^{\pm i\phi}  (\alpha\pm \delta )) 
= \mp 2 m^2_s e^{\pm i\phi} \alpha
,\\
\hat Q^{\mp\pm}_{-}&=& (l^w_{\mp\pm},h^{w}_{++})- (l^w_{\mp\pm},h^{w}_{--}) 
= \mp m^2_s e^{\pm i\phi} (\alpha\mp \delta) -(\mp m^2_s  e^{\pm i\phi}  (\alpha\pm \delta )) = +2  m^2_s e^{\pm i\phi} \delta 
.\end{eqnarray*}
Then for four auxiliary quantities from (\ref{eq:43chiA-CrossSection-via-ScalarProducts-CohCS-factors})
one has
\begin{eqnarray*}
Q_f^{\mp\mp}&=& \hat Q^{\mp\mp}_{+} + \dfrac{\Delta A_f}{A_f} \hat Q^{\mp\mp}_{-}
= 2 m^2_c \alpha_f +\dfrac{\Delta A_f}{A_f} (\pm  2 m^2_c \delta_f)
= 2 m^2_c \Big[\alpha_f \pm   \delta_f \dfrac{\Delta A_f}{A_f} \Big] 
,\\ Q_f^{\mp\pm}&=&\hat Q^{\mp\pm}_{+} + \dfrac{\Delta A_f}{A_f} \hat Q^{\mp\pm}_{-}
= \mp 2 m^2_s e^{\pm i\phi} \alpha_f  +\dfrac{\Delta A_f}{A_f} 2  m^2_s e^{\pm i\phi} \delta_f
= \mp 2 m^2_s e^{\pm i\phi} \Big[\alpha_f  \mp \delta_f \dfrac{\Delta A_f}{A_f} \Big] 
.\end{eqnarray*}
As a result, based on (\ref{eq:43chiA-CrossSection-via-ScalarProducts-both-CS-main}), 
we obtain a set of coherent $\chi A$ cross sections for the nonrelativistic interaction of weak currents
\begin{eqnarray}\nonumber\label{eq:43chiA-CrossSections-via-ScalarProducts-Weak-nonrel-CohSC-general}
\frac{d\sigma^{\mp\mp}_\text{coh}(\bm{q})}{g_\text{c} d T_A}&=&  \cos^2\frac{\theta}{2} 
\dfrac{G^2_F  m_A}{4\pi}\dfrac{ m^2_\chi }{|\bm{k}^l_\chi|^2 } 
 \Big[ \sum^{}_{f=p,n} F_{f}(\bm{q}) {A_f} \Big(\alpha_f \pm   \delta_f \dfrac{\Delta A_f}{A_f} \Big)\Big]^2
,\\
\frac{d\sigma^{\mp\pm}_\text{coh}(\bm{q})}{ g_\text{c} d T_A}&=&\sin^2\frac{\theta}{2}
\dfrac{G^2_F  m_A}{4\pi} \dfrac{m^2_\chi}{ |\bm{k}^l_\chi|^2 }    
 \Big[ \sum^{}_{f=p,n} F_{f}(\bm{q}) A_f \Big( \alpha_f  \mp \delta_f \dfrac{\Delta A_f}{A_f} \Big)  \Big]^2
,\\
 \dfrac{d\sigma^{\text{total}}_\text{coh}(\bm{q})}{ g_\text{c} d T_A}&=&  \dfrac12 \sum^{}_{s's}\dfrac{d\sigma^{s's}_\text{coh}}{ g_\text{c} d T_A} =  
  \dfrac{G^2_F  m_A}{4\pi}\dfrac{m^2_\chi}{|\bm{k}^l_\chi|^2 } 
  \Big(\big[\sum^{}_{f=p,n}\alpha_f A_f F_{f}(\bm{q}) \big]^2+ \big[\sum^{}_{f=p,n}\delta_f {\Delta A_f} F_{f}(\bm{q}) \big]^2  \Big)
 \nonumber
.\end{eqnarray}
The last formula in (\ref{eq:43chiA-CrossSections-via-ScalarProducts-Weak-nonrel-CohSC-general})
corresponds to the total coherent $\chi A$ cross section 
(averaged over the initial lepton helicities and summed over the final helicities).

It is evident from formulas (\ref{eq:43chiA-CrossSections-via-ScalarProducts-Weak-nonrel-CohSC-general}) 
that the common factor $\cos^2\dfrac{\theta}{2}\simeq 1-\dfrac{T_A}{T_A^{\max}}$ from the first formula 
(\ref{eq:43chiA-CrossSections-via-ScalarProducts-Weak-nonrel-CohSC-general})
typical of coherent (anti)neutrino scattering corresponds to the case where helicity of the incident 
$\chi$ particle remains unchanged after interaction (which is always the case for neutrinos and antineutrinos). 
When the $\chi$-particle helicity changes (which is possible owing to the nonzero mass), in the second formula 
(\ref{eq:43chiA-CrossSections-via-ScalarProducts-Weak-nonrel-CohSC-general})
there arises another common factor $\sin^2\dfrac{\theta}{2}\simeq \dfrac{T_A}{T_A^{\max}}$, 
which cancels out $\cos^2\dfrac{\theta}{2} $
in the total coherent cross section (last formula) in 
(\ref{eq:43chiA-CrossSections-via-ScalarProducts-Weak-nonrel-CohSC-general}), 
negating the kinematic dependence of the common factor of the total coherent cross section 
on the external variable $T_A$, which occurs in this place.
\par 
The following “simplifications” of formulas (\ref{eq:43chiA-CrossSections-via-ScalarProducts-Weak-nonrel-CohSC-general}) 
for coherent $\chi A$ cross sections are possible.
\par (i) Nuclear form factors of protons and neutrons are identical, i.e., $F_p(\bm{q})=F_n(\bm{q})=F(\bm{q})$. 
\begin{eqnarray}\nonumber\label{eq:43chiA-CrossSections-via-ScalarProducts-Weak-nonrel-CohSC-Fp=Fn}
\frac{d\sigma^{\mp\mp}_\text{coh}(\bm{q})}{g_\text{c} d T_A}&=&  
\dfrac{G^2_F  m_A}{4\pi}\dfrac{ m^2_\chi }{|\bm{k}^l_\chi|^2 }\cos^2\frac{\theta}{2}  F^2(\bm{q})  
 \Big[ \sum^{}_{f=p,n} {A_f} \Big(\alpha_f \pm   \delta_f \dfrac{\Delta A_f}{A_f} \Big)\Big]^2
,\\ \frac{d\sigma^{\mp\pm}_\text{coh}(\bm{q})}{g_\text{c} d T_A}&=&
\dfrac{G^2_F  m_A}{4\pi} \dfrac{m^2_\chi}{ |\bm{k}^l_\chi|^2 }  \sin^2\frac{\theta}{2} F^2(\bm{q})
 \Big[ \sum^{}_{f=p,n} A_f \Big( \alpha_f  \mp \delta_f \dfrac{\Delta A_f}{A_f} \Big)  \Big]^2
,\\  \nonumber
 \dfrac{d\sigma^{\text{total}}_\text{coh}(\bm{q})}{g_\text{c} d T_A}&=&  \dfrac12 \sum^{}_{s's}\dfrac{d\sigma^{s's}_\text{coh}}{g_\text{c} d T_A} =   \dfrac{G^2_F  m_A}{4\pi}\dfrac{m^2_\chi}{|\bm{k}^l_\chi|^2 }  F^2(\bm{q})
  \Big(\big[\sum^{}_{f=p,n}\alpha_f A_f \big]^2+ \big[\sum^{}_{f=p,n}\delta_f {\Delta A_f}\big]^2  \Big)
=\\ &=&   \dfrac{G^2_F  m_A}{4\pi}\dfrac{m^2_\chi}{|\bm{k}^l_\chi|^2 }  F^2(\bm{q})
  \Big[ \sum^{}_{f=p,n} A_f^2 \big\{\alpha^2_f  +\delta^2_f \dfrac{\Delta A^2_f}{A_f^2}\big\}  
 +2A_p A_n \big\{\alpha_p \alpha_n  +\delta_p \delta_n \dfrac{\Delta A_p}{A_p} \dfrac{\Delta A_n}{A_n} \big\}   \Big] \nonumber
.\end{eqnarray}
(ii)
The target nucleus has a zero total spin (more specifically, $\Delta A_f=0$)
Then nonrelativistic weak coherent $\chi A$ cross sections  (at $\Delta A_f=0$)
 have the form
\begin{eqnarray} \nonumber \label{eq:43chiA-CrossSections-via-ScalarProducts-Weak-nonrel-CohSC-spinless}
\frac{d\sigma^{\mp\mp}_\text{coh}(\bm{q})}{g_\text{c} d T_A}&=&\cos^2\frac{\theta}{2}  
 \dfrac{d\sigma^{\text{total}}_\text{coh}(\bm{q})}{g_\text{c} d T_A}
, \quad  \frac{d\sigma^{\mp\pm}_\text{coh}(\bm{q})}{g_\text{c} d T_A}=  \sin^2\frac{\theta}{2}
 \dfrac{d\sigma^{\text{total}}_\text{coh}(\bm{q})}{g_\text{c} d T_A}
, \quad \text{where }\\ \dfrac{d\sigma^{\text{total}}_\text{coh}(\bm{q})}{g_\text{c} d T_A}&=& 
\dfrac{G^2_F  m_A}{4\pi} \dfrac{m^2_\chi}{ |\bm{k}^l_\chi|^2 } [\alpha_p A_p F_p (\bm{q}) +\alpha_n A_n F_n (\bm{q})  ]^2
.\end{eqnarray}
It thus follows that in the nonrelativistic case only the “vector component” of the weak current 
proportional to parameters $\alpha_f$ “works” in the coherent $\chi A$ cross section on the nucleus with the zero total spin 
while dependence on $\delta_f$ entirely disappears.
 Note also that in the nonrelativistic approximation there is no contribution from interference 
 of pure vector and pure axial vector components of the weak interaction.
\par
To calculate the nonrelativistic incoherent cross section for the $\chi A$ scattering due to the weak interaction, 
one should use general expressions (\ref{eq:43chiA-CrossSection-via-ScalarProducts-both-CS-main}),  where
the key factors $Q^{s's}_\pm$ from (\ref{eq:43chiA-CrossSection-via-ScalarProducts-InCohCS-factors}) 
are defined in terms of squares of scalar products of the weak current in the nonrelativistic approximation 
(\ref{eq:42chiA-CrossSection-ScalarProducts-ChiEta-All-Weak-Nonrel}). 
With these squares of scalar products, the factors $Q^{s's}_\pm$  are as follows:
\begin{eqnarray*}
 Q^{\mp\mp}_+  &=&  \sum^{}_{r'r=\pm} |(l_{\mp\mp}, h^{f}_{r'r})|^2 
 = 2(4m_\chi m)^2\Big[\cos^2\frac{\theta}{2}\alpha^2 + (1+\sin^2\frac{\theta}{2})\delta^2\Big]
,\\Q^{\mp\pm}_+ &=& \sum^{}_{r'r=\pm} |(l_{\mp\pm}, h^{f}_{r'r})|^2 
= 2(4m_\chi m)^2\Big[\sin^2\frac{\theta}{2}\alpha^2 + (1+\cos^2\frac{\theta}{2} )\delta^2\Big]
,\\ Q^{++}_- &=& - Q^{--}_-  = \sum^{}_{r'=\pm} |(l_{++}, h^{f}_{r'+})|^2-\sum^{}_{r'=\pm} |(l_{++}, h^{f}_{r'-})|^2
 = - 4 \delta (4m_\chi m)^2 \Big[\cos^2\frac{\theta}{2} \alpha  +  \sin^2\frac{\theta}{2}\delta \Big]
\\ Q^{+-}_- &=& - Q^{-+}_- = \sum^{}_{r'=\pm} |(l_{+-}, h^{f}_{r'+})|^2-\sum^{}_{r'=\pm} |(l_{+-}, h^{f}_{r'-})|^2 
= 4 \delta (4m_\chi m)^2 \Big[\sin^2\frac{\theta}{2} \alpha  +  \cos^2\frac{\theta}{2}\delta \Big] 
 .\end{eqnarray*}
As a result, after substitution of the above expressions into 
(\ref{eq:43chiA-CrossSection-via-ScalarProducts-both-CS-main}) 
the cross sections of the incoherent $\chi A$ scattering in the nonrelativistic approximation take the form
\begin{eqnarray} \nonumber \label{eq:43chiA-CrossSections-via-ScalarProducts-Weak-nonrel-InCohSC-general}
  \frac{d\sigma^{\mp\mp}_\text{inc}(\bm{q})}{g_\text{i} d T_A}&=&
\dfrac{ G^2_F  m_A}{4\pi}\dfrac{m^2_\chi }{|\bm{k}^l_\chi|^2} \sum_{f=p,n}\big[1-F^2_f(\bm{q})\big] A_f 
\times\\&&\times  \nonumber 
\Big\{\cos^2\frac{\theta}{2}\alpha^2_f + (1+\sin^2\frac{\theta}{2})\delta^2_f
\pm\dfrac{2\Delta A_f }{A_f}\delta_f\Big[\cos^2\frac{\theta}{2} \alpha_f +\sin^2\frac{\theta}{2}\delta_f \Big]\Big\}
, \\   \frac{d\sigma^{\mp\pm}_\text{inc}(\bm{q})}{g_\text{i} d T_A}&=&
\dfrac{ G^2_F m_A}{4\pi}\dfrac{m^2_\chi }{|\bm{k}^l_\chi|^2}\sum_{f=p,n}\big[1-F^2_f(\bm{q})\big]  A_f 
\times\\&&\times \nonumber
\Big\{\sin^2\frac{\theta}{2}\alpha^2_f + (1+\cos^2\frac{\theta}{2} )\delta^2_f\mp\dfrac{2\Delta A_f}{A_f }\delta_f\Big[\sin^2\frac{\theta}{2} \alpha_f+ \cos^2\frac{\theta}{2}\delta_f\Big]\Big\}
,\\  \frac{d\sigma^{\text{total}}_\text{inc}(\bm{q})}{g_\text{i} d T_A} &=&\frac12 \sum^{}_{s's} \frac{d\sigma^{s's}_\text{inc}}{g_\text{i} d T_A}   = \dfrac{ G^2_F m_A}{4\pi}\dfrac{m^2_\chi }{|\bm{k}^l_\chi|^2}
 \sum_{f=p,n} A^f \big[1-F^2_f(\bm{q})\big]  \big[\alpha^2_f + 3 \delta^2_f \big]
 \nonumber   .\end{eqnarray}
It is seen that there are not any “interference” terms proportional to the weak-current coupling constants 
$\beta$ and $\gamma$. 
Only dependence on the pure vector and pure axial vector coupling constants $\alpha$
and $\delta$  “survives”. 
Further, in the total incoherent $\chi A$ cross section, dependence on the nuclear spin factor $\Delta A_f$
disappears even it is not zero.
Note that for $\alpha_f=0$ and $\delta_f=1$ 
(i.e., in the absence of the vector contribution) these expressions entirely coincide with the similar formulas for pure axial-axial interactions (\ref{eq:43chiA-CrossSectionsEvaluations-Axial-ChiEta-nonrel-InCohCS}).
\par
The experimentally measured total cross section for the $\chi A$ scattering due to interaction of weak 
leptonic and nucleon currents in the nonrelativistic approximation is
 \begin{eqnarray}\label{eq:43chiA-CrossSections-via-ScalarProducts-Weak-nonrel-Full}
\dfrac{d\sigma^{\text{weak}}_{\text{nonrel}}}{d T_A}(\chi A \to \chi A^*)&=&
\dfrac{G^2_F  m_A }{4\pi}  \dfrac{m^2_\chi  }{|\bm{k}^l_\chi|^2}
\Big\{g_\text{i} \sum_{f=p,n} A^f \big[1-F^2_f(\bm{q})\big]  \big[\alpha^2_f + 3 \delta^2_f \big]
\\&& +g_\text{c} \Big(\big[\sum^{}_{f=p,n}\alpha_f A_f F_{f}(\bm{q}) \big]^2+ \big[\sum^{}_{f=p,n}\delta_f {\Delta A_f} F_{f}(\bm{q}) \big]^2  \Big)\Big\} 
\nonumber. \end{eqnarray}
Let us consider two “simplifications” of the incoherent $\chi A$ cross sections 
(\ref{eq:43chiA-CrossSections-via-ScalarProducts-Weak-nonrel-InCohSC-general}). 
\par
(i) Nuclear form factors of neutrons and protons are identical. 
Unlike the coherent case, here it is not much of simplification, since dependence on the form factor is simply taken 
outside the sign of summation over $f$, and the following formulas are obtained:
\begin{eqnarray} \nonumber  \label{eq:43chiA-CrossSections-via-ScalarProducts-Weak-nonrel-InCohSC-Fp=Fn}
\frac{d\sigma^{\mp\mp}_\text{inc}(\bm{q})}{g_\text{i} d T_A}&=&
\dfrac{ G^2_F  m_A}{4\pi}\dfrac{m^2_\chi }{|\bm{k}^l_\chi|^2}[1-F^2(\bm{q})]  \sum_{f=p,n}A_f 
\times\\&&\times  \nonumber 
\Big\{\cos^2\frac{\theta}{2}\alpha^2_f + (1+\sin^2\frac{\theta}{2})\delta^2_f
\pm\dfrac{2\Delta A_f }{A_f}\delta_f\Big[\cos^2\frac{\theta}{2} \alpha_f +\sin^2\frac{\theta}{2}\delta_f \Big]\Big\}
, \\  \frac{d\sigma^{\mp\pm}_\text{inc}(\bm{q})}{g_\text{i} d T_A}&=&
\dfrac{ G^2_F m_A}{4\pi}\dfrac{m^2_\chi }{|\bm{k}^l_\chi|^2}[1-F^2(\bm{q})]   \sum_{f=p,n}A_f 
\times\\&&\times \nonumber
\Big\{\sin^2\frac{\theta}{2}\alpha^2_f + (1+\cos^2\frac{\theta}{2} )\delta^2_f\mp\dfrac{2\Delta A_f}{A_f }\delta_f\Big[\sin^2\frac{\theta}{2} \alpha_f+ \cos^2\frac{\theta}{2}\delta_f\Big]\Big\}
,\\   \frac{d\sigma^{\text{total}}_\text{inc}(\bm{q})}{g_\text{i} d T_A} &=&
\dfrac{ G^2_F m_A}{4\pi}\dfrac{m^2_\chi }{|\bm{k}^l_\chi|^2}[1-F^2(\bm{q})]  
 \sum_{f=p,n} A^f \big[\alpha^2_f + 3 \delta^2_f \big]
 \nonumber   .\end{eqnarray}
(ii) If the target nucleus has a zero total spin,  $\Delta A_f =0$, the formulas are as follows:
\begin{eqnarray} \nonumber  \label{eq:43chiA-CrossSections-via-ScalarProducts-Weak-nonrel-InCohSC-spinless}
  \frac{d\sigma^{\mp\mp}_\text{inc}(\bm{q})}{g_\text{i} d T_A}&=&
\dfrac{ G^2_F  m_A}{4\pi}\dfrac{m^2_\chi }{|\bm{k}^l_\chi|^2} \sum_{f=p,n}\big[1-F^2_f(\bm{q})\big] A_f 
\big\{\cos^2\frac{\theta}{2}\alpha^2_f + (1+\sin^2\frac{\theta}{2})\delta^2_f\big\}
, \\   \frac{d\sigma^{\mp\pm}_\text{inc}(\bm{q})}{g_\text{i} d T_A}&=&
\dfrac{ G^2_F m_A}{4\pi}\dfrac{m^2_\chi }{|\bm{k}^l_\chi|^2}\sum_{f=p,n}\big[1-F^2_f(\bm{q})\big]  A_f 
\big\{\sin^2\frac{\theta}{2}\alpha^2_f + (1+\cos^2\frac{\theta}{2} )\delta^2_f\big\}
,\\  \frac{d\sigma^{\text{total}}_\text{inc}(\bm{q})}{g_\text{i} d T_A} &=&
\dfrac{ G^2_F m_A}{4\pi}\dfrac{m^2_\chi }{|\bm{k}^l_\chi|^2}\sum_{f=p,n} A^f \big[1-F^2_f(\bm{q})\big]  \big[\alpha^2_f + 3 \delta^2_f \big]
 \nonumber   .\end{eqnarray}
The ratio of the total nonrelativistic cross sections for the incoherent 
 (\ref{eq:43chiA-CrossSections-via-ScalarProducts-Weak-nonrel-InCohSC-general}) and coherent 
 (\ref{eq:43chiA-CrossSections-via-ScalarProducts-Weak-nonrel-CohSC-general})
$\chi A$ scattering due to the weak interaction is 
 {\small
 \begin{eqnarray}\label{eq:43chiA-CrossSectionsEvaluations-Weak-ChiEta-nonrel-InCoh-2-Coh-total-ratio}
\dfrac{\dfrac{d\sigma^{\text{total}}_\text{inc}}{d T_A}(\chi A\to \chi A^{*})}{\dfrac{d\sigma^{\text{total}}_\text{coh}}{d T_A}(\chi A\to \chi A^{})}=  \dfrac{g_\text{i}  [1-F^2(\bm{q})]  \sum_{f=p,n} A^f \big[\alpha^2_f + 3 \delta^2_f \big]}
{g_\text{c} F^2(\bm{q})\Big[ \sum^{}_{f=p,n} A_f^2 \big\{\alpha^2_f  +\delta^2_f \dfrac{\Delta A^2_f}{A_f^2}\big\}  
+2A_p A_n \big\{\alpha_p \alpha_n  +\delta_p \delta_n \dfrac{\Delta A_p}{A_p} \dfrac{\Delta A_n}{A_n} \big\} \Big]}
 .\end{eqnarray}}%
 This ratio “goes into” a pure axial case of the $\chi A$ interaction 
 (when $\alpha_f=0$, see (\ref{eq:43chiA-CrossSectionsEvaluations-Axial-ChiEta-nonrel-InCoh-2-Coh-total-ratio}))
 and a pure scalar case of the $\chi A$ interaction (when $\delta_f=0$, see 
(\ref{eq:43chiA-CrossSectionsEvaluations-Scalar-ChiEta-nonrel-InCoh-2-Coh-total-ratio})) 
coinciding with a pure vector case in the nonrelativistic approximation.
\par
This section concludes the main part of this work. 
Here expressions are obtained for the cross sections of the elastic and inelastic scattering 
of the massive neutral (weakly interacting) $\chi$ particle off the nucleus $\chi A\to \chi A^{(*)}$
when the latter retains its integrity 
(i.e., at $\chi$ particle energies that are usually below 100 MeV). 
The pure scalar, axial vector, and generalized weak interaction of the $\chi$ particle with nucleons 
of the nucleus is considered in the nonrelativistic approximation.
\par 
Practically, these new formulas are of interest for correctly understanding the balance of coherence and 
incoherence in problems of the direct search for galactic dark-matter particles 
when the signature of the desired events is the acts of interaction of these particles with nuclear targets. 
This very issue is investigated in a separate work \cite{Bednyakov:2021dmc}. 
Theoretically, new expressions for the scattering of massive leptons are of interest as 
a generalization of the above formulas
\cite{Bednyakov:2018mjd,Bednyakov:2019dbl,Bednyakov:2021ppn} 
to the case of coherent and incoherent scattering of massless neutrinos and antineutrinos from nuclei.

\section{\large Numerical estimations and discussions of ${\chi}$ particle–nucleus scattering} 
\label{50chiA-ResultsAndDiscussion}
A total set of expressions for cross sections of the coherent and incoherent scattering due to weak 
${\chi A}$ interaction in the nonrelativistic approximation is given by formulas 
(\ref{eq:43chiA-CrossSections-via-ScalarProducts-Weak-nonrel-CohSC-Fp=Fn}) 
and (\ref{eq:43chiA-CrossSections-via-ScalarProducts-Weak-nonrel-InCohSC-Fp=Fn}).
For numerical estimation, we will use the expression for the experimentally measured 
$\chi A$ scattering cross section, which is a sum of two cross sections averaged over the initial and
summed over the final $\chi$-lepton spin projections
 \footnote{For simplicity, we consider in all further calculations that   ${g_\text{i}} \simeq {g_\text{c}} \simeq 1$.}:
 \begin{eqnarray}\nonumber\label{eq:50chiA-Results-CrossSection-Weak-nonrel-Full}
\dfrac{d\sigma^{\text{weak}}_{\text{nonrel}}}{d T_A}(\chi A \to \chi A^*)&=&
\dfrac12 \sum^{}_{s's} \frac{d\sigma^{s's}_\text{inc}}{d T_A}(\chi A\to \chi A^{*}) +
\dfrac12 \sum^{}_{s's}\dfrac{d\sigma^{s's}_\text{coh}}{d T_A}(\chi A\to \chi A^{})
=\\&=& \dfrac{G^2_F  m_A }{4\pi}  \dfrac{m^2_\chi  }{|\bm{k}^l_\chi|^2}\Big\{\big[1-F^2(\bm{q})\big]  \sum_{f=p,n} A_f  \big[\alpha^2_f + 3 \delta^2_f \big]
\\&&\qquad +F^2(\bm{q})\Big[ \big[\sum^{}_{f=p,n}\alpha_f A_f \big]^2+ \big[\sum^{}_{f=p,n}\delta_f {\Delta A_f} \big]^2  \Big]\Big\} 
\nonumber. \end{eqnarray}
We will use the form of the Helm nuclear form factor \cite{PhysRev.104.1466} for protons and neutrons:
$$F_{p/n}(\bm{q})=F(\bm{q})  = 3 \dfrac{j_1(qR_0)}{qR_0}e^{-\dfrac{(s q)^2}{2}},  
\quad  j_1(x) = \dfrac{\sin x}{x^2} - \dfrac{\cos x}{x},  \ s = 0.9 \text{~fm}, \  R_0 = 1.14 A^{1/3} \text{~fm},$$
and the following designation and values
\footnote{Constant quantities:  $1 \text{~fm} =  \dfrac{1}{0.1973 \text{~GeV} } =  \dfrac{1}{197.3 \text{~MeV}}$,  
$\mbox{~GeV} = 10^3 \mbox{~MeV}, \  \mbox{MeV} = 1$.}:
$$ \bm{q}^2 \simeq 2 m_A T_A,  \quad 
G_{\text{F}}=1.166 378 \times 10^{-5}/\mbox{GeV}^2 \text{~~and~~}m = 0.938  \mbox{~GeV} \simeq 1 \mbox{~GeV}.
$$
According to the definition of the scalar product for weak currents of the general form
\begin{equation}\label{eq:50chiA-Results-WeakScalarProdict-Definotion}
(l^w_{s's}, h^{w,f}_{r'r}) = \alpha_f (l^v_{s's}\, h^v_{r'r}) + \beta_f (l^v_{s's}\, h^a_{r'r}) 
+ \gamma_f (l^a_{s's}\, h^v_{r'r})  +  \delta_f  (l^a_{s's}\, h^a_{r'r}),
\end{equation}
the effective coupling constants having the form
$$\alpha_f = \chi_V h^f_V,\quad  \beta_f = \chi_V h^f_A, \quad \gamma_f = \chi_A  h^f_V ,\quad \delta_f = \chi_A h^f_A,
$$
are free parameters. In the SM, for neutrinos there is $\chi_V  = - \chi_A = 1$, and for nucleons there is
$$ 
h_V^p=g_V^p =  \frac12 -2 \sin^2\theta_W, \quad h_V^n=g_V^n = - \frac12; 
\quad h_A^p = g_A^p =   \frac{g_A}{2}, \quad h_A^n = g_A^n =  - \frac{g_ A}{2}. 
$$
Then the effective parameters in the SM are
$$\alpha_f = \chi_V h^f_V  = + g_V^f, \quad \delta_f = \chi_A h^f_A = + g_A^f,
\quad \gamma_f = \chi_A  h^f_V = - g_V^f, \quad  \beta_f = \chi_V h^f_A = - g_A^f .$$
Considering that $\sin^2\theta_W = 0.23865$ and $g_A = 1.27$,
the numerical values obtained in the SM are
\begin{eqnarray}\nonumber \label{eq:50chiA-Results-SM-alpha+delta}
\alpha_p  &=&  + \frac12 -2 \sin^2\theta_W  = - \gamma_p  \simeq 0.02  
 ,\quad \delta_p  =  \frac{g_A}{2} = - \beta_p  \simeq 0.64
; \\  \alpha_n  &=& - \frac12  = - \gamma_n  = - 0.5
,\qquad \qquad\qquad \delta_n  =  - \frac{g_ A}{2} = - \beta_n  \simeq  - 0.64 
. \end{eqnarray}
In the nuclear rest frame, calculations of cross sections and their ratios involve only variation of the
kinetic (detected) nuclear recoil energy $T_A$ in the interval  from $T_A^{\min}= \epsilon_A^{\text{thre}} > 0$ 
to a certain maximal value $T_A^{\max}$ for this nucleus, 
which is determined by the condition of the square of the nuclear form factor $|F(T_A)|^2$ 
becoming zero.
\par
The energy of the $\chi$ particle incident on the nucleus (at rest)  $T_0 = \dfrac{|\bm{k}^l_\chi|^2}{2 m_\chi}$, 
or the square of the momentum $|\bm{k}^l_\chi|^2$, 
which enters into  (\ref{eq:43chiA-CrossSections-via-ScalarProducts-Weak-nonrel-CohSC-Fp=Fn}) 
and (\ref{eq:43chiA-CrossSections-via-ScalarProducts-Weak-nonrel-InCohSC-Fp=Fn}), 
 is obtained from the relation between $T_0$ and $T_A$  
 through the $\chi$ particle escape angle $\theta$ in the lab frame 
 (\ref{eq:2chiA-Kinematics-CosT-from-k2prime-and-T_A}):
\begin{eqnarray}\label{eq:50chiA-Kinematics-CosT-from-k2prime-and-T_A}
 \cos\theta(T_A) &=&   \dfrac{m_\chi( 2 T_0- \Delta\varepsilon_{mn})  - T_A (m_\chi +m_A)}{2 m_\chi\sqrt{ T_0 (T_0-\Delta\varepsilon_{mn}-T_A)}} 
.\end{eqnarray}
The maximum $T_A$  is achieved when the incident $\chi$ particle is “reflected” strictly in the opposite 
direction, i.e., when  $\cos\theta(T_A)=-1$. 
Then, from this condition there arises the minimum value for the elastic case (when $\Delta\varepsilon_{mn}=0$)
 \begin{equation}\label{eq:50chiA-Kinematics-T_0-from-T_A}
 T_0= T_A  \dfrac{(m_\chi+m_A)^2} { 4m_\chi m_A }  = T_A  \dfrac{(1+r)^2} { 4r }
 , \quad \text{where}\quad  r\equiv  \dfrac{m_\chi}{m_A}
, \end{equation}
at which this value of $T_A$  (maximum efficiency of momentum–energy transfer) is still possible.
\par
\begin{figure}[h!] 
\includegraphics[width=\linewidth]{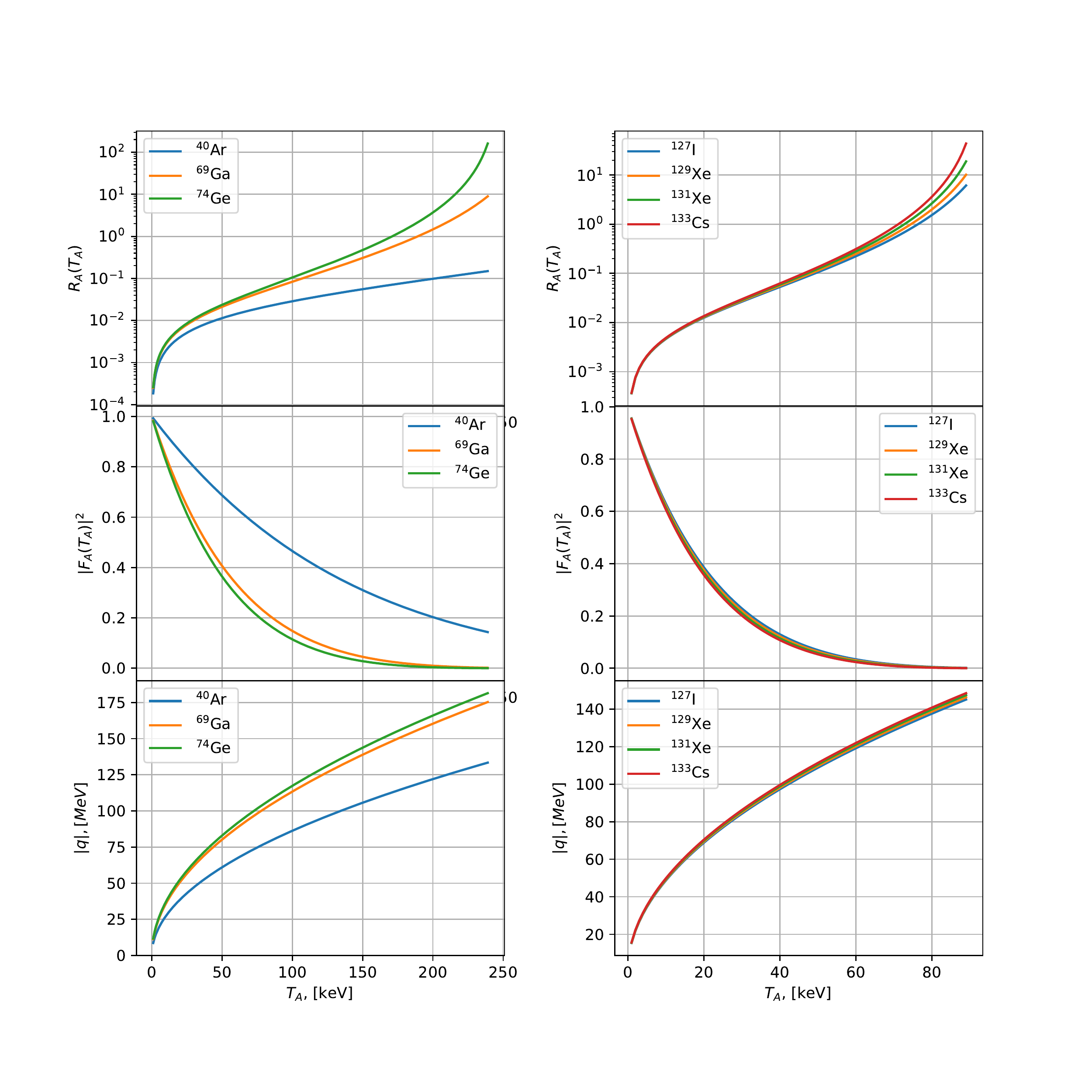}
\vspace*{-50pt}
\caption{“Nuclear” ratios $R_{A}(T_A)$ (\ref{eq:50chiA-Results-Inc2CohRatio-A}) (top row), 
squares of nuclear form factors $|F_A(T_A)|^2$  (middle row), and absolute values of the momentum transferred 
to the nucleus $|\bm{q}|$  (bottom row) for a few target nuclei as a function of nuclear recoil energy $T_A$.}
\label{fig:50chiA-Ratios-RA+FF+q-vs-T_A}
\end{figure}

The ratio of the total nonrelativistic cross sections for the incoherent 
 (\ref{eq:43chiA-CrossSections-via-ScalarProducts-Weak-nonrel-InCohSC-general}) and coherent 
 (\ref{eq:43chiA-CrossSections-via-ScalarProducts-Weak-nonrel-CohSC-general})
 $\chi A$ scattering due to weak interaction is as follows:
 \begin{eqnarray}\nonumber \label{eq:50chiA-Results-Inc2CohRatio-Weak}
 R^{\text{total}}_w(A,\bm{q}) &\equiv&
\dfrac{\dfrac{d\sigma^{\text{total}}_\text{inc}}{d T_A}(\chi A\to \chi A^{*})}{\dfrac{d\sigma^{\text{total}}_\text{coh}}{d T_A}(\chi A\to \chi A^{})}
=\dfrac{ [1-F^2(\bm{q})]  \sum_{f=p,n} A^f \big[\alpha^2_f + 3 \delta^2_f \big]}{F^2(\bm{q})\Big[
 \big[\sum^{}_{f=p,n}\alpha_f A_f \big]^2+ \big[\sum^{}_{f=p,n}\delta_f {\Delta A_f} \big]^2 \Big]}
=\\&=& A R_A(\bm{q}) \dfrac{ A^p \big[\alpha^2_p+ 3 \delta^2_p \big] + A^n \big[\alpha^2_n + 3 \delta^2_n \big]}
{\big[\alpha_p A_p+\alpha_n A_n \big]^2+ \big[\delta_p {\Delta A_p}+\delta_n {\Delta A_n} \big]^2}
 .\end{eqnarray}
 Here the following designation for the ratio of the pure “nuclear parts” of 
 formula \ref{eq:50chiA-Results-Inc2CohRatio-Weak})  is introduced:
\begin{equation} \label{eq:50chiA-Results-Inc2CohRatio-A}
R_A(\bm{q})=R_A(T_A)\equiv \dfrac{1-F^2_A(\bm{q}) }{A F^2_A(\bm{q})}=\dfrac{1-F^2_A(T_A) }{A F^2_A(T_A)}.
\end{equation}
These quantities together with the square of the corresponding form factors and the modulus 
of the transferred momentum are shown in 
Fig.~\ref{fig:50chiA-Ratios-RA+FF+q-vs-T_A}  as a function of the nuclear recoil energy $T_A$. 
It is seen in Fig.~\ref{fig:50chiA-Ratios-RA+FF+q-vs-T_A}  that the square of the nuclear form factor  $|F(T_A)|^2$ 
monotonically decreases with increasing energy transfer to the nucleus (kinematic nuclear recoil energy) $T_A$.
Once $|F(T_A)|^2$  is sufficiently close to zero, the “coherence effect” (together with the coherent contribution 
to the total cross section) weakens fast and totally disappears when 
$|F(T_A)|^2\simeq 0$
\footnote{As was already mentioned, for each of the nuclei this moment determines $T_A^{\max}$ 
Tmax in the corresponding plots.}. 
Simultaneously, the ratio $R_{A}(T_A)$  rapidly increases, which actually means a 
transition to total domination of the “incoherence effect” 
(together with the incoherent contribution to the total cross section). 
For different nuclei this occurs at different $T_A$  and $q$ that are, 
roughly speaking, inversely proportional to the atomic number of the nucleus $A$. 
This is how the nucleus “works” in the “pure form,” regulating the transition from coherence to incoherence.
\par
The second factor in (\ref{eq:50chiA-Results-Inc2CohRatio-Weak}) written in the form
  \begin{eqnarray}\label{eq:50chiA-Results-Inc2CohRatio-WeakCharge}
R_I(A,\alpha,\delta) &\equiv& (A_p+A_n) \dfrac{ A^p (\alpha^2_p+ 3 \delta^2_p) + A^n (\alpha^2_n + 3 \delta^2_n )}
{(\alpha_p A_p+\alpha_n A_n)^2+ (\delta_p {\Delta A_p}+\delta_n {\Delta A_n} )^2}
 ,\end{eqnarray}
 accumulates the entire dependence on the character of interaction of the $\chi$ lepton with nucleons. 
 It is determined by the balance of the weak-current coupling constants $\alpha_{p/n}$ 
 and $\delta_{p/n}$, “weighted” with the proton-neutron structure of a particular nucleus. 
 It can both enhance and weaken the “pure” nuclear structure effect shown in Fig. ~\ref{fig:50chiA-Ratios-RA+FF+q-vs-T_A}. 
Its SM values for a few target nuclei are listed in Table~\ref{tab:50chiA-Ratios-RIad-vs-A}.
\begin{table}[h!] \begin{center}\begin{tabular}{|r | c c c c c c | c | c|}
$^A$Nucleus$(Z,N)$& $Z_+$ &$ Z_-$& $\Delta Z$ & $N_+$  &$N_-$ &$\Delta N$ &  Spin &  
$R_I(A,\alpha,\delta)$ \\ \hline
${}^{4}\text{He}(2,2)$ &1 & 1 & 0 & 1 & 1 & 0 &  &  23.44 \\ 
${}^{12}\text{C}(6,6)$ &3 & 3 & 0 & 3 & 3 & 0 &  &  23.44 \\ 
${}^{19}\text{F}(9,10)$ &5 &4 &1 & 5 & 5 & 0   &  $\frac12=\frac{ \Delta Z}{2}$ &20.71 \\ 
${}^{40}\text{Ar}(18,22)$ & 9 & 9 & 0 & 11 & 11 & 0  & &19.22\\ 
${}^{74}\text{Ge}(32,42)$ & 16 & 16 & 0 & 21 & 21 & 0   & &18.00 \\ 
${}^{127}\text{I}(53,74)$&27&26&1&39&35&4&$\frac{5}{2}=\frac{\Delta Z}{2}+\frac{\Delta N}{2}$&17.01 \\ 
 ${}^{129}\text{Xe}(54,75)$&27&27&0&38&37&1& $\frac{1}{2} =\frac{\Delta Z}{2}+\frac{\Delta N}{2}$ &17.13 \\ 
${}^{131}\text{Xe}(54,77)$&27&27&0&40&37&3&$\frac{3}{2}=\frac{\Delta Z}{2}+\frac{\Delta N}{2}$ & 16.72 \\ 
${}^{133}\text{Cs}(55,78)$&28&27&1&46&32&6&$\frac{7}{2}=\frac{\Delta Z}{2}+\frac{\Delta N}{2}$ & 16.07
\end{tabular}\end{center}
\caption{Characteristics of some target nuclei and their corresponding quantities $R_I(A,\alpha,\delta)$ 
with $\alpha_{p/n},\delta_{p/n}$ within the SM from (\ref{eq:50chiA-Results-SM-alpha+delta}).
Here $Z=A_p,\ N=A_n$,  $\Delta A_p \equiv \Delta Z = Z_+ - Z_-$, and $ \Delta A_n \equiv \Delta N= N_+ - N_-$.
The subscripts $\pm$ indicate direction of nucleon spins relative to a chosen quantization axis.} \label{tab:50chiA-Ratios-RIad-vs-A}
\end{table} 
\begin{figure}[h!] 
\centering\includegraphics[width=0.6\linewidth]{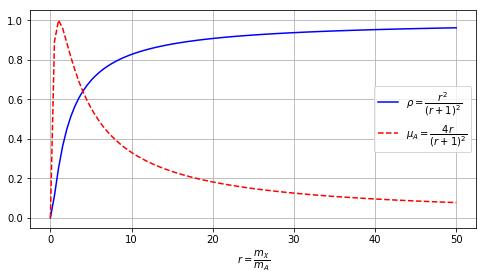}  
\caption{Illustrated dependence of the kinematic factors  $\rho$ and $\mu_A$ on the ratio 
$r=\dfrac{m_\chi}{m_A}$.} \label{fig:50chiA-SimplePolot-r}
\end{figure} 
\par
\begin{figure}[h!] 
\includegraphics[width=\linewidth]{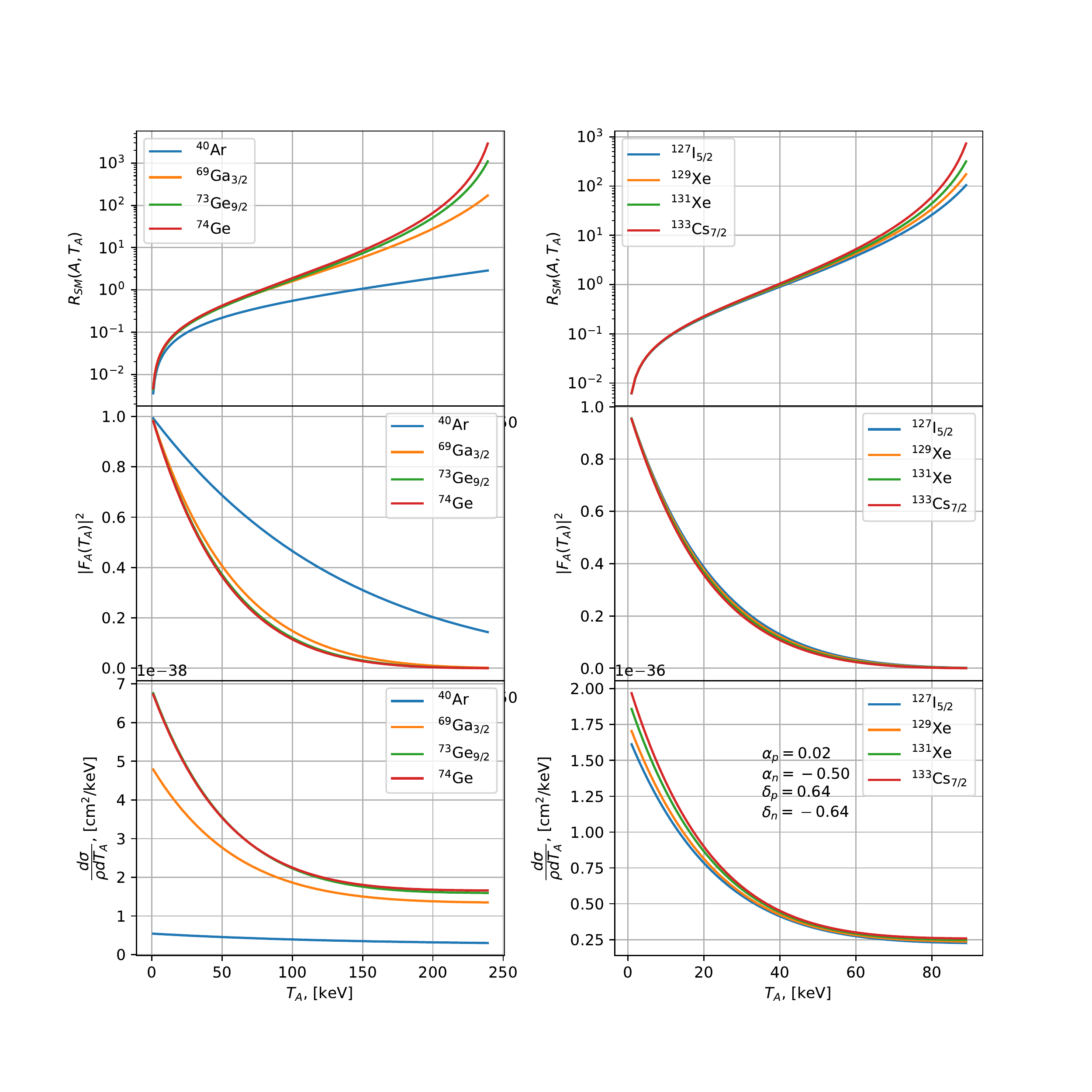} 
\vspace*{-60pt}
\caption{Ratio $ R^{\text{total}}_{\text{SM}}(A,T_A)$ (\ref{eq:50chiA-Results-Inc2CohRatio-Weak}) 
of the total nonrelativistic cross sections for the incoherent and coherent $\chi A$ scattering due to weak
SM interaction with the parameters from (\ref{eq:50chiA-Results-SM-alpha+delta}) (top row), 
squares of nuclear form factors $|F(T_A)|^2$ (middle row), 
and measured total differential cross sections  (\ref{eq:50chiA-Results-CrossSection-Weak-nonrel-via-Coh+R}) 
(bottom row) for a few target nuclei as a function of the recoil energy $T_A$.}
\label{fig:50chiA-Ratios+FF+DS-vs-T_A-4All}
\end{figure}

The influence of the “nuclear charge” factor (\ref{eq:50chiA-Results-Inc2CohRatio-WeakCharge})  
from Table ~\ref{tab:50chiA-Ratios-RIad-vs-A}, which enhances the “incoherence effect” more than tenfold, 
is demonstrated in Fig~\ref{fig:50chiA-Ratios+FF+DS-vs-T_A-4All}, 
(upper panel of two plots), where ratios  (\ref{eq:50chiA-Results-Inc2CohRatio-Weak})  for the
effective coupling constants $\alpha_{p/n}$ and  $\delta_{p/n}$ from the SM are shown. 
Dependences of the nuclear form factors on $T_A$ in the middle panel of Fig.~\ref{fig:50chiA-Ratios+FF+DS-vs-T_A-4All} 
are given as “reference plots.” 
They allow judging the maximum value of  $T_A$ for a particular nucleus.
Two lower pots in Fig.~\ref{fig:50chiA-Ratios+FF+DS-vs-T_A-4All} 
illustrate $T_A$ dependence of the measured total cross sections
(\ref{eq:50chiA-Results-CrossSection-Weak-nonrel-Full}) calculated by the formula
\begin{eqnarray}\label{eq:50chiA-Results-CrossSection-Weak-nonrel-via-Coh+R}
\dfrac{d\sigma^{\text{weak}}_{\text{nonrel}}}{\rho d T_A}(\chi A \to \chi A^*)&=&
\dfrac{G^2_F  m_A^2 }{2\pi  T_A^{\max} } 
\big\{ F^2(T_A)\big[(\alpha_p A_p + \alpha_n A_n)^2+ (\delta_p \Delta A_p +\delta_n \Delta A_n)^2 \big] \quad
\\&&\qquad \qquad 
 +(1-F^2(T_A))\big[ A^p (\alpha^2_p+ 3 \delta^2_p) + A^n (\alpha^2_n + 3 \delta^2_n )\big]\big\} 
\nonumber
,  \end{eqnarray}
where $\rho \equiv  \dfrac{r^2 }{(1+r)^2}$ (see. Fig.~\ref{fig:50chiA-SimplePolot-r}) 
considering the definition $r=\dfrac{m_\chi}{m_A}$ from (\ref{eq:50chiA-Kinematics-T_0-from-T_A}),
“hides” the entire kinematic dependence on the $\chi$-lepton mass. 
Indeed, the right-hand side of (\ref{eq:50chiA-Results-CrossSection-Weak-nonrel-via-Coh+R})
does not depend on $m_\chi$.
The explicit form of this expression is due to the fact that the calculation of cross sections 
(\ref{eq:50chiA-Results-CrossSection-Weak-nonrel-Full})
(110) requires knowing not only $T_A$ but also $T_0$,  whose minimum value for the particular  $T_A$
is given by formula (\ref{eq:50chiA-Kinematics-T_0-from-T_A}).
Therefore, in expression (\ref{eq:50chiA-Results-CrossSection-Weak-nonrel-via-Coh+R}), 
the key dimensional factor of all cross sections from (\ref{eq:50chiA-Results-CrossSection-Weak-nonrel-Full})
$$
\dfrac{G^2_F  m_A }{4\pi}  \dfrac{m^2_\chi  }{|\bm{k}^l_\chi|^2} = \dfrac{G^2_F  m_A }{4\pi}  \dfrac{m_\chi  }{2 T_0}=\dfrac{G^2_F  m_A^2 }{2\pi T_A } \rho
$$
involves one universal one universal $T_0^{\min 4\max}$, that allows achieving the maximum value $T_A^{\max}$ presented in the
plots and defined from the condition $F_A(T_A^{\max})\simeq 0$: 
\begin{equation}\label{eq:50chiA-Kinematics-T_0min-from-T_A}
 T_0^{\min 4\max}=  \dfrac{T_A^{\max}}{\mu_A}, \quad \text{where}\quad  \mu_A \equiv  \dfrac{ 4r }{  (1+r)^2} 
\quad \text{(see. Fig.~\ref{fig:50chiA-SimplePolot-r})}
. \end{equation}
This choice of the initial energy $T_0$ guaranteeing
achievability of $T_A^{\max}$ seems quite sufficient for illustrating the behavior of $\chi A$-cross sections as a function of  $T_A$.
\par 
From Fig.~\ref{fig:50chiA-Ratios+FF+DS-vs-T_A-4All}, 
obtained for the SM effective coupling constants (\ref{eq:50chiA-Results-SM-alpha+delta})
at the (auxiliary simplifying) condition (\ref{eq:50chiA-Kinematics-T_0min-from-T_A}),
one can see the following. 
When squares of nuclear form factors of heavy nuclei decay to zero with increasing kinetic nuclear recoil energy $T_A$
(coherence is almost entirely lost, right middle panel), 
the measured total cross section (sum of coherent and incoherent ones, right bottom panel) 
decreases by about an order of magnitude and is saturated mainly by the incoherent contribution (top right panel). 
Note also that in the case of these very nuclei, in already the half of the $T_A$ 
 interval (at 40 keV) the incoherent contribution reaches the coherent one as 
 the measured total cross section decreases by about a factor of five, which seems insignificant.
\par
\begin{figure}[h!] 
\includegraphics[width=\linewidth]{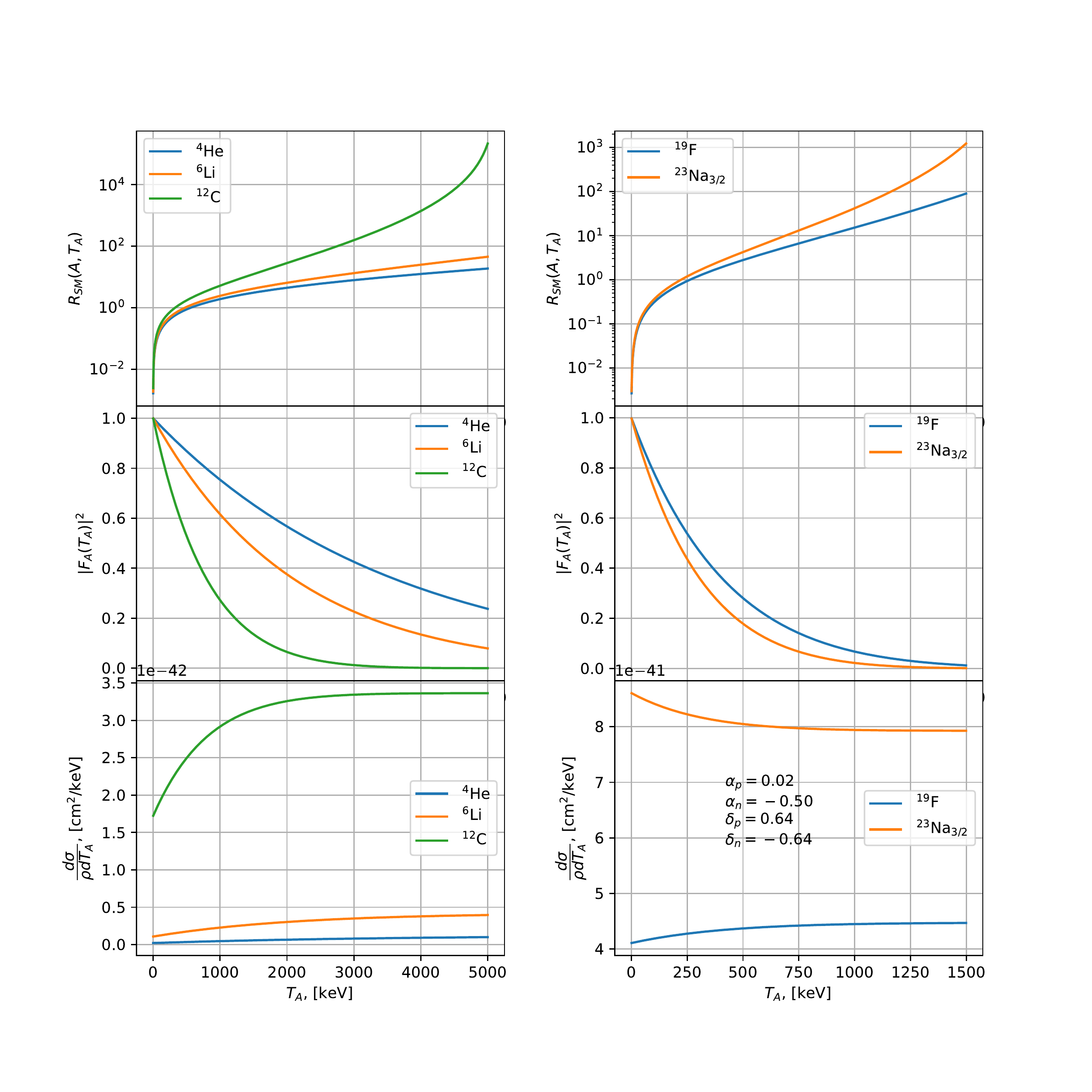}
 \vspace*{-40pt}
\caption{Ratio $ R^{\text{total}}_{\text{SM}}(A,T_A)$ (\ref{eq:50chiA-Results-Inc2CohRatio-Weak}) 
of the total nonrelativistic cross sections for the incoherent and coherent $\chi A$ scattering due to weak
SM interaction with the parameters from (\ref{eq:50chiA-Results-SM-alpha+delta}) 
(top row), squares of nuclear form factors $|F(T_A)|^2$ (middle row), and measured
total differential cross sections  (\ref{eq:50chiA-Results-CrossSection-Weak-nonrel-via-Coh+R}) 
(bottom row) for a few light target nuclei as a function of the recoil energy $T_A$.}
\label{fig:50chiA-Ratios+FF+DS-vs-T_A-4light}
\end{figure}

 
The situation is quite different for light nuclei (Fig.~\ref{fig:50chiA-Ratios+FF+DS-vs-T_A-4light}). 
Here “coherence” is preserved long enough (the lighter the nucleus, the longer). 
Squares of nuclear form factors decay slowly, reaching the region $F_A(T_A)\simeq 0$  
at rather high recoil energies $T_A$ (the lighter the nucleus, the higher $T_A^{\max}$). 
On the one hand, these high $T_A$ should probably be readily observable, but on the
other hand, they could already be incompatible with preservation of nuclear integrity ($\bm{q}\ge 300$ MeV/$c$) 
The total cross sections are noticeably smaller than for heavy nuclei and vary weakly. 
An intermediate situation is for nuclei like Ga–Ge ($A\simeq 70$). 
It is similar in terms of the character of dependence on $T_A$ to the case
of heavy nuclei with the main difference being the
magnitude of the recoil energy. Here it is noticeably higher ($T_A^{\max}\simeq 200~$keV).
 \par
 Figures~\ref{fig:50chiA-Ratios+FF+DS-vs-T_A-4All} and \ref{fig:50chiA-Ratios+FF+DS-vs-T_A-4light}
 illustrate the main property of the approach 
\cite{Bednyakov:2018mjd,Bednyakov:2019dbl,Bednyakov:2021ppn}: 
as the nuclear recoil energy $T_A$ increases, a smooth, well-controlled transition takes
place from domination of elastic $\chi A$ interaction to domination of inelastic 
$\chi A$ interaction in nonrelativistic scattering of a massive $\chi$ particle off a nucleus.
\par
Since the results of the experiments on the search for dark matter are usually interpreted in terms of 
spin-independent and spin-dependent cross sections of interaction of a dark-matter particle with nucleons, we
will consider in more detail consequences of the approach 
\cite{Bednyakov:2018mjd,Bednyakov:2019dbl,Bednyakov:2021ppn}  for these two types of interaction.
\par
\begin{figure}[h!] 
\hspace*{-20pt}
\includegraphics[width=1.1\linewidth]{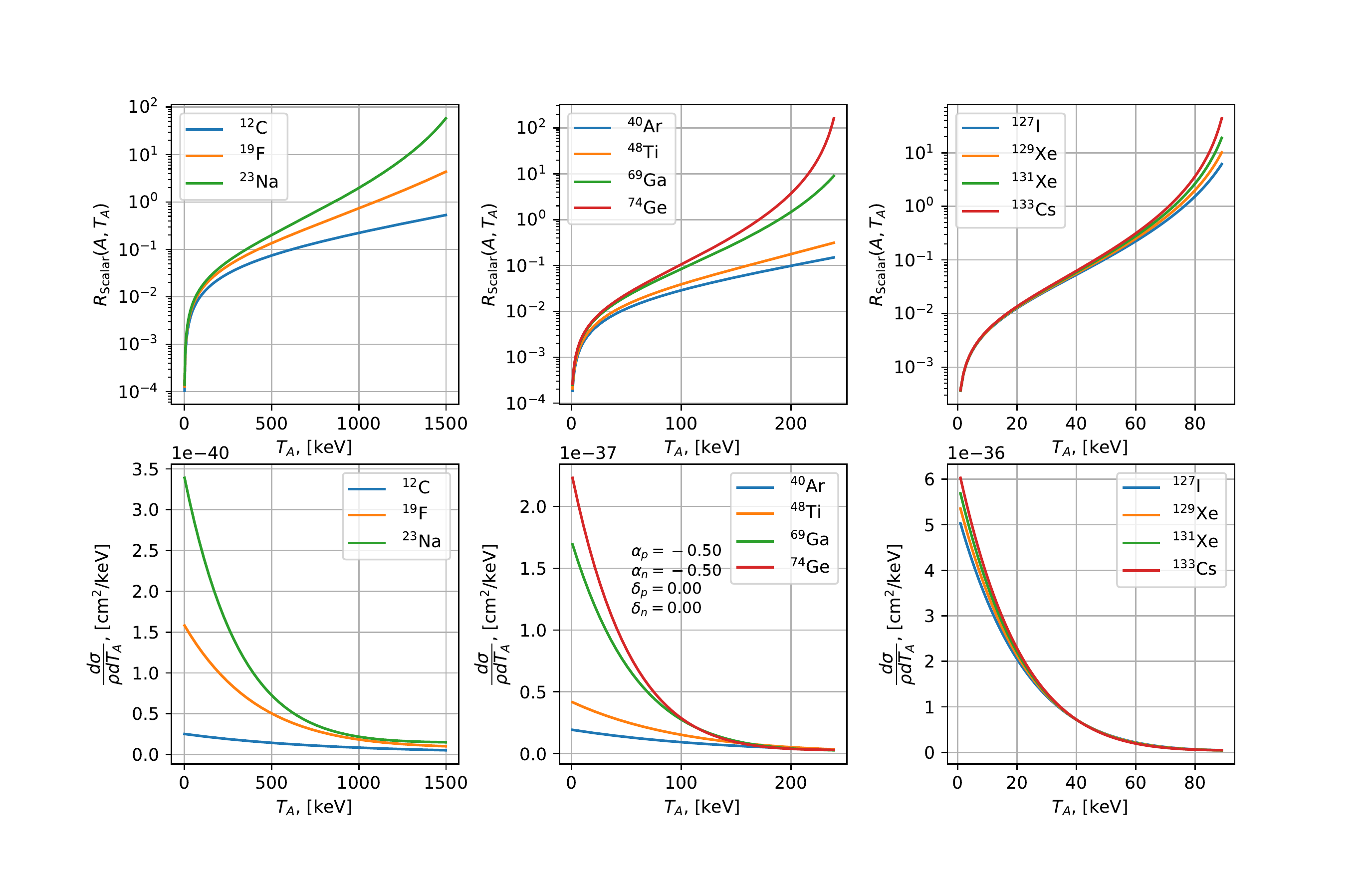}
 \vspace*{-40pt}
\caption{Ratio $ R^{\text{total}}_{\text{Scalar}}(A,T_A)$ 
(\ref{eq:50chiA-Results-Inc2CohRatio-Scalar})  of the total nonrelativistic cross sections for the incoherent and $\chi A$ coherent  scattering due to weak scalar interaction when $a_p= a_n$, 
(dependence on the square of these parameters is preserved only in the cross section) (top row) 
and measured total differential cross sections (\ref{eq:50chiA-Results-CrossSection-Weak-nonrel-via-Coh+R})
(bottom row) for a few target nuclei as a function of the recoil energy $T_A$.}
\label{fig:50chiA-Ratios-RA+DS_rho-vs-T_A-ScalarIsoScalar}
\end{figure}

{\em Scalar}\/, or {\em spin-independent}\/ 
in terms of the direct search for dark matter, interaction in the nonrelativistic
approximation has the same form as a pure vector interaction, when only the effective coupling constants
$\alpha_p$ and $\alpha_n$ remain different from zero in the general formula for the scalar product 
(\ref{eq:50chiA-Results-WeakScalarProdict-Definotion}). 
Then both axial coupling constants should be set equal to zero, $\delta_p=\delta_n=0$, 
in expression (\ref{eq:50chiA-Results-Inc2CohRatio-Weak}), 
which, after a change of notation  $a_p \equiv \alpha_p$ and $a_n \equiv \alpha_n$, 
takes the form
\begin{eqnarray} \label{eq:50chiA-Results-Inc2CohRatio-Scalar}
 R^{\text{total}}_{\text{scalar}}(A,T_A) &=&R_A(T_A) (A_p+A_n)\dfrac{A_p a^2_p+ A_n a^2_n}{(A_p a_p+A_n a_n  )^2} 
 .\end{eqnarray}
In the isoscalar case, where $a_p= a_n$, dependence on these constants is completely cancelled, 
 and this formula takes the form of expression  (\ref{eq:50chiA-Results-Inc2CohRatio-A}):
$R^{\text{total}}_{\text{scalar}}(A,T_A) =  R_A(T_A)$.
It is shown in Fig.~\ref{fig:50chiA-Ratios-RA+DS_rho-vs-T_A-ScalarIsoScalar} 
for different nuclei with the corresponding total cross sections.
\par 
If one of the parameters is much smaller than the other, as is, for example, in the SM, 
where $\alpha_p \ll \alpha_n$ (or for simplicity, $\alpha_p \simeq 0$), then
$$ R^{\text{total}}_{\text{scalar}}(A,T_A)= R_A(T_A) \dfrac{A_p+A_n}{A_n}=\dfrac{1-F^2_A(T_A) }{A_n F^2_A(T_A)}, $$
which does not depend on $\alpha_n $ either and only differs from $R_A(T_A)$ by the factor of about 2.  
\par
\begin{figure}[h!] 
\vspace*{-10pt}\hspace*{-20pt}
\includegraphics[width=1.1\linewidth]{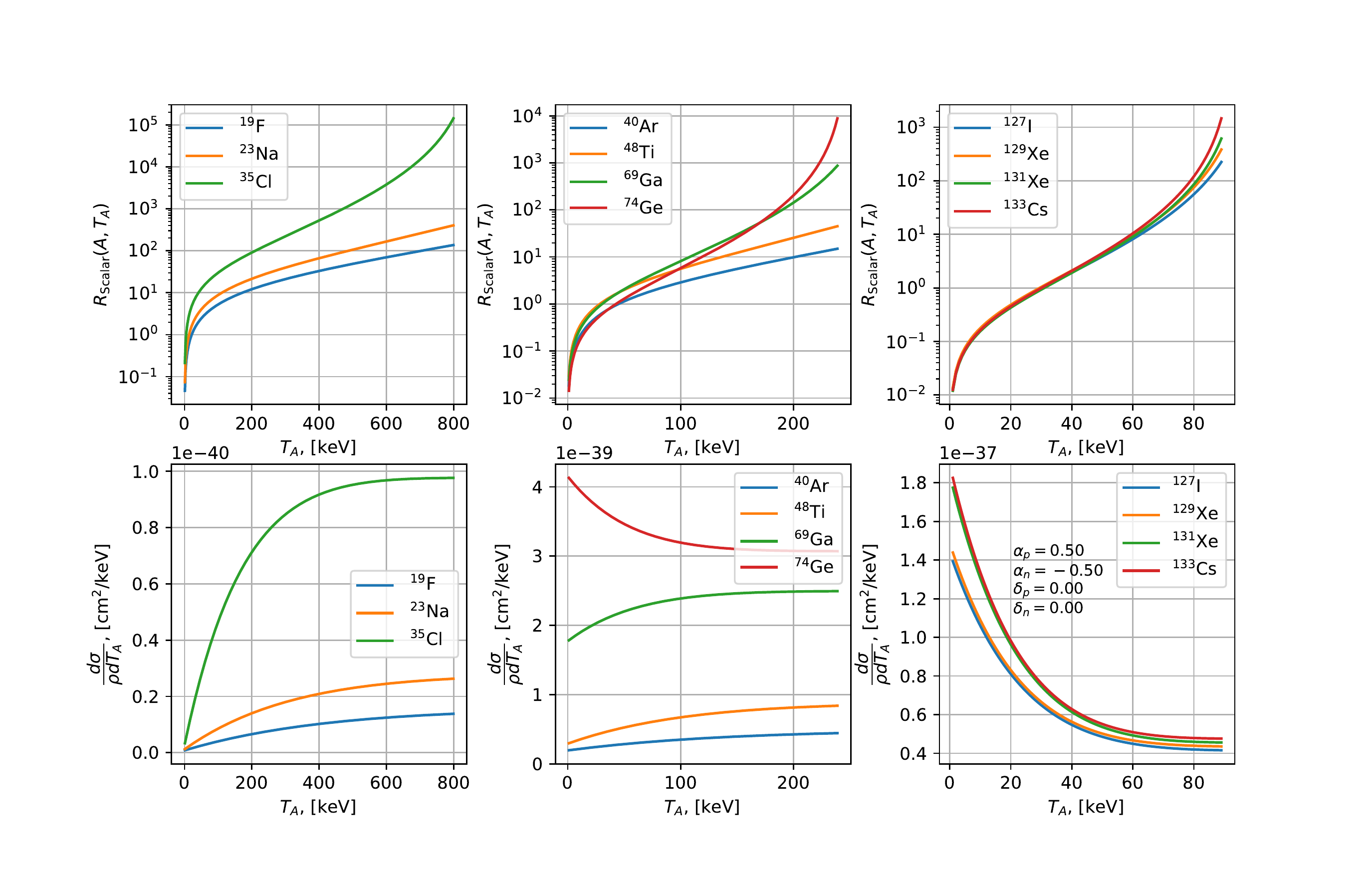}  \vspace*{-40pt}
\caption{Ratio $ R^{\text{total}}_{\text{Scalar}}(A,T_A)$ 
(\ref{eq:50chiA-Results-Inc2CohRatio-Scalar}) of the total nonrelativistic cross sections for the incoherent and coherent $\chi A$ scattering due to a weak scalar interaction when $a_p= - a_n$  (top row) 
and measured total differential cross sections  (\ref{eq:50chiA-Results-CrossSection-Weak-nonrel-via-Coh+R}) (bottom row) for a few target nuclei as a function of the nuclear recoil energy $T_A$.}
\label{fig:50chiA-Ratios-RA+DS_rho-vs-T_A-ScalarAntiIsoScalar}
\end{figure}

In the {\em anti}\/-isoscalar case, where $a_n= - a_p$ and $p =\dfrac{A_n}{A_p}$, one obtains the following:
$$ R^{\text{total}}_{\text{scalar}}(A,T_A)=R_A(T_A)(A_p+A_n)\dfrac{A_p a^2_p+ A_n a^2_p}{(A_p a_p-A_n a_p  )^2}=R_A(T_A)\Big(\dfrac{p+1}{p-1}\Big)^2. 
$$
It is evident from Table~\ref{tab:50chiA-Ratios-RIad-vs-A}  that in this case there is not any coherent contribution for the lightest nuclei, since the number of protons coincides with the number of neutrons, and $p=1$. 
Further, the heavier the target nucleus, the larger $p$, and the smaller the difference
of $R^{\text{total}}_{\text{scalar}}(A,T_A)$ from $R_A$ presented in 
Fig.~\ref{fig:50chiA-Ratios-RA+DS_rho-vs-T_A-ScalarAntiIsoScalar}.
\par 
Figures~\ref{fig:50chiA-Ratios-RA+DS_rho-vs-T_A-ScalarIsoScalar} and 
\ref{fig:50chiA-Ratios-RA+DS_rho-vs-T_A-ScalarAntiIsoScalar} reveal the “main trend”: as $T_A$ 
increases, the “content” of the total measured cross section smoothly changes --- 
the elastic (coherent) contribution is replaced by the inelastic (incoherent) one.
The cross section itself decreases, but by no more than an order of magnitude, which, to our mind, barely
reduces the probability of its measurement (detection).
That is, the total expected number of events to be detected falls within the capability limits of a typical experiment. 
However, these events change their origin.
If a detector is set up to detect only elastic scattering, it grows incapable of sensing anything at all as
$T_A$ increases (elastic processes become fewer and fewer in number). 
At the same time, the number of inelastic processes noticeably increases, while being undetectable
by the detector (say, because of its inability to detect photons from nuclear deexcitation). 
A situation arises where desired interactions occur but the detector cannot see them. 
Note, that contrary to the statements in 
\cite{Sahu:2020kwh} and \cite{McCabe:2015eia}
\footnote{Where it is stated that the scalar interaction is noticeably weaker
than the spin-dependent one, because the initial and final
nuclear states change, and only the spin-dependent interaction
should be considered.}, 
the scalar $\chi A$ interaction in the considered version may well make a noticeable
contribution to inelastic $\chi A$ scattering.
 \par
\begin{figure}[h!] 
\vspace*{-20pt}
\hspace*{-20pt}
\includegraphics[width=1.1\linewidth]{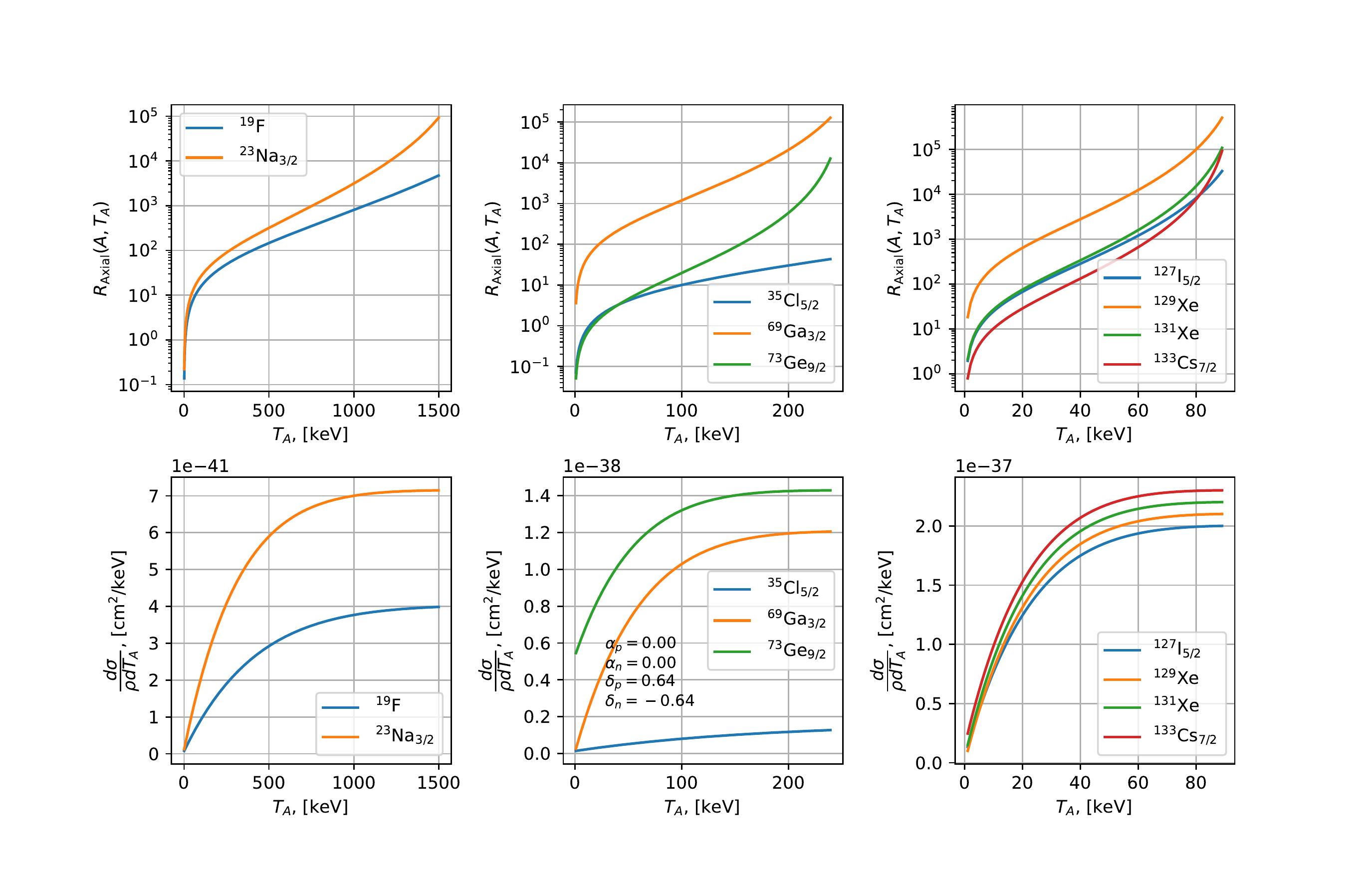}
 \vspace*{-40pt}
\caption{Ratio $ R^{\text{total}}_{\text{SM,axial}}(A,T_A)$ 
(\ref{eq:50chiA-Results-Inc2CohRatio-Axial})  
of the total nonrelativistic cross sections for the incoherent and coherent $\chi A$ scattering due to
a weak axial interaction when $\delta_n=-\delta_p$, 
(top row) and
measured total differential cross sections  (\ref{eq:50chiA-Results-CrossSection-Weak-nonrel-via-Coh+R}) (bottom row) for a few target nuclei as a function of the nuclear recoil energy $T_A$.}
\label{fig:50chiA-Ratios-RA+DS_rho-vs-T_A-AxialSM}
\end{figure}

 Let us turn to the pure {\em axial vector}\/ $\chi A$ interaction
referred to as {\em  spin-dependent}\/ in terms of the direct search for dark matter. 
 Then, the proton and neutron vector effective coupling constants in general formulas 
 (\ref{eq:50chiA-Results-Inc2CohRatio-Weak}) and (\ref{eq:50chiA-Results-CrossSection-Weak-nonrel-via-Coh+R})  are $\alpha_p=\alpha_n=0$,  and formula (\ref{eq:50chiA-Results-Inc2CohRatio-Weak})  takes the form
 \begin{eqnarray} \label{eq:50chiA-Results-Inc2CohRatio-Axial}
R^{\text{total}}_{\text{axial}}(A,T_A)  &=& 3 R_A(T_A)(A^p + A^n)
\dfrac{A^p \delta^2_p + A^n \delta^2_n} {(\delta_p \Delta A_p + \delta_n \Delta A_n)^2}
 . \end{eqnarray}
From the above and from formula (\ref{eq:50chiA-Results-CrossSection-Weak-nonrel-via-Coh+R})
 it is evident that for spin-zero nuclei (more specifically, when $\Delta A_{p/n}=0$) 
 there is no coherent contribution to the total cross section from the pure axial vector interaction,
and this ratio loses its meaning. 
Since the axial constants are connected by the relation $\delta_n=-\delta_p$,
in the SM, the ratio
 $$  R^{\text{total}}_{\text{SM, axial}}(A,T_A)= \dfrac{3 R_A(T_A) A^2} {(\Delta A_p - \Delta A_n)^2}
=\dfrac{1-F^2_A(T_A) }{F^2_A(T_A)}\dfrac{3 A} {(\Delta A_p - \Delta A_n)^2}
 $$
 does not depend on the axial coupling constant $\delta_p$ and, 
 which is more significant, is not suppressed by $A$, but, on the contrary, is directly proportional to it 
 if $\Delta A_p \ne \Delta A_n$ (see Fig.~\ref{fig:50chiA-Ratios-RA+DS_rho-vs-T_A-AxialSM}). 
 Otherwise, relation (\ref{eq:50chiA-Results-Inc2CohRatio-Axial}) loses its meaning again.
\par
\begin{figure}[h!] 
\hspace*{-20pt}
\includegraphics[width=1.1\linewidth]{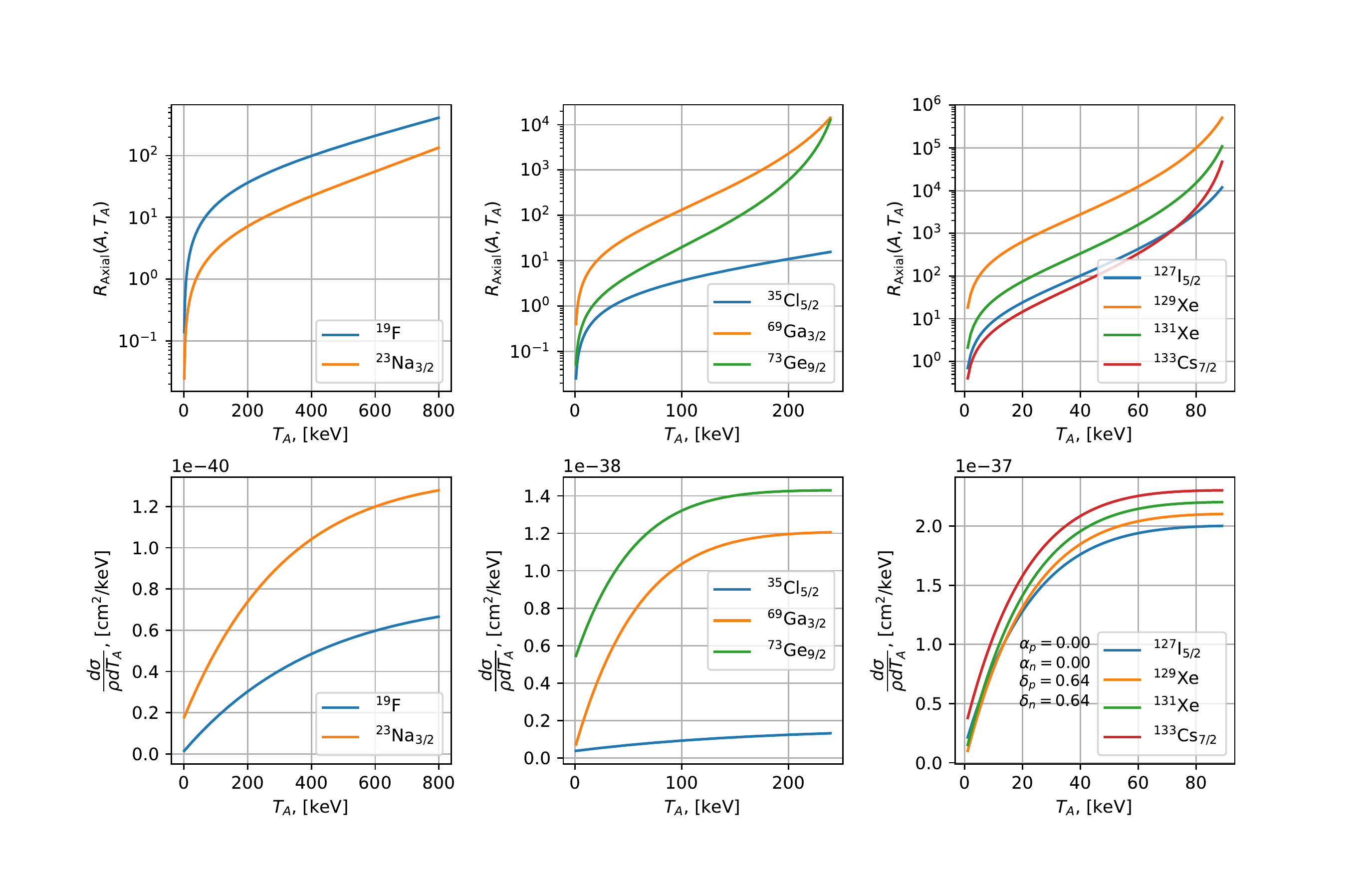}
 \vspace*{-40pt}
\caption{Ratio $ R^{\text{total}}_{\text{spin,axail}}(A,T_A)$ 
(\ref{eq:50chiA-Results-Inc2CohRatio-Axial}) 
of the total nonrelativistic cross sections for the incoherent and coherent $\chi A$ scattering due to
a weak axial interaction when $\delta_n=\delta_p$
(top row) and measured total differential cross sections  
(\ref{eq:50chiA-Results-CrossSection-Weak-nonrel-via-Coh+R}) (bottom row) for a few target
nuclei as a function of the nuclear recoil energy $T_A$.}
\label{fig:50chiA-Ratios-RA+DS_rho-vs-T_A-AxialIsoAxial}
\end{figure}

  
When $\delta_n=\delta_p$, dependence on the particular value of the axial coupling constant also drops out of the ratio (\ref{eq:50chiA-Results-Inc2CohRatio-Axial}),  and it takes the (similar) form
 \begin{eqnarray*}
R^{\text{total}}_{\text{spin,axial}}(A,T_A)&=&  \dfrac{3 R_A(T_A)A^2} {( \Delta A_p + \Delta A_n)^2}
=\dfrac{1-F^2_A(T_A) }{F^2_A(T_A)}\dfrac{3 A} {|\Delta A|^2}
, \end{eqnarray*}
where $\Delta A \equiv \Delta A_p +\Delta A_n$ plays the role of the total spin
of the nucleus $A$. 
This case can be treated as a complete analog of the traditional spin-dependent interaction
of the (dark-matter) $\chi$ particle with the detector material since the coherent cross section 
is here proportional to the square of the nuclear spin. 
Figure~\ref{fig:50chiA-Ratios-RA+DS_rho-vs-T_A-AxialIsoAxial} shows the corresponding plots.
\par
Figures ~\ref{fig:50chiA-Ratios-RA+DS_rho-vs-T_A-AxialSM} and
\ref{fig:50chiA-Ratios-RA+DS_rho-vs-T_A-AxialIsoAxial} reveal that the “general trend” in
the balance of coherence and incoherence mentioned above for the scalar interaction becomes a “dominant.”
Indeed, if a $\chi$ particle interacts with nucleons of the nucleus in a pure axial vector way, the target 
consisting of spin-zero nuclei does not allow observing interactions of this particle in the coherent (elastic)
channel because it is simply absent. 
Detection is only possible through observation of its inelastic (incoherent) interaction with nucleons. 
Moreover, even if the target consists of nonzero-spin nuclei, the coherent (elastic, proportional to square of nuclear spin) contribution to the measured cross section is “invisible” against the background of the incoherent (inelastic, proportional to atomic mass of nucleus $A$) contribution
in the almost entire interval of all possible  $T_A$. 
\par
Thus, in the case of the pure axial vector $\chi A$ interaction, detectors traditionally searching for the spin dependent signal of dark matter through detection of the elastic scattering channel depending on the nonzero
nuclear spin are doomed to see nothing at almost any $T_A$, 
since the entire “signal-producing” scattering proceeds via the inelastic channel to which detectors
of this kind are usually insensitive. 
Normally, detectors aimed at the direct search for dark matter have the maximum sensitivity in the region of the energy threshold for detection of the nuclear recoil energy. 
As a rule, they are unable to adequately detect a highenergy (as compared to the indicated threshold) 
radiation arising from deexcitation of nuclei excited due
the inelastic nature of the $\chi A$ interaction. 
A situation may arise again where desired interactions can have noticeable (potentially detectable) intensity, but the detector used to search for them is incapable of detecting them.
 \par 
 The character of domination of inelastic over elastic spin-dependent interaction between dark-matter
particles and   $^{129,131}$Xe nuclei was studied in  \cite{Baudis:2013bba}, 
 where the contribution of the entire inelastic channel was governed by excitation of one low-lying level
 \footnote{The first excited $3/2^+$ state of $^{129}$Xe is at 39.6 keV (half-life 0.97 ns) above the ground 
 $1/2^+$ state \cite{TIMAR2014143}. 
 The first excited $1/2^+$ state of $^{131}$Xe has the energy of 80.2 keV above the ground
$3/2^+$ state (0.48 ns) \cite{KHAZOV20062715}. 
}, 
and the detected (scintillation) signal was a sum of the recoil energy contribution and the contribution from nuclear deexcitation photons.
 \par
Methods for observing the incoherent signal from the neutrino–nucleus scattering were discussed in
\cite{Bednyakov:2018mjd,Bednyakov:2021ppn,Bednyakov:2021bty}. 
For a massive nonrelativistic $\chi$ lepton, there is nothing fundamentally different. 
After interacting with it, the nucleus either remains in the same quantum
state (elastic scattering) or its internal quantum state changes (inelastic scattering). 
When the experimental setup can measure only the kinetic nuclear recoil energy, it is impossible to know from it whether the nucleus has remained in its initial state or the transition
to the excited level occurred.
\par
If the transition to the excited nuclear state is energetically possible and has occurred, i.e., the inelastic
interaction did take place
\footnote{This “classical” inelastic approach should not be mixed up with “inelasticity” caused by the transition 
of the incident  $\chi_1$ lepton (of dark matter) to a more massive  $\chi_2$ lepton (also from dark sector). 
The nucleus is thought of as being unchanged, i.e., interacting coherently. See, for example
\cite{Giudice:2017zke,XENON:2020fgj,Zurowski:2020dxe,Baryakhtar:2020rwy,Feng:2021hyz,Filimonova:2022pkj,Bell:2022yxn,Aboubrahim:2022lwb}.},  
the nucleus should eventually return to its initial (ground) state. 
This unavoidable deexcitation of the nucleus should be accompanied
by energy release, for example, in the form of $\gamma$ radiation. 
Interestingly, the possibility of detecting gamma quanta from this inelastic interaction was suggested
as far back as 1975~\cite{Donnelly:1975ze}.
The energy spectrum of these photons is dictated by the structure of nuclear excitation levels and is strictly fixed for each target nucleus. 
These photons can produce a detectable signal \cite{Donnelly:1975ze}, 
 which, generally speaking, will be correlated with the target irradiation time if the source of
 $\chi$ leptons is, for example, an accelerator. 
 According to the above figures (see, for example, Fig.~\ref{fig:50chiA-Ratios-RA+DS_rho-vs-T_A-AxialIsoAxial}), the prospects for detecting a noticeable (for axial case, overwhelming)
number of events with $\gamma$ quanta from the incoherent $\chi A$ interaction do not seem 
 futile provided that the $\chi A$ interaction itself has detectable intensity for modern
detectors.
\par
It should be stressed that for each specific target nucleus these photons will be characterized by three
important parameters. 
First, their energy is often noticeably higher than the kinetic nuclear recoil
energy ($T_{\text{Cs}}\le $ 40 keV in Fig.~\ref{fig:50chiA-Ratios-RA+DS_rho-vs-T_A-AxialIsoAxial}). 
Second, emission of photons with the energy dictated by the difference of the nuclear excitation levels will be shifted in time (relative to the beginning of the interaction specified by one or another method) by the deexcitation time typical of the level of a particular nucleus. 
For example, emission of these $\gamma$ quanta for the ${}^{133}\text{Cs}$ nucleus occurs within an interval from a few picoseconds to nanoseconds, and their energies are in the region of a few hundreds of keV. 
Third, the counting rate of these $\gamma$ quanta will be proportional to the ratio
$N_\text{inc}/N_\text{coh}$, where 
$$N_\text{inc/coh} = \int dE_\nu \Phi(E_\nu)\int_{dT_A^\text{min}}^{dT_A^\text{max}} dT_A\frac{d\sigma_\text{inc/coh}}{dT_A} \varepsilon(T_A),$$ 
and $\varepsilon(T_A)$  is the detector efficiency. 
However, the possibility of constructively using these properties to discriminate the desired 
$\chi A$ inelastic interaction from the background seems to be a topic for a separate consideration.
\par 
Let us discuss the accuracy of the estimated inelastic $\chi A$ cross section. 
As shown earlier for the (anti)neutrino-nucleus scattering 
\cite{Bednyakov:2018mjd,Bednyakov:2019dbl,Bednyakov:2021ppn}, this (combined)
inelastic cross section is the upper boundary for contributions to the total measured $\chi A$ cross section
from the totality of all allowed inelastic subprocesses (in the given kinematic region). 
This upper limit only follows from the probability conservation rule: the sum of the probabilities of all inelastic processes and the probability of the elastic process is equal to unity.
\par
The degree of “saturation” of this combined inelastic cross section by individual inelastic channels
(contributions from transitions to different allowed levels) depends on the structure of excitation levels of
a particular nucleus and on the incident particle energy. 
Clearly, this saturation cannot be 100\% reproduced numerically. 
Inelastic cross sections for the $\nu(\bar{\nu}) A$ scattering due to weak neutral currents 
were calculated within modern nuclear models for specific excitation levels of particular nuclei, for example, in
 \cite{Divari:2012zz,Divari:2012cj,Lykasov:2007iy}. 
 Recently, similar calculations have been generalized to inelastic scattering of neutral dark-matter
particles by nuclei \cite{Sahu:2020kwh,Sahu:2020cfq,Dutta:2022tav}.
\par
In \cite{Sahu:2020kwh}, detailed calculations were performed for cross sections and counting rates of elastic and inelastic scattering events on the  $^{73}$Ge, $^{127}$I, $^{133}$Cs and $^{133}$Xe nuclei. 
Similar calculations for $^{23}$Na and $^{40}$Ar were performed in \cite{Sahu:2020cfq}. 
In the case of the neutrino-nucleus scattering, the authors used formulas from
\cite{Bednyakov:2018mjd,Bednyakov:2019dbl,Bednyakov:2021ppn}. 
Nuclear effects were considered within the deformed shell model. 
With the lowest levels taken as an example
\footnote{In $^{127}$I, the first excited state $7/2^+$ is 57.6 keV above the ground state. 
In $^{133}$Cs, the lowest state $5/2^+$ has the energy of 81 keV.}, 
 it was shown within this nuclear model that incoherent neutrino–nuclear processes can noticeably
enhance the expected signal (above the energy threshold) in a modern dark-matter detector. 
 Similar enhancement was shown for inelastic WIMP-nucleus scattering. 
Thus, the above works confirmed importance of the inelastic channel at high recoil energies
first indicated in \cite{Bednyakov:2018mjd,Bednyakov:2019dbl,Bednyakov:2021ppn}.
Still, it is worth noting that expressions obtained in this work are necessary for correct analysis of the balance of coherence and incoherence in the WIMP-nucleus scattering.
\par 
In \cite{Dutta:2022tav}, it was pointed out that neutrinos with an energy of several tens of MeV can excite many various levels in nuclei of detector construction materials in different experiments. 
The authors believe that though the inelastic scattering cross section is much smaller that the coherently enhanced elastic scattering cross section, investigation of inelastic processes is an important source of additional information in searching for new physics, revealing background conditions for the search, etc. 
In addition, the understanding of inelastic neutrino-nucleus scattering is particularly important for detecting a supernova signal by the newgeneration neutrino detectors, such as DUNE 
 \cite{DUNE:2020zfm} and Hyper-K \cite{Hyper-Kamiokande:2016srs}. 
 There are also interesting models with higher significance of inelastic interaction  \cite{Arcadi:2019hrw}.
Discussing earlier estimations of inelastic $\nu A$ scattering, the authors of 
 \cite{Dutta:2022tav} hold that in our work \cite{Bednyakov:2018mjd}
nuclear structure details were neglected, 
 and in  \cite{Sahu:2020kwh} only the lowest nuclear levels were considered.
According to them, the free nucleon approximation used in
 \cite{Bednyakov:2018mjd}  is especially inadequate for estimation of
inelastic scattering in the discussed energy region,
because it entirely ignores the nuclear structure\footnote{Note that nucleons in \cite{Bednyakov:2018mjd}  are not free, and the nuclear structure is considered in terms of nuclear form factors.}. 
\par
\begin{figure}[h!] 
\includegraphics[width=0.9\linewidth]{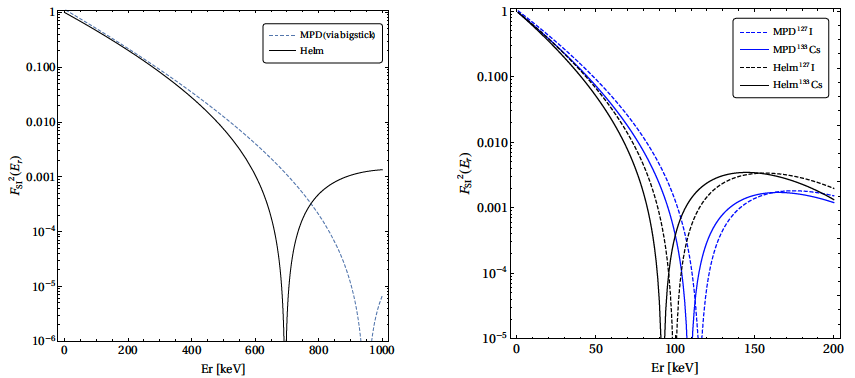} 
\caption{Difference between the parametrization of the Helm elastic form factors and the calculations within the shell model \cite{Dutta:2022tav}.
Left panel: for $^{40}$Ar; right panel: for  $^{133}$Cs and $^{127}$I. From \cite{Dutta:2022tav}.}
\label{fig:50chiA-Dutta2022tav-Fig12}
\end{figure}

 
In \cite{Dutta:2022tav}, on the basis of the formalism of the semileptonic electroweak theory 
\cite{DeForest:1966ycn,Serot:1978vj,Donnelly:1979ezn}, inelastic cross sections for scattering of neutrinos and dark matter particles (due to neutral weak currents) off the $^{40}$Ar, $^{133}$Cs and $^{127}$I.
nuclei were calculated within the nuclear shell model. 
In this formalism, the hadronic current is expanded in multipoles for obtaining irreducible
tensor operators that act on single-particle states. 
The (total) inelastic cross section for the neutrino-nucleus scattering at rather low neutrino energies
($E_\nu$ < 20 MeV) obtained by the authors was in agreement with other calculations 
(including \cite{Bednyakov:2018mjd}), but at $E_\nu \simeq$  40 MeV 
their cross sections were about an order of magnitude smaller than the upper limits obtained in \cite{Bednyakov:2018mjd}) for all inelastic channels. 
There seems to be no contradiction. 
In Fig.~\ref{fig:50chiA-Dutta2022tav-Fig12} from \cite{Dutta:2022tav}, 
it is seen that the Helm parametrization for nuclear form factors (used in \cite{Bednyakov:2018mjd}) works quite well in the entire significant kinematic region where the square of the form factor is not smaller than $0.01$.
For argon, differences in the behavior of squares of form factors arise at the nuclear recoil  energy 
$E_r \simeq 600$~keV (positions of minima are noticeably different), 
but to this difference there corresponds the value of the square of the form factor $0.001$, 
which makes this difference almost imperceptible. 
For heavier $^{133}$Cs and $^{127}$I nuclei, this difference at  $E_r \simeq 80\div 100$  keV seems even more insignificant than for $^{40}$Ar due to smallness of the form factor values themselves.
 \par
 In this connection, two factors are noteworthy.
First, in this work, as in the approach \cite{Bednyakov:2018mjd,Bednyakov:2019dbl,Bednyakov:2021ppn}, 
 the Helm parametrization is used to calculate form factors defined in 
(\ref{eq:3chiA-ScatteringAmplitude-f^k_mn-definition})
 in terms of the matrix elements of the shift operator $\langle m|e^{i\bm{q}\hat{\bm{X}}_{k}} |n\rangle$, 
 without recourse to the expansion in spherical functions and multipoles like 
\cite{SajjadAthar:2022pjt}
$\displaystyle e^{i\bm{q}{\bm{x}}_{}}= \sum^{\infty}_{l=0}\sqrt{4\pi (2l+1)} i^l j_l(qx) Y_{l0}(\theta,\phi).$ 
\par
In other words, it can be stated that in the approach
\cite{Bednyakov:2018mjd,Bednyakov:2019dbl,Bednyakov:2021ppn}, 
unlike the case in \cite{Dutta:2022tav}, entirely all multipoles “work,” and none of their contributions is ignored.
Second, in the approach \cite{Bednyakov:2018mjd,Bednyakov:2019dbl,Bednyakov:2021ppn} nothing prevents
replacing the simple and convenient Helm parametrization with more complicated expressions from 
 \cite{Dutta:2022tav} (see Fig.~\ref{fig:50chiA-Dutta2022tav-Fig12}) in calculations of form factors. 
 However, these parametrizations are not much different in the
significant region; the difference becomes more or less
noticeable only when their squares drop below 0.01,
and therefore, this replacement obviously gives a correction
at the level of  $O(0.01)$.
\par 
Thus, it is seen that calculations of inelastic cross sections using modern nuclear models yield results
that are always smaller than estimates obtained in this work. 
As was already mentioned, this is because the formalism 
\cite{Bednyakov:2018mjd,Bednyakov:2019dbl,Bednyakov:2021ppn} 
gives the upper bound for the total inelastic cross section 
(i.e., where all possible inelastic channels are included), 
since the probability of all inelastic scattering channels taken together is obtained
by subtracting the probability of a single elastic reaction channel from unity.
\par
Note that results of precision calculations of inelastic cross sections within advanced nuclear models do
not allow the transition from “coherence to incoherence” to be quantitatively controlled. 
This means that formulas to be used for description and analysis of the data obtained in each kinematic region are chosen by the discretionary decision based on preliminary estimations of coherence condition satisfiability. 
In the formalism \cite{Bednyakov:2018mjd,Bednyakov:2019dbl,Bednyakov:2021ppn} it is not necessary to do any estimations before using formulas, since the transition from one regime to the other is smooth here.
\par 
To complete the picture, recall the results of searching for inelastic interaction of weakly interacting
neutral massive (dark-matter) particles by the XENON1T Collaboration 
\cite{XENON:2020fgj}, who consider the $^{129}$Xe nucleus as being most appropriate for the search
for inelastic WIMP-nucleus interaction, the signature of which is a 39.6-keV deexcitation photon (recoil
electron) detected simultaneously with the nuclear recoil energy. 
Since no evidence for this inelastic process has been found yet, the most significant upper
limit of $3.3\times 10^{-39}$ cm$^2$ was obtained for dark-matter
particles with the mass of 130 GeV/$c^2$. 
Nevertheless, the authors stress that detection of inelastic scattering
would give stronger limits on properties of dark matter
than observation of only elastic interaction \cite{XENON:2020fgj}.

\section{\large Conclusion}
\label{60chiA-Conclusions} 
Let us briefly formulate the key issues discussed.
\par 
(i) The formalism \cite{Bednyakov:2018mjd,Bednyakov:2019dbl,Bednyakov:2021ppn}, 
first proposed for describing neutrino–nucleus interaction was generalized to the case of nonrelativistic\footnote{
The relativistic version of this description is considered separately \cite{Bednyakov:2022rel}.}
 weak interaction of a massive neutral particle ($\chi$ lepton) with a nucleus as a compound system. 
 This interaction is parametrized in the form of four free parameters—effective coupling constants—that determine contributions to the probability amplitude from scalar products of leptonic and nucleon currents.
\par 
(ii) Within this approach, by considering all possible initial and final nuclear quantum states (completeness condition), it becomes possible to obtain a rather general (i.e., independent of details of the nuclear model) unified description of elastic (coherent) and inelastic (incoherent) processes of neutral $\chi$-lepton scattering off nuclei.
\par 
(iii) As in the case of (anti)neutrino–nucleus scattering, the behavior of the elastic (coherent) and inelastic (incoherent) 
$\chi A$ cross sections is respectively governed by the factors  $F_{p/n}(\bm{q})|^2$ and  $(1-|F_{p/n}(\bm{q})|^2)$, where $F_{p/n}(\bm{q})$ is the proton/neutron form factor normalized to unity and averaged over the initial nuclear state. 
These form factors govern a smooth transition from the elastic (coherent) to the inelastic (incoherent) scattering regime. 
Domination of the elastic (coherent) or inelastic (incoherent) term in the observed cross section is determined by the relation
between the values of the factors  $A^2_{p/n}|F_{p/n}(\bm{q})|^2$ and $A_{p/n} {\left(1-|F_{p/n}(\bm{q})|^2\right)}$.
depending on the momentum $\bm{q}$ transferred to the nucleus.
\par
(iv) Figures~\ref{fig:50chiA-Ratios+FF+DS-vs-T_A-4All} and \ref{fig:50chiA-Ratios+FF+DS-vs-T_A-4light}
(obtained with the effective coupling constants of weak interaction in the Standard Model) illustrate the main property of the approach
\cite{Bednyakov:2018mjd,Bednyakov:2019dbl,Bednyakov:2021ppn}. 
As the detected nuclear recoil energy $T_A$  increases, a well-controlled transition occurs from the domination of the 
elastic $\chi A$-interaction to the domination of the inelastic $\chi A$ interaction in nonrelativistic scattering 
of a {\em massive}\/ $\chi$ particle off a nucleus. 
 \par 
In other words, as $T_A$ increases, detected events will change their origin. 
If the detector is set up to detect only elastic scattering events, it starts to lose capability of “seeing” anything with 
increasing $T_A$  (elastic processes become fewer and fewer in number). 
The number of inelastic processes undetectable by this detector noticeably increases. 
For example, it cannot detect photons from nuclear deexcitation or with energies beyond the detection zone of the detector. 
A situation arises where desired interactions occur, but the detector is “blind” to them. 
In its most critical form, this “phenomenon” can manifest itself in the problem of direct search for dark matter, 
the results of which are usually interpreted in terms of spin-independent and spin-dependent cross sections 
for interaction of a particle of the dark matter halo of our galaxy with nucleons.
\par
(v) For pure scalar (spin-independent) interaction, the above mentioned smooth change in the “content”
of the total measured cross section with increasing $T_A$ is seen in 
Figs.~\ref{fig:50chiA-Ratios-RA+DS_rho-vs-T_A-ScalarIsoScalar} and 
\ref{fig:50chiA-Ratios-RA+DS_rho-vs-T_A-ScalarAntiIsoScalar}, while the cross section itself
decreases insignificantly, remaining within the capability limits of a typical experiment.
\par 
Further, if a (nonrelativistic) $\chi$ particle interacts with nucleons of the nucleus {\em only}\/ 
in the axial vector way (spin-dependent), interaction of this particle with spin-zero nuclei via the coherent (elastic) channel is impossible to observe. 
Detection is only possible through revealing its inelastic (incoherent) interaction with nucleons. 
Moreover, even if the nucleus has a nonzero spin, the coherent (elastic, proportional to the square of the nuclear spin) contribution to the measured cross section is poorly distinguishable against the background of the incoherent (inelastic, proportional to the atomic mass of the nucleus $A$) contribution in the almost entire interval of possible values of  $T_A$ 
(Figs.~\ref{fig:50chiA-Ratios-RA+DS_rho-vs-T_A-AxialSM} and \ref{fig:50chiA-Ratios-RA+DS_rho-vs-T_A-AxialIsoAxial}).
\par
Thus, in the case of pure axial vector $\chi A$ interaction, detectors traditionally aimed at detecting the elastic spin-dependent 
signal of dark matter are doomed to see nothing, since the entire potentially “signal-producing” scattering proceeds via 
the inelastic channel, to which detectors of this kind are normally insensitive. 
Again, one cannot exclude a situation where the desired interactions may well have noticeable (potentially detectable) intensity, 
but the detector used to search for them is incapable of observing them.
\par
(vii) Another situation is also possible. 
If the nuclear recoil energy detected after the $\chi A$ scattering is below the detection threshold of the detector, i.e.,
$T_A< T_A^{\min}$, the elastic signal can by no means be observed. 
With these “invisible”  $T_A$  the only evidence of the $\chi A$ interaction that occurred is radiation from deexcitation of a nucleus, 
that is, the inelastic signal, though its intensity at $T_A< T_A^{\min}$ can be an order of
magnitude lower than the intensity of the elastic signal. 
Generally, when only the nuclear recoil energy $T_A$ is detected, 
it is impossible to identify whether the process was elastic or inelastic. 
Worse yet, when the inelastic signal falls within the expected detection region of the elastic one, its origin is even more obscure.
\par
Thus, for the above reasons, experiments on direct detection of dark-matter particles should be planned to be such as to 
allow detection of two signals, the nuclear recoil energy and the nuclear deexcitation  $\gamma$ quanta. 
Recall that the energy of these $\gamma$ quanta is strictly specific for each nucleus in question, 
being equal to the difference between levels of excitation energies. 
The result of this experiment will give the fullest information on the $\chi A$ interaction.
\par
(viii) The set of expressions for $\chi A$ cross sections obtained in this work 
(formulas (\ref{eq:43chiA-CrossSections-via-ScalarProducts-Weak-nonrel-CohSC-Fp=Fn}), 
(\ref{eq:43chiA-CrossSections-via-ScalarProducts-Weak-nonrel-InCohSC-Fp=Fn}), and others) 
should be used for the adequate description of scattering of weakly interacting 
neutral $\chi$ particles off target nuclei when there are ways for increasing velocities of these particles by an order (or two) 
of magnitude compared to velocities of dark-matter particles in the halo of our galaxy, e.g., to the level of
$ |\bm{v}|/c={|\bm{k}|}/{c m_\chi} = 10^{-2} \div 10^{-1}$. 
A few such possibilities were suggested in the literature. For example, one is
generation of dark-matter particles with quite high energies (and even with relativistic velocities at the level of the Earth 
\cite{Feng:2021hyz}) due to various acceleration mechanisms in space 
\cite{Bardhan:2022ywd,CDEX:2022fig,Xia:2022tid,Granelli:2022ysi,Wang:2021jic}.
Another is production of rather energetic candidates for dark-matter particles at modern accelerators 
\cite{Krnjaic:2022ozp,Boos:2022gtt,Kim:2017qdi}.
Note that in this case, radiation from deexcitation of nuclei can be not only $\gamma$ quanta but also 
can have a richer structure involving, for example, neutrons or other particles from the breakup of the target nucleus.

\section*{Acknowledgments}
The author is grateful to V. Naumov, E. Yakushev, N. Russakovich, and I. Titkova for discussions and important comments.


\bibliographystyle{JHEP}
\bibliography{CohPapers.bib}

\end{document}